\documentclass[pra,aps,reprint,twocolumn,superscriptaddress,10pt,nofootinbib,longbibliography,floatfix]{revtex4-2}

\usepackage[utf8]{inputenc}
\usepackage[T1]{fontenc}
\usepackage[english]{babel}
\usepackage{graphicx,epsfig}
\graphicspath{{images/}}
\usepackage{color}
\usepackage[usenames,dvipsnames]{xcolor}
\usepackage{rotating}

\usepackage{longtable}
\usepackage{graphicx}

\usepackage{physics}
\usepackage{amsmath,amssymb,amsthm}
\usepackage{bbm,dsfont} 
\usepackage{stmaryrd}
\usepackage{enumitem}
\usepackage[caption=false]{subfig}
\usepackage[normalem]{ulem}
\usepackage{tikz}
\usetikzlibrary{quantikz2}

\newtheorem{theorem}{Theorem}
\newtheorem{definition}{Definition}

\usepackage{dsfont}
\usepackage{physics}

\usepackage{mathtools}
\usepackage{cancel}

\definecolor{linkcolor}{rgb}{0,0,0.6}	
\usepackage[colorlinks=true,pdfstartview=FitV,linkcolor=ForestGreen,citecolor=linkcolor,urlcolor=linkcolor,hyperindex=true,hyperfigures=false]{hyperref}

\usepackage{numprint}

\usepackage[bb=boondox]{mathalfa}

\usepackage{upgreek}
\usepackage{soul}

\newcommand{\kket}[1]{{|#1 \rangle \!\rangle}}
\newcommand{\bbra}[1]{{\langle \!\langle #1 |}}
\def\kketbra#1#2{\mathinner{|{#1}\rangle\!\rangle\!\langle\!\langle{#2}|}}

\newcommand{\id}{\mathbbm{1}}

\renewcommand\L{\mathcal{L}}

\renewcommand\L{\mathcal{L}}
\renewcommand{\H}{\mathcal{H}}
\newcommand{\M}{\mathcal{M}}
\newcommand{\N}{\mathcal{N}}

\newcommand{\A}{\mathcal{A}}

\newcommand{\map}[1]{\mathcal{#1}}

\newcommand{\QCFO}{\textup{\textsf{QC-FO}}}

\makeatletter 
    
\renewcommand\onecolumngrid{
\do@columngrid{one}{\@ne}%
\def\set@footnotewidth{\onecolumngrid}
\def\footnoterule{\kern-6pt\hrule width 1.5in\kern6pt}%
}

\renewcommand\twocolumngrid{
        \def\footnoterule{
        \dimen@\skip\footins\divide\dimen@\thr@@
        \kern-\dimen@\hrule width.5in\kern\dimen@}
        \do@columngrid{mlt}{\tw@}
}%

\makeatother    

\begin{document}

\title{Efficient quantum-circuit simulation of classical control of computational order}

\author{Raphaël Mothe}
\email{raphael.mothe@uni-siegen.de}
\affiliation{Naturwissenschaftlich-Technische Fakultät, Universität Siegen, Walter-Flex-Straße 3, 57068 Siegen, Germany}

\author{Jessica Bavaresco}
\email{jessica.bavaresco@lip6.fr}
\affiliation{Sorbonne Université, CNRS, LIP6, F-75005 Paris, France}

\date{\today}

\begin{abstract}
We show that quantum circuits with classical control of computational order admit an efficient simulation by fixed-order quantum circuits: any such transformation acting on $N$ independent quantum channels can be deterministically, exactly, and universally simulated using at most $N^2$ queries. We generalize these constructions to extended, bipartite quantum channels and quantum instruments, and demonstrate that they can be implemented catalytically, without consuming the additional $N^2-N$ queries required for the simulation. In the case of quantum circuits with quantum control of computational order, we prove a similar simulation result, but restricted to unitary input channels (hence non-universal). These results show the limitations of the computational power of these classes of transformations in the studied scenarios.
\end{abstract}

\maketitle

The quantum circuit model is one of the most widely applied models of quantum computing. A well-known such example is the quantum circuit that implements Shor’s algorithm, a paramount result in quantum computing about the efficient factoring of integer numbers~\cite{shor1994algorithms}. Quantum channels are central to this framework: traditionally regarded as devices transporting, and in the process, transforming quantum states---such as in communication tasks---they are also key characters of quantum circuits, which are composed of a sequence of quantum gates and measurements. The more recent formalism of higher-order quantum computing~\cite{Chiribella08,Chiribella09,Chiribella08supermaps,taranto2025higher} places quantum channels center stage, not only as the means to execute a computation through a quantum circuit, but as the input and output of the computation itself.

Formally, a higher-order transformation called a \textit{superchannel} is the most general linear map that transforms several quantum channels into another quantum channel \cite{Chiribella08,Chiribella08supermaps,Chiribella09,oreshkov12}. Hence, a quantum circuit with open slots, where arbitrary quantum channels can be plugged in, can be seen as a higher-order transformation. Since quantum channels carry an inherent input–output structure, the question of causal order naturally emerges when transforming multiple channels. Strikingly, this formalism allows superchannels that act on the input channels in a manner that goes beyond what can be described with the standard paradigm of quantum circuits, where quantum gates are implemented on a fixed, pre-defined order. These transformations play with the order in which the input channels are acted upon: they can be described with classical control of computational order~\cite{Chiribella13,wechs21} or even with quantum control of computational order~\cite{Chiribella13,wechs21}, leading to computations that can be understood as carried out in a dynamical and/or in an indefinite causal order \cite{BaumelerWolf,Oreshkov16,wechs21,mothe25}. Such transformations raise immediate fundamental questions about the role of causality in quantum theory, as well as more practical questions, concerning their implementability and potential advantages.

Theoretical advantages of quantum control of computational order have been confirmed in a wide range of quantum information processing tasks, such as quantum channel discrimination~\cite{chiribella12,bavaresco2021strict,bavaresco2022unitary}, quantum metrology~\cite{zhao2020quantum,Liu23,mothe23}, quantum computational and query complexity~\cite{araujo14,Abbott24}, among others. As for classical control, a so far less explored class of transformations, an advantage has been recently reported~\cite{bavaresco25} with respect to fixed-order strategies in a task of approximately simulating quantum control of computational order, and another was obtained with respect to convex mixtures of fixed-order strategies in a causal game \cite{mothe25}. Such information-processing advantages have mostly been shown via comparison with fixed-order quantum circuits that act on a single call of each input channel. However, their true practical significance depends on whether these advantages would still hold when comparing superchannels with quantum and classical control of computational order against fixed-order quantum circuits that use a larger number of queries to the input channels. Recent results have shown that a universal, deterministic, and exact simulation of the quantum switch, a superchannel with quantum control of computational order, is at least computationally hard, if at all possible~\cite{bavaresco25}.

In contrast, in this work, we show that classical control of computational order can be simulated in a quantum query-efficient manner: we prove that any quantum circuit with classical control of computational order that acts on $N$ independent quantum channels can be deterministically, exactly, and universally simulated by a fixed-order quantum circuit that acts on at most $N^2$ queries to the input quantum channels. Our proof is constructive and builds on top of a known quantum circuit that can simulate the action of the quantum $N$-switch on unitary channels~\cite{Colnaghi12,araujo14,facchini15}. We generalize these results to show that, at no extra cost in terms of channel queries, these simulations can be adapted to work for extended quantum channels and quantum instruments, and can furthermore be made catalytically, i.e., without consuming the $N^2-N$ extra queries of the input channels. Moreover, we show that quantum circuits with quantum control of computational order can also be simulated in a similar manner, albeit not universally: their action on $N$ \textit{unitary} quantum channels can be deterministically and exactly simulated with at most $N^2$ queries to the input unitary channels. Such simulations can also be generalized to extended unitary channels and be made catalytically, similarly to the classical control case. Beyond establishing efficient simulations, these results help to clarify the computational power in terms of quantum query complexity of the largest currently known physically interpretable classes of higher-order transformations.

\vspace*{5pt}

\noindent\textit{---Quantum channels and superchannels.}
In quantum theory, quantum channels are deterministic transformations of quantum states, described by completely positive, trace-preserving (CPTP) linear maps $\map{A}:\L(\H_I)\to\L(\H_O)$ that map all quantum states $\rho\in\L(\H_I\otimes\H_{I'})$ acting on a joint input and auxiliary Hilbert space to quantum states $\map{A}\otimes\map{I}(\rho)\in\L(\H_O\otimes\H_{I'})$ acting on a joint output and auxiliary Hilbert space, where $\map{I}$ is the identity map on the auxiliary system, even when acting only on part of the input state.  Unitary channels $\map{U}(\rho)=U\rho U^\dagger$, characterized by unitary operators $U:\H_I\to\H_O$, $\H_I\cong\H_O$, satisfying $U^\dagger U=\id$, are a particular case of quantum channel.

Superchannels, on the other hand, are higher-order transformations whose inputs and outputs are quantum channels themselves. That is, they are linear higher-order maps $\boldsymbol{S}:\bigotimes_{i=1}^N[\L(\H_{I_i})\to\L(\H_{O_i})]\to[\L(\H_{P})\to\L(\H_{F})]$ that act on part of $N$ independent quantum channels $\map{A}_i:\L(\H_{I_i}\otimes\H_{I'_i})\to\L(\H_{O_i}\otimes\H_{O'_i})$, where $\H_{I'_i}$ and $\H_{O'_i}$ are auxiliary input and output Hilbert spaces, and transform them into another channel $\boldsymbol{S}\otimes\boldsymbol{I}(\map{A}_1\otimes\ldots\otimes \map{A}_N):\L(\bigotimes_{i=1}^N\H_{I'_i}\otimes\H_{P})\to\L(\bigotimes_{i=1}^N\H_{O'_i}\otimes\H_{F})$, where $\boldsymbol{I}$ is the higher-order identity map acting on the auxiliary systems. To guarantee that the superchannels $\boldsymbol{S}$ transform quantum channels into quantum channels, even when acting only on part of the input channels, they must correspond to completely CPTP preserving linear maps.

\begin{figure}[t]
\centering
    \includegraphics[width=\columnwidth]{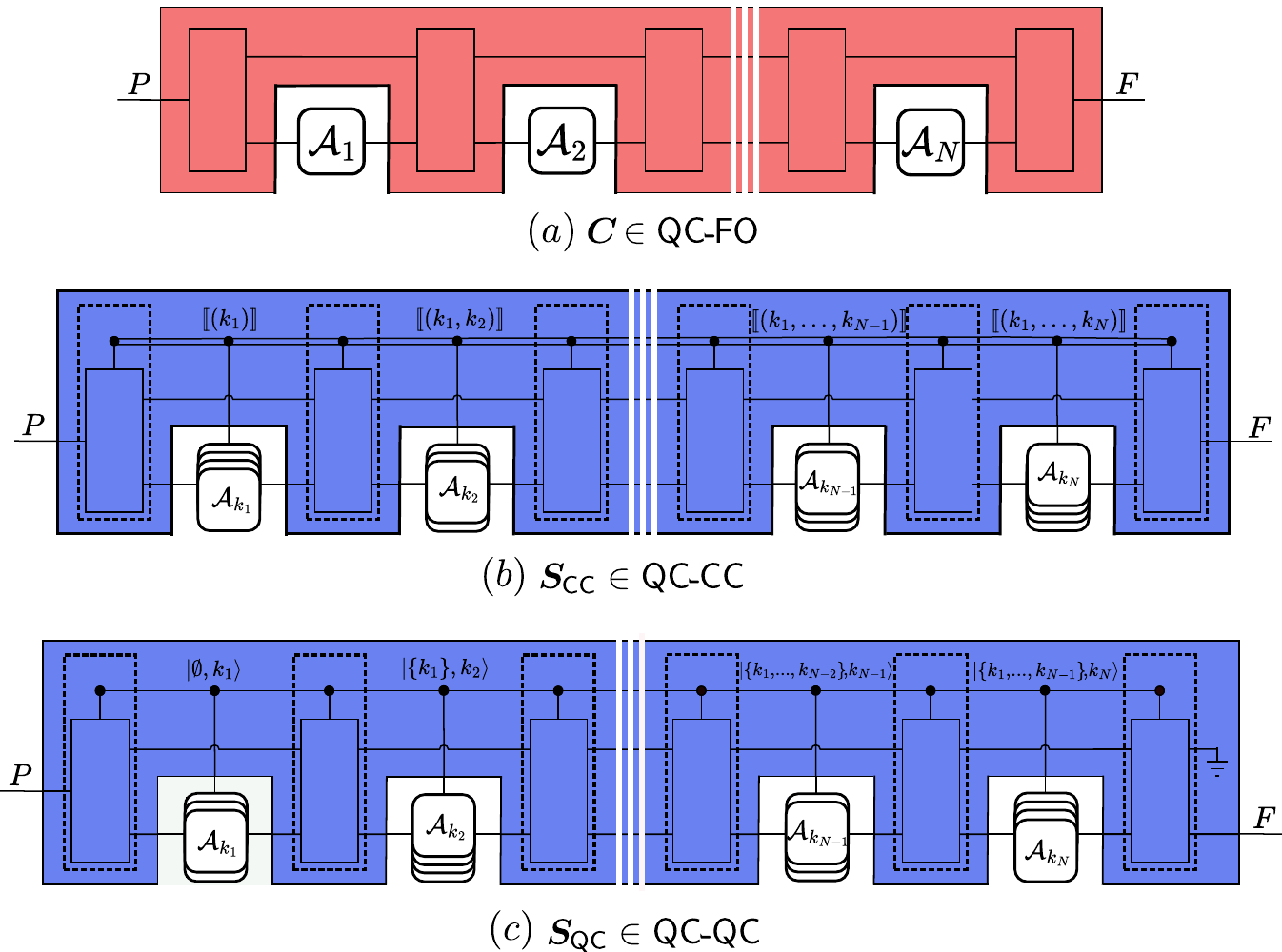}
    \caption{\textbf{Generalized quantum circuits.} Graphical representation of $(a)$ a quantum circuit with a fixed computational order, defined by the superchannels $\boldsymbol{C}\in\textup{\textsf{QC-FO}}$, $(b)$ a quantum circuit with a classical control of computational order, defined by the superchannels $\boldsymbol{S}_{\mathsf{CC}}\in\textup{\textsf{QC-CC}}$, and $(c)$ a quantum circuit with a quantum control of computational order, defined by the superchannels $\boldsymbol{S}_{\mathsf{QC}}\in\textup{\textsf{QC-QC}}$.}
\label{fig::circuits}
\end{figure}

Superchannels have been studied in the literature of higher-order quantum operations with regards to the mathematical characterization of their Choi operators, their implementation within the quantum circuit model, and their physical and causality properties. Here, our focus is on three sets of superchannels: the set of those that can be described by quantum circuits with a fixed-order of computation (QC-FO)~\cite{Chiribella08,Chiribella08supermaps,Chiribella09}; those that can be described by quantum circuits with a classical control of computational order (QC-CC), in which the order of application of each input channel is determined by the state of a classical control system~\cite{Chiribella13,wechs21}; and finally those that can be described by quantum circuits with a quantum control of computational order (QC-QC), in which the order of application of each input channel is determined by the state of a quantum control system~\cite{Chiribella13,wechs21}. While a precise mathematical definition of each of these sets is provided in the Appendix, here we discuss their defining physical features.

\vspace*{5pt}

\noindent\textit{Quantum circuits with fixed computational order} (QC-FO). Also known as quantum combs, these superchannels act on their input channels in a fixed and pre-defined order. They are characterized by quantum circuits with open slots, where the input channels can be plugged in, which apply intermediary gates between each input quantum channel that may act jointly on auxiliary quantum systems, often referred to as a quantum memory~\cite{Chiribella08,Chiribella08supermaps,Chiribella09}. Here, they will be represented by higher-order maps $\boldsymbol{C}\in\textup{\textsf{QC-FO}}$ (see Fig.~\ref{fig::circuits}(a)). A precise definition is given in App.~\ref{subsec::QCFO}. 

\vspace*{5pt}

\noindent\textit{Quantum circuits with classical control of computational order} (QC-CC). 
These superchannels are a generalization of standard quantum circuits, which act on their input channels in an order that is determined by the state of a classical control system. QC-CC superchannels have open slots, in which a single quantum channel can be plugged, however, the choice of which quantum channel will be applied in each slot is not pre-determined: the order in which the channels are called may thus be probabilistic or dynamical, but it can nevertheless be interpreted as a well-defined causal order. They apply intermediary operations that act jointly on auxiliary systems---including, in particular, on the classical control system that can be internally updated to probabilistically determine ``on the fly'' which input quantum channel will be acted upon next~\cite{Chiribella13,wechs21}. QC-CC coincides with the class of causally separable superchannels~\cite{wei2026extensibly,wechs2026causally}, making it the largest class of superchannels with a clear physical interpretation in terms of a definite causal order. Here, they will be represented by higher-order maps $\boldsymbol{S}_{\mathsf{CC}}\in\textup{\textsf{QC-CC}}$ (see Fig.~\ref{fig::circuits}(b)).
A precise definition is given in App.~\ref{subsec::QCCC}. The canonical example of a QC-CC superchannel is the classical switch~\cite{Chiribella13}.

\vspace*{5pt}

\noindent\textit{Quantum circuits with quantum control of computational order} (QC-QC).
These superchannels are a generalization of quantum circuits with classical control, which allows for quantum control of computational order. This order may be dynamical and, moreover, it may be causally indefinite. In particular, it allows the possibility of entanglement between the state of the quantum control system and the order of application of the input quantum channels, in what can be interpreted as a superposition of different fixed orders~\cite{Chiribella13,wechs21}. This is the largest class of superchannels that have a clear interpretation in terms of generalized quantum circuits. Here, they will be represented by higher-order maps $\boldsymbol{S}_{\mathsf{QC}}\in\textup{\textsf{QC-QC}}$ (see Fig.~\ref{fig::circuits}(c)). A precise definition is given in App.~\ref{subsec::QCQC}. The canonical example of a $\textup{\textsf{QC-QC}}$ superchannel is the quantum switch~\cite{Chiribella13}. 

We remark that, while $\textup{\textsf{QC-FO}}\subset\textup{\textsf{QC-CC}}\subset\textup{\textsf{QC-QC}}$, the set $\textup{\textsf{QC-QC}}$ is itself a strict subset of the set of all admissible superchannels~\cite{wechs21}.

\vspace*{5pt}

\noindent\textit{---Framework for quantum-circuit simulation of higher-order quantum computing}.
%
\begin{figure}[t]
\centering
    \includegraphics[width=\columnwidth]{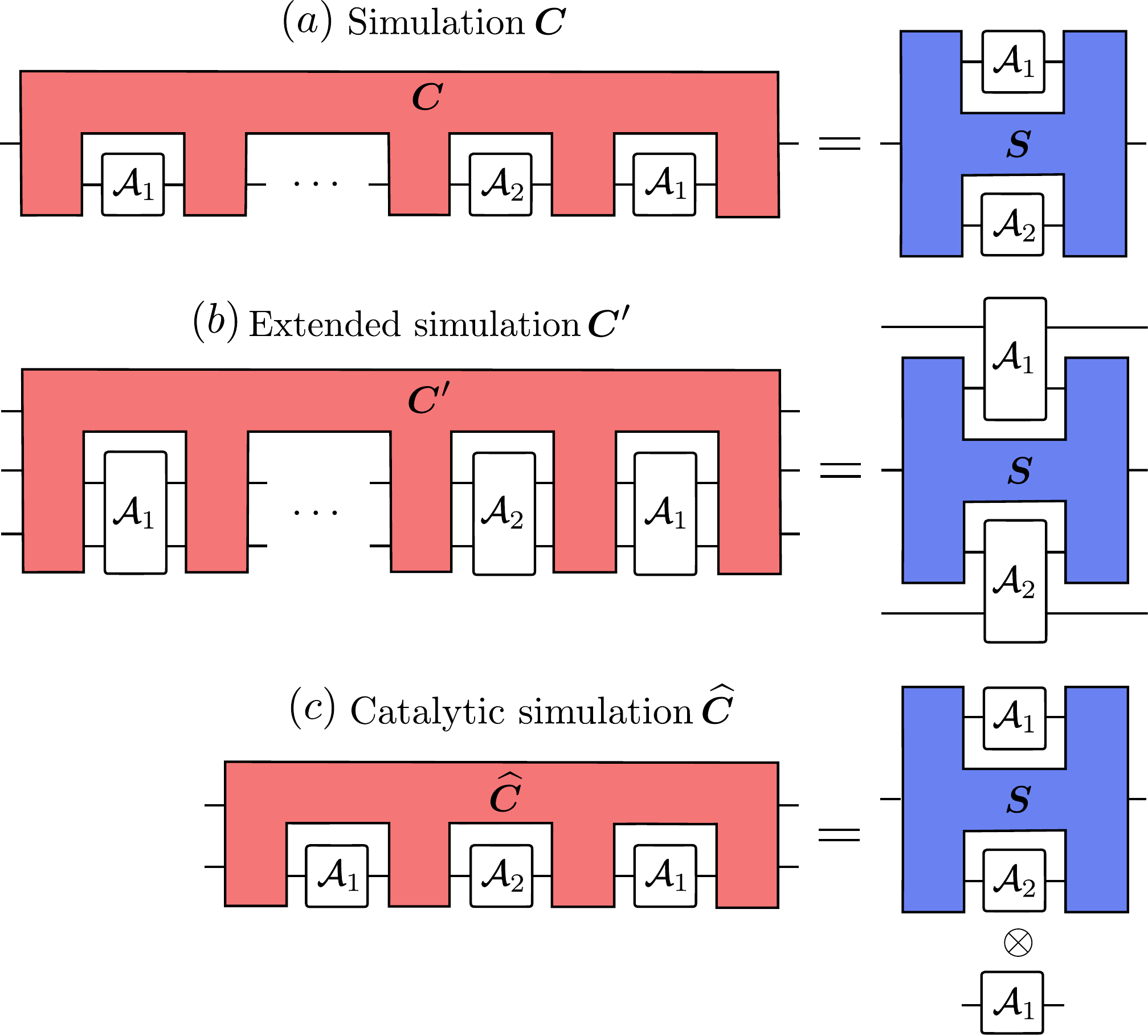}
    \caption{\textbf{Simulation of a superchannel.} Graphical representation of different scenarios for the simulation of a superchannel $\boldsymbol{S}$ (which may be in \textup{\textsf{QC-CC}} or a \textup{\textsf{QC-QC}}, but is not limited to these) that acts on $N=2$ input quantum channels, $\A_1$ and $\A_2$, by a fixed-order quantum circuit $\boldsymbol{C}$, $\boldsymbol{C}^\prime$, or $\widehat{\boldsymbol{C}}\in\textup{\textsf{QC-FO}}$ that acts on more queries of the input quantum channels: $(a)$ depicts the superchannel simulation of Def.~\ref{def:simulation}, where the action of $\boldsymbol{S}$ on all quantum channels is reproduced by $\boldsymbol{C}$, $(b)$ depicts the extended superchannel simulation of Def.~\ref{def:simulation_extended}, where the action of $\boldsymbol{S}$ on all extended quantum channels is reproduced by $\boldsymbol{C}^\prime$, and $(c)$ depicts the catalytic superchannel simulation of Def.~\ref{def:simulation_catalytic}, where the action on all quantum channels is reproduced by $\widehat{\boldsymbol{C}}$ and, furthermore, the extra queries needed in the simulation are recovered as an output of $\widehat{\boldsymbol{C}}$.
    }
\label{fig::simulation}
\end{figure}
%
The main question we address is what is the cost, in terms of the necessary number of queries to the input quantum channels, of simulating the action of a given superchannel on a given set of inputs using standard quantum circuits, which constitute a universal model of quantum computation. Notice that this question is inherently different from what is typically considered as a simulation question in the literature of quantum computing: generally, one is interested in the cost of simulating a \textit{quantum computation} using a \textit{classical computer}. Here, on the other hand, we are interested in the cost of simulating a \textit{higher-order quantum computation} using a \textit{quantum computer}. For this purpose, we follow the simulation framework developed in Ref.~\cite{bavaresco25}. We begin by defining a simulation of a superchannel.

\begin{definition}[Simulation of a superchannel]\label{def:simulation}
    A universal, deterministic, and exact simulation of a given superchannel $\boldsymbol{S}:\bigotimes_{i=1}^N[\L(\H_{I_i})\to\L(\H_{O_i})]\to[\L(\H_{P})\to\L(\H_{F})]$ is a quantum circuit with fixed order $\boldsymbol{C}\in\QCFO$, where $\boldsymbol{C}:\bigotimes_{i=1}^N[\L(\H_{I_i})\to\L(\H_{O_i})]^{\otimes n_i}\to[\L(\H_{P})\to\L(\H_{F})]$, that acts on finite numbers $n_i\in\mathbb{N}$ of queries to the respective quantum channels $\map{A}_i$, where $\map{A}_i:\L(\H_{I_i})\to\L(\H_{O_i})$ with $i\in\{1,\ldots,N\}$, such that
    \begin{align}
        \boldsymbol{C}(\map{A}_{1}^{\otimes n_1}\otimes\ldots\otimes\map{A}_{N}^{\otimes n_N}) = \boldsymbol{S}(\map{A}_1\otimes\ldots\otimes\map{A}_N),
    \end{align}
    for all possible quantum channels $\map{A}_1,\ldots,\map{A}_N$.
\end{definition}

In all of the simulation scenarios approached here, when the action of the superchannel $\boldsymbol{S}$ on only a subset of possible input channels can be simulated (e.g.~only on unitary channels), the simulation is called \textit{non-universal}.

Our first definition of simulation captures the notion of reproducing the action of a superchannel on any set of input channels, when this superchannel acts on the ``entire'' input channel. However, the full behavior of a superchannel is given by its action on part of input channels that act jointly on auxiliary systems, which in turn are not accessible to the superchannel. Therefore, to fully reproduce the action of a superchannel, a fixed-order quantum circuit must reproduce its action on any set of extended input channels. As shown in Ref.~\cite{bavaresco25}, this is the most general notion of simulation of a superchannel, which includes the simulation of its action on sets of quantum instruments---that is, on probabilistic transformations of quantum states that are described by sets of completely positive, trace-nonincreasing maps, which sum up to a trace-preserving map. In particular, sets of quantum instruments can be seen as a bipartite quantum channel in which one of the inputs and one of the outputs is a classical system, and hence, a simulation that works for extended input channels also works for input sets of quantum instruments.

\begin{definition}[Extended simulation of a superchannel]\label{def:simulation_extended}
    A universal, deterministic, and exact simulation of a given superchannel $\boldsymbol{S}:\bigotimes_{i=1}^N[\L(\H_{I_i})\to\L(\H_{O_i})]\to[\L(\H_{P})\to\L(\H_{F})]$ on extended quantum channels is a quantum circuit with fixed order $\boldsymbol{C}^\prime\in\QCFO$, where $\boldsymbol{C}^\prime:\bigotimes_{i=1}^N[\L(\H_{I_i}\otimes\H_{I'_i})\to\L(\H_{O_i}\otimes\H_{O'_i})]^{\otimes n_i}\to[\L(\bigotimes_{i=1}^N\H_{I'_i}\otimes\H_{P})\to\L(\bigotimes_{i=1}^N\H_{O'_i}\otimes\H_{F})]$, that acts on $n_i\in\mathbb{N}$ queries to the respective extended quantum channels $\map{A}^\prime_i$, where $\map{A}^\prime_i:\L(\H_{I_i}\otimes\H_{I'_i})\to\L(\H_{O_i}\otimes\H_{O'_i})$, such that
    \begin{align}
        \boldsymbol{C}^\prime({\map{A}^\prime_{1}}^{\otimes n_1}\otimes\ldots\otimes{\map{A}^\prime_{N}}^{\otimes n_N}) = \boldsymbol{S}\otimes\boldsymbol{I}(\map{A}^\prime_1\otimes\ldots\otimes\map{A}^\prime_N),
    \end{align}
    for all possible quantum channels $\map{A}^\prime_1,\ldots,\map{A}^\prime_N$.
\end{definition}

Finally, we formalize another notion of superchannel simulation, that of a catalytic simulation of a superchannel. Catalytic transformations of quantum states have been vastly studied in the quantum information theory~\cite{lipka2024catalysis}. More recently, catalytic transformations of quantum channels have also been explored~\cite{Yoshida23,bavaresco25}. We begin by defining a catalytic simulation:

\begin{definition}[Catalytic simulation of a superchannel]\label{def:simulation_catalytic}
    A universal, deterministic, and exact catalytic simulation of a given superchannel $\boldsymbol{S}:\bigotimes_{i=1}^N[\L(\H_{I_i})\to\L(\H_{O_i})]\to[\L(\H_{P})\to\L(\H_{F})]$ is a quantum circuit with fixed order $\widehat{\boldsymbol{C}}\in\QCFO$, where $\widehat{\boldsymbol{C}}:\bigotimes_{i=1}^N[\L(\H_{I_i})\to\L(\H_{O_i})]^{\otimes n_i}\to[\L(\bigotimes_{i=1}^N\H_{I_i}^{\otimes (n_i-1)}\otimes\H_{P}) \to \L(\bigotimes_{i=1}^N\H_{O_i}^{\otimes (n_i-1)}\otimes\H_{F})]$, that acts on finite numbers $n_i\in\mathbb{N}$ of queries to the respective quantum channels $\map{A}_i$, where $\map{A}_i:\L(\H_{I_i})\to\L(\H_{O_i})$ with $i\in\{1,\ldots,N\}$, such that
    \begin{align}
    \begin{split}
        \widehat{\boldsymbol{C}}&(\map{A}_{1}^{\otimes n_1}\otimes\ldots\otimes\map{A}_{N}^{\otimes n_N}) \\
        &= \boldsymbol{S}(\map{A}_1\otimes\ldots\otimes\map{A}_N)\otimes \map{A}_{1}^{\otimes(n_1-1)}\otimes\ldots\otimes \map{A}_{N}^{\otimes (n_N-1)},
    \end{split}        
    \end{align}
    for all possible input channels $\map{A}_1,\ldots,\map{A}_N$.
\end{definition}
While additional queries of the input channel may be needed for a fixed-order quantum circuit $\widehat{\boldsymbol{C}}\in\QCFO$ to simulate a more general superchannel $\boldsymbol{S}$, in the catalytic notion of simulation these extra queries are not consumed in the process of the simulation, and are returned, unchanged, as part of the output of the simulation $\widehat{\boldsymbol{C}}$. Notice that the definition of catalytic transformation proposed here requires (i) the catalyst to be composed of exact copies of the input channels and (ii) the catalyst to factorize from the rest of the output of the transformation. While this notion of catalysis satisfies the basic requirements of a catalytic transformation, a less restrictive definition where the catalyst channel is arbitrary and where the catalyst is preserved as a marginal of the global output, which does not necessarily factorize, could also be considered catalytic. More details on the formulation of catalytic simulations and considerations about the reusability of the catalyst queries to the quantum channels are presented in App.~\ref{sec::catalysis}.

We remark that a simulation can be simultaneously an extended and a catalytic simulation.

\vspace*{5pt}

We are now ready to present our main results.

\begin{theorem}[Universal simulation of classical control of computational order]\label{thm::QCCC}
    For every quantum circuit with classical control of computational order $\boldsymbol{S}_{\mathsf{CC}}\in\textup{\textsf{QC-CC}}$ that acts on $N$ quantum channels, there exists a fixed-order quantum circuit $\boldsymbol{C}\in\textup{\textsf{QC-FO}}$ using at most $N$ queries of each input channel, and hence, at most $N^2$ total queries, that universally, deterministically, and exactly simulates $\boldsymbol{S}_{\mathsf{CC}}$.

    Moreover, these fixed-order simulations can be adapted to perform simultaneously extended and catalytic simulations of any given $\boldsymbol{S}_{\mathsf{CC}}\in\textup{\textsf{QC-CC}}$ at no additional overhead in query complexity, and in the catalytic case, recovering all $N^2-N$ extra queries.
\end{theorem}

\begin{theorem}[Non-universal simulation of quantum control of computational order]\label{thm::qcqc}
    For every quantum circuit with quantum control of computational order $\boldsymbol{S}_{\mathsf{QC}}\in\textup{\textsf{QC-QC}}$ that acts on $N$ quantum channels, there exists a fixed-order quantum circuit $\boldsymbol{C}\in\textup{\textsf{QC-FO}}$ using at most $N$ queries of each input channel, and hence, at most $N^2$ total queries, that deterministically and exactly simulates the action of $\boldsymbol{S}_{\mathsf{QC}}$ on unitary channels.

    Moreover, these fixed-order simulations can be adapted to perform simultaneously extended and catalytic simulations of the action of any given $\boldsymbol{S}_{\mathsf{QC}}\in\textup{\textsf{QC-QC}}$ on unitary channels at no additional overhead in query complexity, and in the catalytic case, recovering all $N^2-N$ extra queries.
\end{theorem}

The proof of both theorems is constructive and the explicit constructions of the simulations are presented in App.~\ref{subsec::theoremQCCC} for Thm.~\ref{thm::QCCC} and App.~\ref{subsec::theoremQCQC} for Thm.~\ref{thm::qcqc}.

Hence, quantum circuits with classical control of computational order can be simulated by standard, fixed-order quantum circuits, universally, exactly, and deterministically, with at most a quadratic overhead in quantum query complexity, and similarly for quantum control---albeit in the latter case non-universally. These results provide an upper bound on the potential computational advantages of classical and quantum control of computational order, establishing at most a quadratic advantage in quantum query complexity. 

\vspace*{5pt}

\noindent\textit{---Efficiency and optimality}. 
Having established that these simulations are possible, a natural question is what resources they require. From the perspective of query complexity, the answer is favorable: all our constructions require at most $N^{2}$ queries to the input channels, hence, they are efficient. As discussed in detail in App.~\ref{sec::optimality}, the additional quantum circuit resources required by our simulation also scale efficient, with a gate overhead of $O(N^{2}\log N\log d)$ for the standard and extended simulations, and $O(N^{3}\log N\log d)$ for the catalytic simulations and a space overhead of $O(N\log d)$ for standard and extended simulations, and of $O(N^{2}\log d)$ for catalytic simulations. Hence, from the perspective of quantum query, circuit, and space complexity, our simulations are efficient, in both the number of slots $N$ of the simulated superchannel and the dimension $d$ of the input quantum channels.

The quantum query complexity of our simulations is our central figure of merit, and it is therefore natural to ask whether the quadratic scaling can be improved. As discussed further in App.~\ref{sec::optimality}, we conjecture that it is asymptotically optimal in the worst case, in agreement with the conjecture of Ref.~\cite{facchini15} for the unitary permutation problem. There is nevertheless room for improvement for small values of $N$: our simulations can be implemented with just three total queries for $N=2$ and seven total queries for $N=3$. 

\vspace*{5pt}

\noindent\textit{---Conclusion}.
We proved that any quantum circuit with classical control of computational order, and any quantum circuit with quantum control of computational order acting only on unitary channels, can be simulated by fixed-order quantum circuits with at most a quadratic overhead in quantum query complexity, hence, efficiently. We showed that these fixed-order simulations hold even in the most general scenario of extended quantum channels, and, furthermore, that all extra queries required for the simulation can be recovered in a catalytic manner. We conjecture that our simulations are optimal, in the required number of queries, in the regime of large $N$. That is, we conjecture that our provably attainable upper bound of $O(N^2)$ is asymptotically tight.

While we show that the computational power of classical control of order is limited (as is the power of quantum control of order when only unitary channels are considered), at least in terms of quantum query complexity, to at most a quadratic advantage over fixed-order quantum circuits, it remains an open question to understand for which kinds of computational and information-processing tasks this quadratic advantage can be achieved, if any. At the same time, our results do not forbid greater advantages with respect to other notions of cost/performance. For example, when comparing superchannels of a fixed number of slots in a given task,  classical control of order can provide advantages over fixed-order quantum circuits in terms of measures such as probability of success or approximation error. It would be interesting to find out for which tasks such advantages are possible and how they scale in, e.g., the dimension of the input quantum channels. Our results imply, nevertheless, that any such advantages can always be matched by fixed-order circuits with at most a quadratic cost in number of queries. 

Another interesting avenue of research would be to investigate whether allowing convex mixtures of fixed-order circuits can reduce the simulation cost further. Since these are the most general quantum circuits with non-dynamical classical control of computational order \cite{mothe25}, it would allow us to assess the query complexity of simulating quantum circuits with dynamical classical computational order. It also remains an important open question whether a universal simulation of quantum control of computational order is possible using fixed-order quantum circuits that have access to a finite number of queries to the input quantum channels.

\vspace*{1em}

\noindent\textit{AI disclosure.} All results were derived by the authors, without the assistance of AI. The figures were also made by the authors. GPT-6 Astra (OpenAI) and Claude Opus 5.5 (Anthropic) were used for text proofreading and editing. 
\\

\noindent\textit{Acknowledgments.} 
We are grateful to Mateus Araújo, Cyril Branciard, Fabio Costa, Alex B. Grilo, and Marco Túlio Quintino for useful discussions. R. Mothe acknowledges the support 
from the Alexander von Humboldt Foundation.

\let\oldaddcontentsline\addcontentsline
\renewcommand{\addcontentsline}[3]{} 
\bibliography{biblio_simu_QCCC}
\let\addcontentsline\oldaddcontentsline 

\clearpage
\onecolumngrid
\appendix

\let\oldaddcontentsline\addcontentsline
\renewcommand{\addcontentsline}[3]{} 
\section*{Appendix}
\let\addcontentsline\oldaddcontentsline 

\tableofcontents

\setcounter{theorem}{0}

\section{Preliminaries}
\label{sec::prelimaries}

\subsection{Mathematical tools}
\label{subsec::mathtools}

Throughout this work we use the Choi–Jamio\l{}kowski isomorphism~\cite{jamiolkowski72,Choi}, which allows one to represent linear operators and linear maps as vectors and matrices, respectively. Any linear operator $V:\H_I\to\H_O$ can be equivalently described by a vector $\kket{V}\in\H_I\otimes\H_O$, called its Choi vector, defined as
\begin{align}
    \kket{V} \coloneqq (\id_I\otimes V) \kket{\id} = \sum_i \ket{i}\otimes V\ket{i},
    \label{eq:pure_Choi_def}
\end{align}
where $\{\ket{i}\}_i$ denotes the computational basis of $\H_I$ and $\kket{\id} \coloneqq \sum_i \ket{i}\otimes\ket{i}$ is the Choi vector of the identity operator. Any linear map $\mathcal{M}: \mathcal{L}(\mathcal{H}_{I}) \to \mathcal{L}(\mathcal{H}_{O})$ can be represented as an operator $M\in\mathcal{L}(\mathcal{H}_{I}\otimes\mathcal{H}_{O})$, called its Choi operator, defined as
\begin{align}
    M \coloneqq \map{I}\otimes\map{M}(\kketbra{\id}{\id})= \sum_{i,j} \ketbra{i}{j} \otimes \mathcal{M}\big(\ketbra{i}{j}\big). \label{eq:Choi_def}
\end{align}

The composition of linear operators and of linear maps in their Choi representation is given by the link product~\cite{Chiribella08}.
Given two operators $V_1:\H_X\to\H_Y$ and $V_2:\H_Y\to\H_Z$ that share a common space $\H_Y$, described by their Choi vectors $\kket{V_1}\in\H_{X}\otimes\H_{Y}$ and $\kket{V_2}\in\H_{Y}\otimes\H_{Z}$, we describe their composition $\kket{Q}\in\H_{X}\otimes\H_{Z}$ with the link product according to
\begin{align}
    \kket{Q} = \kket{V_1} * \kket{V_2} \coloneqq & \big(\id_{X} \otimes \bbra{\id}\otimes \id_{Z} \big) \big(\kket{V_1} \otimes \kket{V_2}\big) = \sum_i (\id_X\otimes\bra{i})\kket{V_1}\otimes(\bra{i}\otimes\id_Z)\kket{V_2}.
    \label{eq:def_pure_link_product}
\end{align}
For two linear maps $\M_1:\L(\H_{X})\to \L(\H_{Y})$ and $\M_2:\L(\H_{Y})\to \L(\H_{Z})$, their composition $\N = \M_2\circ\M_1: \L(\H_{X})\to \L(\H_{Z})$ can be expressed in the Choi representation via the link product of their Choi operators $M_1\in\L(\H_{X}\otimes\H_{Y})$ and $M_2\in\L(\H_{Y}\otimes\H_{Z})$ by operator $N\in\L(\H_{X}\otimes\H_{Z})$ defined as
\begin{align}
    N = M_1* M_2 \coloneqq & \left( \id_{XZ} \otimes \bbra{\id} \right) (M_1 \otimes M_2) \left( \id_{XZ} \otimes \kket{\id} \right) \notag \\
    =& \tr_{Y} \left[ \big( M_1^{T_Y} \otimes \id^Z \big) \big( \id^X \otimes M_2 \big) \right], \label{eq:def_mixed_link_product}
\end{align}
where ${(\cdot)}^{T_Y}$ denotes partial transposition on the computational basis of $\H_Y$.

The link product is commutative (up to a re-ordering of the tensor products) and associative. It reduces to a $M_1\otimes M_2$ if $\H_Y$ is trivial, and to $\tr(M_1^TM_2)$ if $\H_X$ and $\H_Z$ are trivial. 
\\

Higher-order linear maps can also be represented using Choi operators. In order to see this, the first step is to represent a higher-order linear map $\boldsymbol{R}:[\L(\H_I)\to\L(\H_O)]\to[\L(\H_P)\to\L(\H_F)]$ that acts on a linear map $\map{M}:\L(\H_I)\to\L(\H_O)$, transforming it into another linear map $\map{N}=\boldsymbol{R}(\map{M}):\L(\H_P)\to\L(\H_F)$ as a linear map $\map{R}:\L(\H_I\otimes\H_O)\to\L(\H_P\otimes\H_F)$ acting on the Choi operator $M\in\L(\H_I\otimes\H_O)$ of linear map $\map{M}$, and transforming it into the Choi operator $N=\map{R}(M)\in\L(\H_P\otimes\H_F)$ of another linear map. Then, since now the higher-order linear map $\boldsymbol{R}$ is itself described by a linear map $\map{R}$, it can in turn be represented by a Choi operator $R\in\L(\H_P\otimes\H_I\otimes\H_O\otimes\H_F)$. Hence, the output of $\boldsymbol{R}$ can be represented in the Choi picture using the link product. Let $\mathbb{J}$ be the map that transforms a linear map into its Choi operator, i.e., $\mathbb{J}(\map{M}) = M$ for all linear maps $\map{M}$. Then,
\begin{align}
    N = \mathbb{J}(\map{N}) = \mathbb{J}\left(\boldsymbol{R}(\map{M})\right) = \map{R}(M) = R * M.
\end{align}

While technically $R$ is the Choi operator of the map $\map{R}$ induced by the higher-order map $\boldsymbol{R}$, we will informally refer to $R$ as the associated Choi operator of the higher-order map $\boldsymbol{R}$.

\subsection{Quantum circuits with fixed computational order (QC-FO)}
\label{subsec::QCFO}

In the higher-order transformations formalism, the notion of a quantum circuit with open slots where arbitrary quantum channels can be plugged in is commonly known as a quantum comb~\cite{Chiribella08,Chiribella09}, or interchangeably as a quantum channel with memory~\cite{Kretschmann05} or a sequential quantum strategy~\cite{Gutoski06}. In the context of this work, they are referred to as quantum circuits with a fixed computational order (QC-FO). They can be abstractly defined as the most general higher-order linear map
\begin{align}
    \boldsymbol{C}:\bigotimes_{i=1}^N[\L(\H_{I_i})\to\L(\H_{O_i})]\to[\L(\H_{P})\to\L(\H_{F})]
\end{align}
that transforms $N$ quantum channels 
\begin{align}
    \mathcal{A}_i:\L(\H_{I_i})\to\L(\H_{O_i}), \ \ \ \forall \ i\in\{1,\ldots,N\}
\end{align}
into a valid quantum channel defined from a global past Hilbert space $\H_P$ to a global future Hilbert space $\H_F$ by acting in a fixed order $\mathcal{A}_{1} \prec \mathcal{A}_{2} \prec \ldots \prec \mathcal{A}_{N}$ on the input channels~\cite{Chiribella08,Chiribella09}.

\begin{figure}[t]
\centering
    \includegraphics[width=\columnwidth]{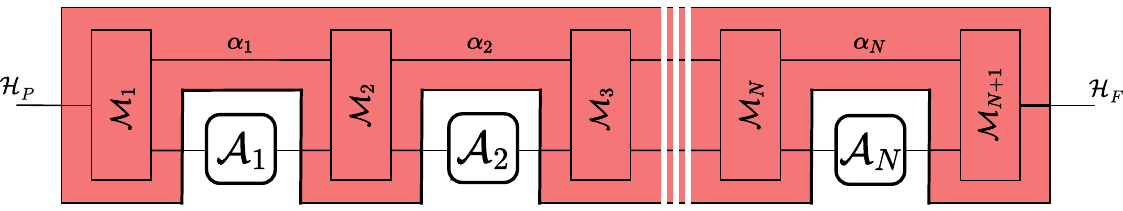}
    \caption{\textbf{Quantum circuits with fixed computational order (QC-FO).} Graphical representation of a quantum circuit with a fixed computational order, adapted from Ref.~\cite{wechs21}.}
\label{fig::circuit_QCFO}
\end{figure}

The architecture of the resulting transformation consists of an alternation between the input quantum channels $\mathcal{A}_i$ and completely positive trace-preserving (CPTP) internal operations $\M_j$ for $j\in\{1,\ldots,N+1\}$. The first internal operation 
\begin{align}
    \M_1:\L(\H_P)\to\L(\H_{I_1}\otimes\H_{\alpha_1})
\end{align}
transforms the input state of the circuit, prepared in the global past Hilbert space $\H_P$, into a joint state in the input space $\H_{I_1}$ of the first quantum channel $\map{A}_1$ and an auxiliary/memory system $\H_{\alpha_1}$. Then, all intermediary operations act according to 
\begin{align}
    \M_i:\L(\H_{O_{i-1}}\otimes\H_{\alpha_{i-1}})\to\L(\H_{I_{i}}\otimes\H_{\alpha_i}) \ \ \ \forall \ i\in\{2,\ldots,N\}. 
\end{align}
Finally, the last internal operation 
\begin{align}
    \M_{N+1}:\L(\H_{O_{N}}\otimes\H_{\alpha_{N}})\to\L(\H_F) 
\end{align}
jointly transforms the state of the last auxiliary system $\H_{\alpha_{N}}$ and the state of the output system $\H_{O_{N}}$ of the last input channel $\map{A}_N$ into the state of the output system $\H_F$ in the global future of the circuit. See Fig.~\ref{fig::circuit_QCFO} for a graphical representation. 

A QC-FO is then formally defined as follows.

\begin{definition}[Quantum circuit with fixed computational order (QC-FO)~\cite{Chiribella09,Chiribella08,Chiribella08supermaps}]
    A higher-order linear map $\boldsymbol{C}$, with associated Choi operator $C$, is a superchannel that corresponds to a quantum circuit with fixed computational order, i.e. $\boldsymbol{C}\in\textup{\textsf{QC-FO}}$, iff there exists a collection of quantum channels $\M_1,\ldots,\M_{N+1}$, with associated Choi operators  $M_1,\ldots,M_{N+1}$, such that $C$ can be decomposed as a function of $M_1,\ldots,M_{N+1}$ according to
    \begin{equation}
        C = M_1 * \ldots * M_{N+1}.
    \end{equation}
\end{definition}

The action of the QC-FO superchannel on the input quantum channels is then expressed in the Choi picture as
\begin{align}
    C*(A_1\otimes \ldots \otimes A_N) & =  M_1 * A_1 * M_2 * A_2 \ldots * M_N * A_N * M_{N+1}
    \label{eq:output_channel_QCFO}
\end{align}

\section{Simulating classical control of computational order}
\label{sec::simulatingQCCC}

\subsection{Quantum circuits with classical control of computational order (QC-CC)}
\label{subsec::QCCC}

We now define quantum circuits with classical control (QC-CC) of computational order. QC-CCs were introduced in Ref.~\cite{wechs21} as a generalization of quantum circuits with fixed order, in order to characterize the most general transformation of $N$ quantum operations $\mathcal{A}_i$ into a valid quantum channel, where the operations are applied in a well-defined, although not necessarily fixed, computational order. Contrary to quantum circuits with fixed order, the computational order between the operations $\mathcal{A}_i$ might not be fixed from the beginning but instead be dynamically established on the fly as the input quantum channels and the internal operations that define the QC-CC are applied, a feature known as dynamical causal order \cite{hardy05,BaumelerWolf,Oreshkov16,Abbott16,mothe25}. The set of QC-CC superchannels was recently shown to be equivalent to the set of causally separable processes, i.e., of superchannels compatible with a well-defined causal order \cite{wei2026extensibly,wechs2026causally}.

Consider a QC-CC superchannel that acts on $N$ quantum channels, and hence, has $N$ slots, labeled $n\in\{1,\ldots,N\}$. Since the order in which these channels are acted upon by the QC-CC may not be fixed, from now on we use the notation $\mathcal{A}_{k_n}$ to denote the quantum channel that is applied at slot $n$ (note that, e.g., the quantum channel $\mathcal{A}_{k_1}$ applied on time step $n=1$ need not be channel $\mathcal{A}_1$).

A QC-CC is a higher-order linear map 
\begin{align}
    \boldsymbol{S}_\mathsf{CC}:\bigotimes_{i=1}^N[\L(\H_{I_i})\to\L(\H_{O_i})]\to[\L(\H_{P})\to\L(\H_{F})]
\end{align}
defined by a collection of internal maps $\{\{\mathcal{M}_{\emptyset}^{\to k_{1}}\}_{k_1},\ldots, \,  \{\mathcal{M}_{(k_1,\ldots,k_n)}^{\to k_{n+1}}\}_{k_{n+1}},\ldots, \, \mathcal{M}_{(k_1,\ldots,k_N)}^{\to F}\}$, such that, for every $n\in\{1,\ldots,N-1\}$, and every sequence $(k_1,\ldots,k_n)$ of distinct elements of $[N]=\{1,\ldots,N\}$ (we denote the empty sequence $\emptyset$), the set $\{\mathcal{M}_{(k_1,\ldots,k_n)}^{\rightarrow{k_{n+1}}}\}_{k_{n+1}\in [N] \setminus \{k_1,\ldots,k_n\}}$ is a valid quantum instrument---that is, a collection of completely positive maps that sum up to a trace-preserving map---while the operations $\mathcal{M}_{(k_1,\ldots,k_N)}^{\to F}$ are valid CPTP maps. 

More concretely, the CP maps 
\begin{align}
    \M_{\emptyset}^{\to k_1}: \L(\H_P) \to \L(\H_{I_{k_1}}\otimes\H_{\alpha_1})
\end{align} 
transform a state from the global past $\H_P$ into a joint state in the target input system $\H_{I_{k_1}}$ of the first slot and the auxiliary system $\H_{\alpha_1}$. For $n\in\{1,\ldots,N-1\}$ the CP maps 
\begin{align}
    \M_{(k_1,\ldots,k_n)}^{\to k_{n+1}}: \L(\H_{O_{k_n}}\otimes\H_{\alpha_n}) \to \L(\H_{I_{k_{n+1}}}\otimes\H_{\alpha_{n+1}})
\end{align}
jointly transform the joint state of the output target system $\H_{O_{k_n}}$ and the auxiliary system $\H_{\alpha_n}$ into the joint state of the  input target system $\H_{I_{k_{n+1}}}$ and auxiliary system $\H_{\alpha_{n+1}}$. Finally, the CPTP map 
\begin{align}
    \M_{(k_1,\ldots,k_N)}^{\to F}: \L(\H_{O_{k_N}}\otimes\H_{\alpha_N}) \to \L(\H_F)
\end{align}
transforms the joint state of the output target system $\H_{O_{k_N}}$ and auxiliary system $\H_{\alpha_N}$ into a state in the global future system $\H_F$. 

The superscripts and subscripts of an intermediary operation $\M_{(k_1,\ldots,k_n)}^{\to k_{n+1}}$ can be interpreted in the following way: The ordered sequence $(k_1,\ldots,k_n)$ in the subscript denotes the record of the labels of the input channels already applied, in that specific order, up to that point in the circuit, while the superscript $\to k_{n+1}$ denotes the instrument outcome selecting $\map{A}_{k_{n+1}}$ as the next channel to be applied in the next slot. 

With these elements, we can define a QC-CC superchannel.

\begin{figure*}[t]
    \centering
    \includegraphics[width=\columnwidth]{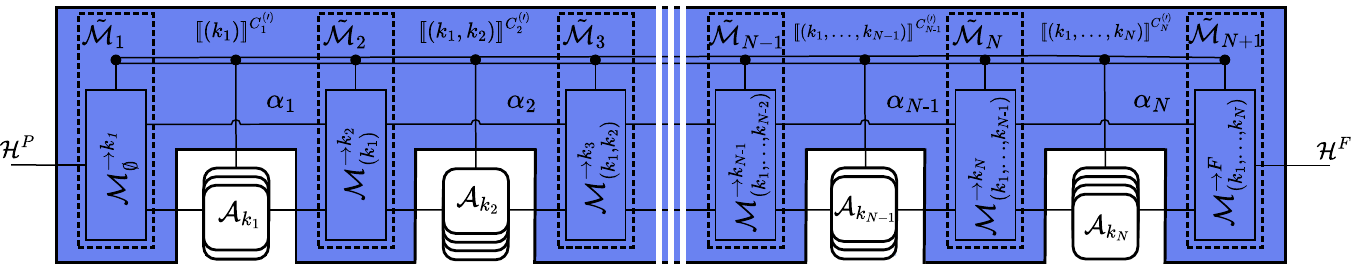}
    \caption{\textbf{Quantum circuits with classical control of computational order (QC-CC).} Graphical representation of a quantum circuit with classical control of computational order, adapted from Ref.~\cite{wechs21}.}
    \label{fig:QCCC}
\end{figure*}

\begin{definition}[Quantum circuit with classical control of computational order ($\textup{\textsf{QC-CC}}$)~\cite{wechs21}]
    A higher-order linear map $\boldsymbol{S}_\mathsf{CC}$, with associated Choi operator $S_{\mathsf{CC}}$, is a quantum circuit with classical control of computational order iff there exists a collection of quantum instruments $\{\M_{\emptyset}^{\to k_1}\}_{k_1}$,
    $\{\mathcal{M}_{(k_1,\ldots,k_n)}^{\rightarrow{k_{n+1}}}\}_{k_{n+1}\in [N] \setminus \{k_1,\ldots,k_n\}}$ for every $n\in\{1,\ldots,N-1\}$ and every ordered sequence $(k_1,\ldots,k_n)$ of distinct elements of $[N]$, and of quantum channels $\mathcal{M}_{(k_1,\ldots,k_N)}^{\to F}$ for every permutation $(k_1,\ldots,k_N)$ of $[N]$, with respective Choi operators $M_{\emptyset}^{\to k_1}$, $M_{(k_1,\ldots,k_n)}^{\to k_{n+1}}$, and $M_{(k_1,\ldots,k_N)}^{\to F}$, such that
    \begin{equation}
        S_{\mathsf{CC}} = \sum_{(k_1,\ldots,k_N)} M_\emptyset^{\to k_1} * M_{(k_1)}^{\to k_2} * \cdots * M_{(k_1,\ldots,k_{N-1})}^{\to k_N} * M_{(k_1,\ldots,k_N)}^{\to F},
    \end{equation}
    where the sum runs over all permutations $(k_1,\ldots,k_N)$ of $[N]$.
\label{def:QC-CC}
\end{definition}

The action of the QC-CC superchannel on the input quantum channels is then expressed in the Choi picture as:
\begin{align}
    S_{\mathsf{CC}}*(A_1\otimes \ldots \otimes A_N) & = \sum_{(k_1,\ldots,k_N)}  M_\emptyset^{\to k_1} * A_{k_1} * M_{(k_1)}^{\to k_2} * A_{k_2} * \cdots * M_{(k_1,\ldots,k_{N-1})}^{\to k_N} * A_{k_N} * M_{(k_1,\ldots,k_N)}^{\to F}.
    \label{eq:output_channel_QCCC}
\end{align}

The classical control of computational order can be made explicit by introducing a register that records which operations have been applied and determines which operation is applied next. In this representation, illustrated in Fig.~\ref{fig:QCCC}, a QC-CC consists of a target system, auxiliary systems $\H_{\alpha_n}$, a classical control register, and internal operations $\tilde{\map{M}}_n$ that jointly process these systems. At each time step, an internal operation prepares the input to the next channel and updates the classical record, allowing subsequent choices of order to depend on the preceding evolution.

We associate each ordered sequence of distinct labels $(k_1,\ldots,k_n)$ with a classical control state given by
\begin{align}
    [\![(k_1,\ldots,k_n)]\!] := \ketbra{(k_1,\ldots,k_n)}{(k_1,\ldots,k_n)} \in \L(\H_{C_n})
\end{align}

Conditioned on this control system, the first $n-1$ channels have been applied in the order $k_1,\ldots,k_{n-1}$, and $\map{A}_{k_n}$ is selected to act at time step $n$. The control register remains diagonal in this history basis, and hence classical, although it may be classically correlated with the joint target–auxiliary state. Thus, the order is well defined along each history without necessarily being determined in advance.

We define quantum channels $\tilde{\M}_{n+1}$, controlled versions of the internal operations, that apply the desired operation $\M_{(k_1,\ldots,k_n)}^{\to k_{n+1}}$ according to the state of the control register. These controlled operations are described in the Choi picture as
\begin{align}
    \tilde{M}_1 = & \sum_{k_1} M_\emptyset^{\to k_1} \otimes [\![(k_1)]\!]^{C_1}, \label{eq:def_tilde_M1_QCCC}\\
    \tilde{M}_{n+1} = & \sum_{(k_1,\ldots,k_n,k_{n+1})} M_{(k_1,\ldots,k_n)}^{\to k_{n+1}} \otimes [\![(k_1,\ldots,k_n)]\!]^{C_n'} \otimes [\![(k_1,\ldots,k_n,k_{n+1})]\!]^{C_{n+1}}\label{eq:Choi_Mn1_QCCC} \\
    \tilde{M}_{N+1} =& \sum_{(k_1,\ldots,k_N)} M_{(k_1,\ldots,k_N)}^{\to F} \otimes [\![(k_1,\ldots,k_N)]\!]^{C_N'}. \label{eq:def_tilde_MN1_QCCC}
\end{align}
For $n\in\{1,\ldots,N\}$, the isomorphic spaces $\H_{C_n}$ and $\H_{C_n'}$ describe the control register immediately before and after the application of $\map{A}_{k_n}$, respectively. In the QC-CC itself, the register is unchanged during this application, so these spaces are connected by an identity channel. We nevertheless use two labels because, in the simulation scheme below, the control register will pass through additional (routing) operations between $\tilde{M}_n$ and $\tilde{M}_{n+1}$.

Note that, strictly speaking, the terms summed in Eqs.~\eqref{eq:def_tilde_M1_QCCC}--\eqref{eq:def_tilde_MN1_QCCC} act on different Hilbert spaces, since the input and output spaces $\H_{O_{k_n}}$ and $\H_{I_{k_{n+1}}}$ depend on the summation indices. To make these sums well defined, one can follow the original QC-CC formulation \cite{wechs21}, and identify at each time step $n$ all the input spaces $\H_{I_k}$ (resp.\ output spaces $\H_{O_k}$) with a common space $\H_{\tilde{I}_n}$ (resp.\ $\H_{\tilde{O}_n}$), e.g., by assuming that they all have the same dimension, or by embedding them into a common space of sufficiently large dimension. The same identification gives a meaning to the condition that $\{\mathcal{M}_{(k_1,\ldots,k_n)}^{\to k_{n+1}}\}_{k_{n+1}}$ is a quantum instrument, whose elements would otherwise have different output spaces. For the sake of conciseness, we employ an abuse of notation and leave this identification implicit.

\subsection{Proof of Theorem~\ref{thm::QCCC} (Universal simulation of classical control of computational order)}
\label{subsec::theoremQCCC}

We are now ready to prove our first theorem.

\begin{theorem}[Universal simulation of classical control of computational order]\label{thm::QCCC_app}
    For every quantum circuit with classical control of computational order $\boldsymbol{S}_{\mathsf{CC}}\in\textup{\textsf{QC-CC}}$ that acts on $N$ quantum channels, there exists a fixed-order quantum circuit $\boldsymbol{C}\in\textup{\textsf{QC-FO}}$ using at most $N$ queries of each input channel, hence at most $N^2$ total queries, that universally simulates $\boldsymbol{S}_{\mathsf{CC}}$, deterministically and exactly.

    Moreover, these fixed-order simulations can be adapted to perform extended and catalytic simulations of any given $\boldsymbol{S}_{\mathsf{CC}}$ at no additional overhead in query complexity, and in the catalytic case, recovering all $N^2-N$ extra queries.
\end{theorem}

The proof is broken down into three parts: simulation, extended simulation, and catalytic simulation.

\subsubsection{Simulation}
\label{subsubsec::proof_simulation_QCCC}

\begin{proof} 
Given a QC-CC superchannel $\boldsymbol{S}_\mathsf{CC}:\bigotimes_{i=1}^N[\L(\H_{I_i})\to\L(\H_{O_i})]\to[\L(\H_{P})\to\L(\H_{F})]$, we construct a fixed-order quantum circuit $\boldsymbol{C}:\bigotimes_{i=1}^N[\L(\H_{I_i})\to\L(\H_{O_i})]^{\otimes N}\to[\L(\H_{P})\to\L(\H_{F})]$ which uses $N$ queries of each input channel $\A_k:\L(\H_{I_k}) \to \L(\H_{O_k})$ to simulate the action of the QC-CC. We start by formally defining $\boldsymbol{C}$ and showing that it is a valid QC-FO superchannel, before showing that $\boldsymbol{C}$ is indeed simulating the action of $\boldsymbol{S}_\mathsf{CC}$.\\

The fixed-order quantum circuit $\boldsymbol{C}$ is an adaptation of the simulation scheme proposed to simulate the quantum $N$-switch   \cite{Colnaghi12,araujo14,facchini15}. It incorporates into this simulation scheme the QC-CC architecture, i.e., the internal CP maps $\mathcal{M}_{(k_1,\ldots,k_n)}^{\to k_{n+1}}$ whose order of application is classically controlled, see Fig.~\ref{fig:simu_QCCC}. 

On top of the $N$ copies of each input channel $\A_k$, the simulation scheme uses $N$ auxiliary registers initialized in state $\ket{a_k} \in \H_{\text{aux}_{k,t_1}}$, where $\H_{\text{aux}_{k,t_1}}$ is isomorphic to the input Hilbert space at the first time step $\H_{I_k}$. The simulation then proceeds in accordance with the structure of a QC-CC, considering $N+1$ time steps labeled by $n \in [N+1]$, each except the last one corresponding to the execution of an operation $\A_k$ on the target system. We will denote the global auxiliary space at the beginning of each time step $n$ as $\H_{\text{aux}_{t_n}}:=\bigotimes_{k=1}^N \H_{\text{aux}_{k,t_n}}$, where each individual Hilbert space $\H_{\text{aux}_{k,t_n}}$ is isomorphic to the generic Hilbert space $\H_{I_k}$.

The action of the simulation at each time step $n \in [N]$ corresponds to the sequential composition of the following four steps: 
\begin{enumerate}
    \item First, the application of the internal QC-CC operation $\tilde M_{n}$ to the target, auxiliary and control systems, as defined in Eqs.~\eqref{eq:def_tilde_M1_QCCC}--\eqref{eq:def_tilde_MN1_QCCC}.
    \item Then, controlled on the last label $k_n$ of the control state $[\![(k_1,\ldots,k_n)]\!]$, a routing operation $\mathcal{S}^I_{k_n}: \mathcal{L}(\mathcal{H}_{\text{aux}_{t_n}} \otimes \mathcal{H}_{I_{k_n}}) \to \mathcal{L}(\mathcal{H}_{\text{aux}_{t_n^I}} \otimes \mathcal{H}_{\bar I_{k_n}})$ swaps the incoming target system in $\H_{I_{k_n}}$ coming from $\M_{(k_1,\ldots,k_{n-1})}^{\to k_n}$ with the auxiliary system in $\mathcal{H}_{\text{aux}_{k_n,t_n}}$, and outputs the target system in $\H_{\text{aux}_{k_n,t_n^I}}$ and the auxiliary system in $\H_{\bar I_{k_n}}$ (which is isomorphic to $\H_{I_{k_n}}$), while it transmits with an identity channel all the other systems. The Hilbert space $\mathcal{H}_{\text{aux}_{t_n^I}}:=\bigotimes_{k=1}^N \H_{\text{aux}_{k,t_n^I}}$ describes the collection of all auxiliary Hilbert spaces $\H_{\text{aux}_{k,t_n^I}}$ (which are isomorphic to $\H_{I_k}$) right after the application of $\mathcal{S}^I_{k_n}$. The overall controlled routing operation $\map{S}_n^I$ applied at time step $n \in [N]$ can be described in the Choi picture as
\begin{align}
    S^I_n = \sum_{(k_1,\ldots,k_n)} S^I_{k_n} \otimes [\![(k_1,\ldots,k_n)]\!]^{C_n} \otimes [\![(k_1,\ldots,k_n)]\!]^{C_n^{\circ}},
    \label{eq:routing_in}
\end{align}
    where $\H_{C_n}$ is the Hilbert space corresponding to the control system stemming from $\tilde{\M}_n$ and input in $S_n^I$, and where $\H_{C_n^{\circ}}$ is the Hilbert space corresponding to the control system output by $S_n^I$ and input in $S_n^O$.
    \item A copy $\A_k:\L(\H_{\text{aux}_{k,t_n^I}})\to \L(\H_{\text{aux}_{k,t_n^O}})$ of each quantum channel\footnote{For the sake of clarity, the $N$ copies of each ``original'' input quantum channel $\A_k:\L(\H_{I_k}) \to \L(\H_{O_k})$ are defined with different input and output Hilbert spaces, $\H_{\text{aux}_{k,t_n^I}}$ and $\H_{\text{aux}_{k,t_n^O}}$ respectively. Whether we are referring to an original quantum input channel or a copy of it, it is denoted by $\A_k$. It must be clear from the context which of the two meanings is intended.} is then applied and transforms the corresponding auxiliary state in $\H_{\text{aux}_{k,t_n^I}}$ into an output state in $\H_{\text{aux}_{k,t_n^O}}$, the latter Hilbert space being isomorphic to $\H_{O_k}$, and where $\mathcal{H}_{\text{aux}_{t_n^O}}:=\bigotimes_{k=1}^N \H_{\text{aux}_{k,t_n^O}}$ describes the collection of all auxiliary Hilbert spaces right after the application of the $N$ external channels.
    \item Finally, the routing is inverted by $\mathcal{S}^O_{k_n}: \mathcal{L}(\mathcal{H}_{\text{aux}_{t_n^O}} \otimes \mathcal{H}_{\bar I_{k_n}}) \to \mathcal{L}(\mathcal{H}_{\text{aux}_{t_{n+1}}} \otimes \mathcal{H}_{O_{k_n}})$, which swaps the target state in $\mathcal{H}_{\text{aux}_{k_n,t_n^O}}$ with the auxiliary state in $\mathcal{H}_{\bar I_{k_n}}$, the target state being eventually output in $\mathcal{H}_{O_{k_n}}$ and sent to $\M_{(k_1,\ldots,k_{n})}^{\to k_{n+1}}$, and the auxiliary state being eventually output in $\mathcal{H}_{\text{aux}_{k_n,t_{n+1}}}$. The overall controlled routing operation $\map{S}_n^O$ applied at time step $n$ can be described in the Choi picture as
    \begin{align}
    S^O_n = \sum_{(k_1,\ldots,k_n)} S^O_{k_n} \otimes [\![(k_1,\ldots,k_n)]\!]^{C_n^{\circ}} \otimes [\![(k_1,\ldots,k_n)]\!]^{C_n'},
     \label{eq:routing_out}
\end{align}
where $\H_{C_n^{\prime}}$ is the Hilbert space corresponding to the control system output by $S_n^O$ and input in $\tilde{\M}_{n+1}$. 

We will denote by $\mathcal{R}_n$ the parallel higher-order transformation with $N$ slots obtained by applying $\map{S}_n^I$, leaving open one slot $\H_{\text{aux}_{k,t_n^I}}\to\H_{\text{aux}_{k,t_n^O}}$ for each channel $\A_k$, $k \in [N]$, and then applying $\map{S}_n^O$. In the Choi picture, it is described by $R_n = S^I_n * S^O_n$.
\end{enumerate}

The last time step $n=N+1$ consists in tracing out all the auxiliary systems in $\H_{\text{aux}_{t_{N+1}}}$ and applying the last QC-CC internal operation $\tilde\M_{N+1}$, which transforms the target, auxiliary and control systems into a state output in the global future Hilbert space $\H_F$.

For any QC-CC superchannel $\boldsymbol{S}_\mathsf{CC}:\bigotimes_{i=1}^N[\L(\H_{I_i})\to\L(\H_{O_i})]\to[\L(\H_{P})\to\L(\H_{F})]$, we can thus formally define the superchannel $\boldsymbol{C}:\bigotimes_{i=1}^N[\L(\H_{I_i})\to\L(\H_{O_i})]^{\otimes N}\to[\L(\H_{P})\to\L(\H_{F})]$ as the alternating composition, step by step, of the procedure described above, except the third step as it corresponds to the application of the quantum channels $\A_k$ which are input in $\boldsymbol{C}$. The Choi operator of the superchannel $\boldsymbol{C}$ is thus described as 

\begin{align}
C & = \rho_{\text{aux}_{t_1}} * \tilde M_1 * R_1 * \tilde M_2 * R_2  * \cdots * \tilde M_N * \tr_{\text{aux}_{t_{N+1}}}\left[R_N \right]* \tilde M_{N+1},
\label{eq:output_state_simu_QCCC}
\end{align}
where $\rho_{\text{aux}_{t_1}}=\bigotimes_{i=1}^N \ketbra{a_i}{a_i} \in \L(\H_{\text{aux}_{t_1}})$ is the initial auxiliary state.

That $\{\tilde M_{n}\}_{n\in[N]}$ are Choi operators of CPTP maps follows from the fact that $\{\mathcal{M}_{(k_1,\ldots,k_{n-1})}^{\rightarrow{k_{n}}}\}_{k_{n}\in [N] \setminus \{k_1,\ldots,k_{n-1}\}}$ is a quantum instrument. The maps $\mathcal{S}_n^I$ and $\mathcal{S}_n^O$ are clearly CPTP, and $\mathcal{R}_n$ is a valid parallel higher-order superchannel. Thus, at each time step $\tilde M_n * R_n$ defines a valid parallel higher-order process, which can be interpreted as a QC-FO between the $N$ input channels $\A_1,\ldots,\A_N$. Finally, by Eq.~\eqref{eq:output_state_simu_QCCC}, the superchannel $\boldsymbol{C}$ is obtained by preparing the auxiliary registers in the state $\rho_{\text{aux}_{t_1}}$, applying sequentially the blocks $\tilde M_1 * R_1, \ldots, \tilde M_N * R_N$ and then $\tilde M_{N+1}$, and tracing out the auxiliary registers in $\H_{\text{aux}_{t_{N+1}}}$. Since each block $\tilde M_n * R_n$ is a QC-FO, while $\tilde M_{N+1}$, the state preparation and the partial trace are CPTP maps, this sequential composition is itself a QC-FO, i.e., $\boldsymbol{C}$ is a valid QC-FO superchannel.

\begin{figure*}[h]
    \centering
    \includegraphics[width=\columnwidth]{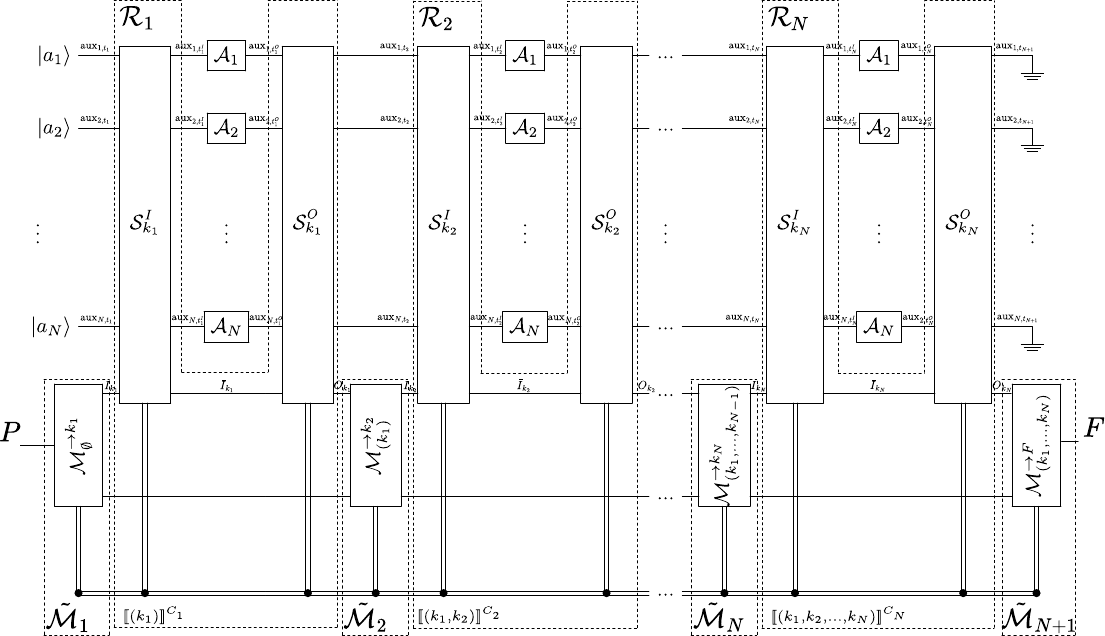}
    \caption{Fixed order simulation of a QC-CC with input quantum channels $\mathcal{A}_1,\ldots,\mathcal{A}_N$ as represented in Fig.~\ref{fig:QCCC}. $N$ queries of each external operation $\mathcal{A}_k$ are made, so that the simulation makes $N^2$ queries to the external operations.}
    \label{fig:simu_QCCC}
\end{figure*}

\vspace{5mm}

We will now show that $\boldsymbol{C}$ applied on $N$ copies of each input channel is simulating the action of $\boldsymbol{S}_\mathsf{CC}$, for any choice of the input channels. More formally, we want to show that  \begin{align}
        \boldsymbol{C}(\A_{1}^{\otimes N}\otimes\ldots\otimes\A_{N}^{\otimes N}) = \boldsymbol{S}_\mathsf{CC}(\A_1\otimes\ldots\otimes\A_N),
    \end{align}
    for all possible quantum channels $\A_1,\ldots,\A_N$. Equivalently, we can rephrase the last equation in the Choi picture as
    \begin{align}
        C*(A_{1}^{\otimes N}\otimes\ldots \otimes A_{N}^{\otimes N}) = S_{\mathsf{CC}}*(A_1 \otimes \ldots \otimes A_N),
    \end{align}
    for all possible quantum channels $A_1,\ldots,A_N$.

The proof consists in computing iteratively the output of the simulation scheme presented in Fig.~\ref{fig:simu_QCCC} and showing that it matches the output of the QC-CC as defined in Eq.~\eqref{eq:output_channel_QCCC}.

Following Definition~\ref{def:QC-CC}, any QC-CC superchannel $\boldsymbol{S}_\mathsf{CC}$ is defined by a collection of internal operations $\{\mathcal{M}_{\emptyset}^{\to k_{1}},\ldots, \,  \mathcal{M}_{(k_1,\ldots,k_n)}^{\to k_{n+1}},\ldots, \, \mathcal{M}_{(k_1,\ldots,k_N)}^{\to F}\}.$

The output of the simulation scheme $\boldsymbol{C}$ can be computed by composing iteratively the operations $\tilde{\M}_n$, $\mathcal{R}_n$ and the quantum channels $\A_1 \otimes \ldots \otimes \A_N$. The auxiliary systems are initialized in the pure state $\rho_{\text{aux}_{t_1}}=\bigotimes_{i=1}^N \rho_i \in \L(\H_{\text{aux}_{t_1}})$, with $\rho_i=\ketbra{a_i}{a_i}$. We describe by $\varrho_{(n)}'$ (resp.~$\varrho_{(n+1)}$) the simulation output at time step $n$, before (resp.~after) the operation $\tilde{\M}_{n+1}$ is applied:

\begin{align}
\varrho_{(n)}' & = (\rho_{\text{aux}_{t_1}} * \tilde M_1 * R_1 *  \tilde{M}_2 * \cdots * \tilde M_n * R_n) * (A_{1}^{\otimes {n}}\otimes\ldots \otimes A_{N}^{\otimes {n}}), \notag 
\end{align}
and
\begin{align}
\varrho_{(n+1)} & = (\rho_{\text{aux}_{t_1}} * \tilde M_1 * R_1 * \tilde M_2 * \cdots * R_n * \tilde M_{n+1}) * (A_{1}^{\otimes {n}}\otimes\ldots \otimes A_{N}^{\otimes {n}}). 
\end{align}
At the first time step, right after the first internal operation $\tilde \M_1$ has been applied, the simulation output is described by
\begin{align}
    \varrho_{(1)} & = \rho_{\text{aux}_{t_1}}* \tilde M_1 \nonumber \\
    &= \sum_{k_1} (\rho_1 \otimes \ldots \otimes \rho_N) \otimes M_{\emptyset}^{\to k_{1}} \otimes [\![(k_1)]\!]^{C_{1}}
\end{align}
Applying the first external operations in parallel along with the swap operations we obtain the Choi operator at the end of the first time step as
\begin{align}
    \varrho_{(1)}' & = \varrho_{(1)} * R_1 * (A_{1}\otimes\ldots \otimes A_{N}) = \varrho_{(1)} * S^I_1*(A_1\otimes \ldots \otimes A_N) * S^O_1 \nonumber \\
    & = \left[\sum_{k_1} \left[ (\rho_1\otimes \ldots \otimes \rho_N) \otimes M_{\emptyset}^{\to k_{1}} \right] * S^I_{k_1}\otimes [\![(k_1)]\!]^{C^{\circ}_{1}} \right] *(A_1\otimes \ldots \otimes A_N) * S^O_1\nonumber \\
    & = \left[\sum_{k_1} \left[ (\rho_1 * A_1\otimes \ldots \otimes \rho_{k_1-1} * A_{k_1-1}\otimes  M_{\emptyset}^{\to k_{1}} * A_{k_1} \otimes \rho_{k_1+1} * A_{k_1+1} \otimes \ldots \otimes \rho_N * A_N) \otimes \rho_{k_1} \right] \otimes [\![(k_1)]\!]^{C^{\circ}_{1}}\right] * S^O_1 \nonumber \\
    & = \sum_{k_1} \left[ (\rho_1 * A_1\otimes \ldots \otimes \rho_{k_1-1} * A_{k_1-1}\otimes \rho_{k_1} \otimes \rho_{k_1+1} * A_{k_1+1} \otimes \ldots \otimes \rho_N * A_N) \otimes (M_{\emptyset}^{\to k_{1}} * A_{k_1}) \right] \otimes [\![(k_1)]\!]^{C'_{1}}
\end{align}
The output of the simulation scheme at the second time step before the external operations are applied is
\begin{align}
     \varrho_{(2)} & = \varrho_{(1)}' * \tilde M_2 \nonumber \\
     &= \sum_{k_1,k_2} \left[(\rho_1 * A_1\otimes  \ldots \otimes \rho_{k_1-1} * A_{k_1-1}\otimes \rho_{k_1} \otimes \rho_{k_1+1} * A_{k_1+1} \otimes \ldots \otimes \rho_N * A_N) \otimes (M_{\emptyset}^{\to k_{1}} * A_{k_1} * M_{(k_1)}^{\to k_{2}}) \right] \otimes [\![(k_1,k_2)]\!]^{C_{2}}.
\end{align}
At the end of the second time step, once the external operations and the swap operation have been applied, we note that each external operation $\mathcal{A}_k$ is applied twice on the corresponding auxiliary system in state $\rho_k$, except for $k=k_1,k_2$. Indeed, $\mathcal{A}_{k_1},\mathcal{A}_{k_2}$ are respectively applied only once on the corresponding auxiliary state since the other copy of this operation was applied on the target system state:
\begin{align}
     \varrho'_{(2)} & = \varrho_{(2)} * R_2 * (A_{1}\otimes\ldots \otimes A_{N})\nonumber \\
     &= \sum_{k_1,k_2} [(\rho_1 * A_1^{*2}\otimes \ldots \otimes \rho_{k_1-1} * A^{*2}_{k_1-1}\otimes \rho_{k_1} * A_{k_1} \otimes \rho_{k_1+1} * A^{*2}_{k_1+1} \otimes \ldots \otimes \rho_{k_2-1} * A^{*2}_{k_2-1} \otimes \ldots \nonumber \\
     & \ \ \ \ \ \ \ \ \ \ \ \ \ \ldots \rho_{k_2} * A_{k_2} \otimes \rho_{k_2+1} * A^{*2}_{k_2+1} \otimes \ldots \otimes \rho_N * A^{*2}_N) \otimes (M_{\emptyset}^{\to k_{1}} * A_{k_1} * M_{(k_1)}^{\to k_{2}} * A_{k_2})] \otimes [\![(k_1,k_2)]\!]^{C_{2}'},
\end{align}
where $A_k^{*2}=A_k*A_k$ is the Choi matrix of the channel $\A_k^{\circ2}=\A_k\circ \A_k$ obtained by composing the channel $\A_k$ with itself.
Deriving iteratively the Choi operator after each time step, we eventually obtain the output of the simulation at the second to last time step $N$:
\begin{align}
     &\varrho'_{(N)} = \sum_{(k_1,\ldots,k_N)} [(\rho_1 * A_1^{*(N-1)}\otimes  \ldots \otimes \rho_N * A_N^{*(N-1)}) \otimes (M_\emptyset^{\to k_1} \!*\! A_{k_1} * \cdots \!*\! M_{(k_1,\ldots,k_{N-1})}^{\to k_N} \!*\! A_{k_N})] \otimes [\![(k_1,\ldots,k_N)]\!]^{C'_{N}}.
\end{align}
Applying the last internal operation $\tilde {\mathcal{M}}_{N+1}$:
\begin{align}
     &\varrho_{(N+1)} = \varrho_{(N)}' * \tilde M_{N+1} \nonumber \\
     &= \sum_{(k_1,\ldots,k_N)} (\rho_1 * A_1^{*(N-1)}\otimes  \ldots \otimes \rho_N * A_N^{*(N-1)}) \otimes (M_\emptyset^{\to k_1} \!*\! A_{k_1} * \cdots \!*\! M_{(k_1,\ldots,k_{N-1})}^{\to k_N} \!*\! A_{k_N} \!*\! M_{(k_1,\ldots,k_N)}^{\to F}) \nonumber \\
     &= (\rho_1 * A_1^{*(N-1)}\otimes  \ldots \otimes \rho_N * A_N^{*(N-1)}) \otimes \left[\sum_{(k_1,\ldots,k_N)} M_\emptyset^{\to k_1} \!*\! A_{k_1} * \cdots \!*\! M_{(k_1,\ldots,k_{N-1})}^{\to k_N} \!*\! A_{k_N} \!*\! M_{(k_1,\ldots,k_N)}^{\to F}\right] 
\end{align}
Tracing out the auxiliary systems in $\mathcal{H}_{\text{aux}_{t_{N+1}}}$, we indeed retrieve the output $S_{\mathsf{CC}}*(A_1\otimes \ldots \otimes A_N)$ of the QC-CC superchannel, as defined in Eq.~\eqref{eq:output_channel_QCCC}, which concludes the proof.

We note that because the correlations with the auxiliary systems are classical, one recovers the intermediate QC-CC state at each single time step of the simulation by tracing out the auxiliary systems. We observe that $\A_k$ is applied exactly $N-1$ times on each auxiliary system $\rho_k = \ketbra{a_k}{a_k}$, thus the simulation is not catalytic. This already hints at the fact that in order to define a catalytic simulation, one needs to introduce extra auxiliary systems and distribute these $N-1$ applications of each $A_k$ input channel on those auxiliary systems.
\end{proof}

\subsubsection{Extended simulation}
\label{subsubsec::proof_extendedsimulation_QCCC}
 
In full generality, the quantum channels inputted in a QC-CC can be extended quantum channels $\A'_k:\L(\H_{I_k}\otimes\H_{I'_k})\to\L(\H_{O_k}\otimes\H_{O'_k})$, where the primed Hilbert spaces $\H_{I'_k}$ and $\H_{O'_k}$ describe auxiliary systems (we will refer to them as extension systems in what follows, not to confuse them with the auxiliary systems of the QC-CC or of the simulation quantum circuit) that do not enter the QC-CC. Extending the QC-CC superchannel by the higher-order identity map $\boldsymbol{I}$ acting on these extension systems defines the higher-order linear map
\begin{align}
    \boldsymbol{S}_\mathsf{CC}\otimes\boldsymbol{I}:\bigotimes_{i=1}^N[\L(\H_{I_i}\otimes\H_{I'_i})\to\L(\H_{O_i}\otimes\H_{O'_i})]\to[\L(\H_{P}\otimes\H_{I'_{[N]}})\to\L(\H_{F}\otimes\H_{O'_{[N]}})],
\end{align}
where $\H_{I'_{[N]}}\coloneqq\bigotimes_{k=1}^N\H_{I'_k}$ and $\H_{O'_{[N]}}\coloneqq\bigotimes_{k=1}^N\H_{O'_k}$. Analogously to Eq.~\eqref{eq:output_channel_QCCC}, the action of the QC-CC superchannel on the extended channels is expressed in the Choi picture as 
\begin{align}
    S_{\mathsf{CC}} *(A'_1\otimes \ldots \otimes A'_N) & = \sum_{(k_1,\ldots,k_N)}  M_\emptyset^{\to k_1} * A'_{k_1} * M_{(k_1)}^{\to k_2} * A'_{k_2} * \cdots * M_{(k_1,\ldots,k_{N-1})}^{\to k_N} * A'_{k_N} * M_{(k_1,\ldots,k_N)}^{\to F}.
    \label{eq:output_channel_QCCC_2}
\end{align}

We highlight that a fixed-order simulation as described in Sec.~\ref{subsubsec::proof_simulation_QCCC}, which reproduces the output of a QC-CC for any choice of the input quantum channels, does not necessarily reproduce it when extended channels are applied. We thus update the simulation scheme of Fig.~\ref{fig:simu_QCCC} by routing the extension systems together with the target system, and provide a fixed-order simulation of any QC-CC that remains valid when extended channels are applied, see Fig.~\ref{fig:simu_QCCC_bipartite_channels}.

\begin{proof}
Given a QC-CC superchannel $\boldsymbol{S}_\mathsf{CC}:\bigotimes_{i=1}^N[\L(\H_{I_i})\to\L(\H_{O_i})]\to[\L(\H_{P})\to\L(\H_{F})]$ and extension spaces $\H_{I'_k}$ and $\H_{O'_k}$ for $k\in[N]$, we construct a fixed-order quantum circuit $\boldsymbol{C}':\bigotimes_{i=1}^N[\L(\H_{I_i}\otimes\H_{I'_i})\to\L(\H_{O_i}\otimes\H_{O'_i})]^{\otimes N}\to[\L(\H_{P}\otimes\H_{I'_{[N]}})\to\L(\H_{F}\otimes\H_{O'_{[N]}})]$ which uses $N$ queries of each extended channel $\A'_k:\L(\H_{I_k}\otimes\H_{I'_k}) \to \L(\H_{O_k}\otimes\H_{O'_k})$ to simulate the action of $\boldsymbol{S}_\mathsf{CC}\otimes\boldsymbol{I}$. Analogously to the proof for the non-extended channels presented in Sec.~\ref{subsubsec::proof_simulation_QCCC}, we start by formally defining $\boldsymbol{C}'$ and showing that it is a valid QC-FO superchannel, before showing that $\boldsymbol{C}'$ is indeed simulating the action of $\boldsymbol{S}_\mathsf{CC}\otimes\boldsymbol{I}$.\\

On top of the $N$ copies of each extended channel $\A'_k$, the simulation scheme uses, for each $k\in[N]$, two auxiliary registers initialized in the states $\ket{a_k} \in \H_{\text{aux}_{k,t_1}}$ and $\ket{a'_k} \in \H_{\text{aux}'_{k,t_1}}$, where $\H_{\text{aux}_{k,t_1}}$ and $\H_{\text{aux}'_{k,t_1}}$ are isomorphic to the input Hilbert spaces $\H_{I_k}$ and $\H_{I'_k}$, respectively. It also uses, for each $k\in[N]$, an extension wire which carries the input extension system of $\A'_k$ until $\A'_k$ is applied on the target system, and its output extension system afterwards. The simulation then proceeds in accordance with the structure of a QC-CC, considering $N+1$ time steps labeled by $n \in [N+1]$, each except the last one corresponding to the execution of an operation $\A'_k$ on the target system. We will here denote the global auxiliary space at the beginning of each time step $n$ as $\H_{\text{aux}_{t_n}}\coloneqq\bigotimes_{k=1}^N (\H_{\text{aux}_{k,t_n}}\otimes\H_{\text{aux}'_{k,t_n}})$, where each individual Hilbert space $\H_{\text{aux}_{k,t_n}}$ (resp.~$\H_{\text{aux}'_{k,t_n}}$) is isomorphic to $\H_{I_k}$ (resp.~$\H_{I'_k}$). Similarly, we denote the global extension space at the beginning of each time step $n$ as $\H_{\text{ext}_{t_n}}\coloneqq\bigotimes_{k=1}^N \H_{\text{ext}_{k,t_n}}$, where each individual Hilbert space $\H_{\text{ext}_{k,t_n}}$, describing the extension wire associated with $\A'_k$, is isomorphic to $\H_{I'_k}$. For consistency with $\boldsymbol{S}_\mathsf{CC}\otimes\boldsymbol{I}$ being a transformation into quantum channels from $\L(\H_{P}\otimes\H_{I'_{[N]}})$ to $\L(\H_{F}\otimes\H_{O'_{[N]}})$, we impose $\H_{\text{ext}_{k,t_1}}\coloneqq\H_{I'_k}$ and $\H_{\text{ext}_{k,t_{N+1}}}\coloneqq\H_{O'_k}$, i.e., $\H_{\text{ext}_{t_1}}=\H_{I'_{[N]}}$ and $\H_{\text{ext}_{t_{N+1}}}=\H_{O'_{[N]}}$.

The action of the simulation at each time step $n \in [N]$ corresponds to the sequential composition of the following four steps:
\begin{enumerate}
    \item First, the application of the internal QC-CC operation $\tilde M_{n}$ to the target, auxiliary and control systems, as defined in Eqs.~\eqref{eq:def_tilde_M1_QCCC}--\eqref{eq:def_tilde_MN1_QCCC}.
    \item Then, controlled on the last label $k_n$ of the control state $[\![(k_1,\ldots,k_n)]\!]$, a routing operation $\mathcal{S}'^I_{k_n}: \mathcal{L}(\mathcal{H}_{\text{aux}_{t_n}} \otimes \mathcal{H}_{\text{ext}_{t_n}} \otimes \mathcal{H}_{I_{k_n}}) \to \mathcal{L}(\mathcal{H}_{\text{aux}_{t_n^I}} \otimes \mathcal{H}_{\text{ext}_{t_n^{\circ}}} \otimes \mathcal{H}_{\bar I_{k_n}})$ performs a double swap. First, it swaps the incoming target system in $\H_{I_{k_n}}$ coming from $\M_{(k_1,\ldots,k_{n-1})}^{\to k_n}$ with the auxiliary system in $\mathcal{H}_{\text{aux}_{k_n,t_n}}$, and outputs the target system in $\H_{\text{aux}_{k_n,t_n^I}}$ and the auxiliary system in $\H_{\bar I_{k_n}}$ (which is isomorphic to $\H_{I_{k_n}}$). Second, it swaps the incoming extension system in $\H_{\text{ext}_{k_n,t_n}}$ with the auxiliary system in $\H_{\text{aux}'_{k_n,t_n}}$, and outputs the extension system in $\H_{\text{aux}'_{k_n,t_n^I}}$ and the auxiliary system in $\H_{\text{ext}_{k_n,t_n^{\circ}}}$. It transmits with an identity channel all the other systems. The Hilbert space $\mathcal{H}_{\text{aux}_{t_n^I}}\coloneqq\bigotimes_{k=1}^N (\H_{\text{aux}_{k,t_n^I}}\otimes\H_{\text{aux}'_{k,t_n^I}})$ describes the collection of all auxiliary Hilbert spaces $\H_{\text{aux}_{k,t_n^I}}$ and $\H_{\text{aux}'_{k,t_n^I}}$ (which are isomorphic to $\H_{I_k}$ and $\H_{I'_k}$, respectively) right after the application of $\mathcal{S}'^I_{k_n}$, while the Hilbert space $\H_{\text{ext}_{t_n^{\circ}}}\coloneqq\bigotimes_{k=1}^N \H_{\text{ext}_{k,t_n^{\circ}}}$ describes the collection of all extension wires $\H_{\text{ext}_{k,t_n^{\circ}}}$ (which are isomorphic to $\H_{I'_k}$) right after the application of $\mathcal{S}'^I_{k_n}$. The overall controlled routing operation $\map{S}'^I_n$ applied at time step $n \in [N]$ can be described in the Choi picture as
\begin{align}
    S'^I_n = \sum_{(k_1,\ldots,k_n)} S'^I_{k_n} \otimes [\![(k_1,\ldots,k_n)]\!]^{C_n} \otimes [\![(k_1,\ldots,k_n)]\!]^{C_n^{\circ}},
    \label{eq:routing_in_ext}
\end{align}
    where, as in Eq.~\eqref{eq:routing_in}, $\H_{C_n}$ is the Hilbert space corresponding to the control system stemming from $\tilde{\M}_n$ and input in $\map{S}'^I_n$, and $\H_{C_n^{\circ}}$ is the Hilbert space corresponding to the control system output by $\map{S}'^I_n$ and input in $\map{S}'^O_n$.
    \item A copy $\A'_k:\L(\H_{\text{aux}_{k,t_n^I}}\otimes\H_{\text{aux}'_{k,t_n^I}})\to \L(\H_{\text{aux}_{k,t_n^O}}\otimes\H_{\text{aux}'_{k,t_n^O}})$ of each extended channel (with the same convention as in Sec.~\ref{subsubsec::proof_simulation_QCCC} regarding the input and output Hilbert spaces of the copies) is then applied and transforms the corresponding joint state in $\H_{\text{aux}_{k,t_n^I}}\otimes\H_{\text{aux}'_{k,t_n^I}}$ into an output state in $\H_{\text{aux}_{k,t_n^O}}\otimes\H_{\text{aux}'_{k,t_n^O}}$, the latter Hilbert spaces being isomorphic to $\H_{O_k}$ and $\H_{O'_k}$, respectively, and where $\mathcal{H}_{\text{aux}_{t_n^O}}\coloneqq\bigotimes_{k=1}^N (\H_{\text{aux}_{k,t_n^O}}\otimes\H_{\text{aux}'_{k,t_n^O}})$ describes the collection of all auxiliary Hilbert spaces right after the application of the $N$ external channels. In particular, the copy of $\A'_{k_n}$ acts jointly on the target system and on the extension system associated with $\A'_{k_n}$, while each copy of $\A'_k$, $k\neq k_n$, acts on the pair of auxiliary systems associated with $\A'_k$.
    \item Finally, the routing is inverted by $\mathcal{S}'^O_{k_n}: \mathcal{L}(\mathcal{H}_{\text{aux}_{t_n^O}} \otimes \mathcal{H}_{\text{ext}_{t_n^{\circ}}} \otimes \mathcal{H}_{\bar I_{k_n}}) \to \mathcal{L}(\mathcal{H}_{\text{aux}_{t_{n+1}}} \otimes \mathcal{H}_{\text{ext}_{t_{n+1}}} \otimes \mathcal{H}_{O_{k_n}})$, which performs a double swap. First, it swaps the target state in $\mathcal{H}_{\text{aux}_{k_n,t_n^O}}$ with the auxiliary state in $\mathcal{H}_{\bar I_{k_n}}$, the target state being eventually output in $\mathcal{H}_{O_{k_n}}$ and sent to $\M_{(k_1,\ldots,k_{n})}^{\to k_{n+1}}$, and the auxiliary state being eventually output in $\mathcal{H}_{\text{aux}_{k_n,t_{n+1}}}$. Second, it swaps the extension state in $\H_{\text{aux}'_{k_n,t_n^O}}$ with the auxiliary state in $\H_{\text{ext}_{k_n,t_n^{\circ}}}$, the extension state being eventually output in the extension wire $\H_{\text{ext}_{k_n,t_{n+1}}}$, and the auxiliary state being eventually output in $\H_{\text{aux}'_{k_n,t_{n+1}}}$. It transmits with an identity channel all the other systems. The overall controlled routing operation $\map{S}'^O_n$ applied at time step $n$ can be described in the Choi picture as
    \begin{align}
    S'^O_n = \sum_{(k_1,\ldots,k_n)} S'^O_{k_n} \otimes [\![(k_1,\ldots,k_n)]\!]^{C_n^{\circ}} \otimes [\![(k_1,\ldots,k_n)]\!]^{C_n'},
     \label{eq:routing_out_ext}
\end{align}
where, as in Eq.~\eqref{eq:routing_out}, $\H_{C_n^{\prime}}$ is the Hilbert space corresponding to the control system output by $\map{S}'^O_n$ and input in $\tilde{\M}_{n+1}$.
 
We will denote by $\mathcal{R}'_n$ the parallel higher-order transformation with $N$ slots obtained by applying $\map{S}'^I_n$, leaving open one slot $\H_{\text{aux}_{k,t_n^I}}\otimes\H_{\text{aux}'_{k,t_n^I}}\to\H_{\text{aux}_{k,t_n^O}}\otimes\H_{\text{aux}'_{k,t_n^O}}$ for each extended channel $\A'_k$, $k \in [N]$, and then applying $\map{S}'^O_n$. In the Choi picture, it is described by $R'_n = S'^I_n * S'^O_n$.
\end{enumerate}
 
The last time step $n=N+1$ consists in tracing out all the auxiliary systems in $\H_{\text{aux}_{t_{N+1}}}$ and applying the last QC-CC internal operation $\tilde\M_{N+1}$, which transforms the target, auxiliary and control systems into a state output in the global future Hilbert space $\H_F$. The extension wires in $\H_{\text{ext}_{t_{N+1}}}=\H_{O'_{[N]}}$ are output as the output extension systems.
 
For any QC-CC superchannel $\boldsymbol{S}_\mathsf{CC}:\bigotimes_{i=1}^N[\L(\H_{I_i})\to\L(\H_{O_i})]\to[\L(\H_{P})\to\L(\H_{F})]$ and extension spaces $\H_{I'_k}$ and $\H_{O'_k}$, we can thus formally define the superchannel $\boldsymbol{C}':\bigotimes_{i=1}^N[\L(\H_{I_i}\otimes\H_{I'_i})\to\L(\H_{O_i}\otimes\H_{O'_i})]^{\otimes N}\to[\L(\H_{P}\otimes\H_{I'_{[N]}})\to\L(\H_{F}\otimes\H_{O'_{[N]}})]$ as the alternating composition, step by step, of the procedure described above, except the third step as it corresponds to the application of the extended channels $\A'_k$ which are input in $\boldsymbol{C}'$. The Choi operator of the superchannel $\boldsymbol{C}'$ is thus described as
\begin{align}
C' & = \rho_{\text{aux}_{t_1}} * \tilde M_1 * R'_1 * \tilde M_2 * R'_2  * \cdots * \tilde M_N * \tr_{\text{aux}_{t_{N+1}}}\left[R'_N \right]* \tilde M_{N+1},
\label{eq:output_state_simu_QCCC_extended}
\end{align}
where $\rho_{\text{aux}_{t_1}}=\bigotimes_{i=1}^N \ketbra{a_i}{a_i}\otimes\ketbra{a'_i}{a'_i} \in \L(\H_{\text{aux}_{t_1}})$ is the initial auxiliary state. Note that the extension spaces $\H_{\text{ext}_{t_1}}=\H_{I'_{[N]}}$ and $\H_{\text{ext}_{t_{N+1}}}=\H_{O'_{[N]}}$ remain open in Eq.~\eqref{eq:output_state_simu_QCCC_extended}, and play, together with $\H_P$ and $\H_F$, the role of the global past and global future of $\boldsymbol{C}'$.
 
That $\tilde M_{n}$ is CPTP for $n\in[N]$ follows from the fact that $\{\mathcal{M}_{(k_1,\ldots,k_{n-1})}^{\rightarrow{k_{n}}}\}_{k_{n}\in [N] \setminus \{k_1,\ldots,k_{n-1}\}}$ is a quantum instrument, and that $\tilde M_{N+1}$ is CPTP follows from the fact that the maps $\mathcal{M}_{(k_1,\ldots,k_N)}^{\to F}$ are CPTP. The maps $\map{S}'^I_n$ and $\map{S}'^O_n$ are clearly CPTP, and $\mathcal{R}'_n$ is a valid parallel higher-order superchannel. Thus, at each time step $\tilde M_n * R'_n$ defines a valid parallel higher-order process, which can be interpreted as a QC-FO between the $N$ input extended channels $\A'_1,\ldots,\A'_N$, in which the extension wires are carried along as memory systems. Finally, by Eq.~\eqref{eq:output_state_simu_QCCC_extended}, the superchannel $\boldsymbol{C}'$ is obtained by preparing the auxiliary registers in the state $\rho_{\text{aux}_{t_1}}$, applying sequentially the blocks $\tilde M_1 * R'_1, \ldots, \tilde M_N * R'_N$ and then $\tilde M_{N+1}$, and tracing out the auxiliary registers in $\H_{\text{aux}_{t_{N+1}}}$. Since each block $\tilde M_n * R'_n$ is a QC-FO, while $\tilde M_{N+1}$, the state preparation and the partial trace are CPTP maps, this sequential composition is itself a QC-FO, i.e., $\boldsymbol{C}'$ is a valid QC-FO superchannel.

\begin{figure*}[h]
    \centering
    \includegraphics[width=\columnwidth]{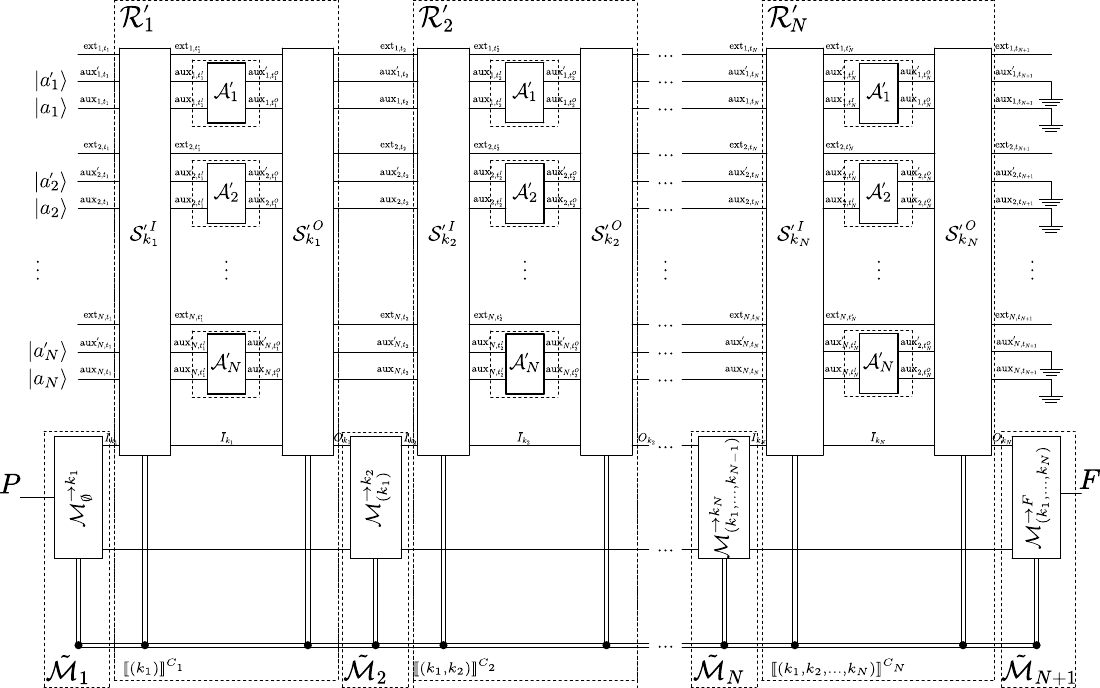}
    \caption{Fixed order simulation of a QC-CC with input extended quantum channels $\A'_1,\ldots,\A'_N$. $N$ queries of each extended channel $\A'_k$ are made, so that the simulation makes $N^2$ queries to the external operations. Compared to Fig.~\ref{fig:simu_QCCC}, the routing operations also swap the extension wires with additional auxiliary registers, so that the copy of $\A'_{k_n}$ applied on the target system at time step $n$ also acts on the extension system associated with $\A'_{k_n}$.}
    \label{fig:simu_QCCC_bipartite_channels}
\end{figure*}

\vspace{5mm}
 
We will now show that $\boldsymbol{C}'$ applied on $N$ copies of each extended channel is simulating the action of $\boldsymbol{S}_\mathsf{CC}\otimes\boldsymbol{I}$, for any choice of the extended channels, i.e., that $\boldsymbol{C}'$ is an extended simulation of $\boldsymbol{S}_\mathsf{CC}$ in the sense of Definition~\ref{def:simulation_extended}. More formally, we want to show that
\begin{align}
    \boldsymbol{C}'(\A_{1}^{\prime\otimes N}\otimes\ldots\otimes\A_{N}^{\prime\otimes N}) = (\boldsymbol{S}_\mathsf{CC}\otimes\boldsymbol{I})(\A'_1\otimes\ldots\otimes\A'_N),
\end{align}
for all possible extended quantum channels $\A'_1,\ldots,\A'_N$. Equivalently, we can rephrase the last equation in the Choi picture as
\begin{align}
    C'*(A_{1}^{\prime\otimes N}\otimes\ldots \otimes A_{N}^{\prime\otimes N}) = S_{\mathsf{CC}}*(A'_1 \otimes \ldots \otimes A'_N),
\end{align}
for all possible extended quantum channels $A'_1,\ldots,A'_N$.
 
The proof consists in computing iteratively the output of the simulation scheme presented in Fig.~\ref{fig:simu_QCCC_bipartite_channels} and showing that it matches the output of the QC-CC applied on the extended channels, as defined in Eq.~\eqref{eq:output_channel_QCCC_2}.
 
Following Definition~\ref{def:QC-CC}, any QC-CC superchannel $\boldsymbol{S}_\mathsf{CC}$ is defined by a collection of internal operations $\{\mathcal{M}_{\emptyset}^{\to k_{1}},\ldots, \,  \mathcal{M}_{(k_1,\ldots,k_n)}^{\to k_{n+1}},\ldots, \, \mathcal{M}_{(k_1,\ldots,k_N)}^{\to F}\}.$
 
The output of the simulation scheme $\boldsymbol{C}'$ can be computed by composing iteratively the operations $\tilde{\M}_n$, $\mathcal{R}'_n$ and the extended channels $\A'_1 \otimes \ldots \otimes \A'_N$. The auxiliary systems are initialized in the pure state $\rho_{\text{aux}_{t_1}}=\bigotimes_{i=1}^N \rho_i\otimes\rho'_i \in \L(\H_{\text{aux}_{t_1}})$, with $\rho_i=\ketbra{a_i}{a_i}$ and $\rho'_i=\ketbra{a'_i}{a'_i}$. We describe by $\varrho_{(n)}'$ (resp.~$\varrho_{(n+1)}$) the simulation output at time step $n$, before (resp.~after) the operation $\tilde{\M}_{n+1}$ is applied:
 
\begin{align}
\varrho_{(n)}' & = (\rho_{\text{aux}_{t_1}} * \tilde M_1 * R'_1 *  \tilde{M}_2 * \cdots * \tilde M_n * R'_n) * (A_{1}^{\prime\otimes {n}}\otimes\ldots \otimes A_{N}^{\prime\otimes {n}}), \notag
\end{align}
and
\begin{align}
\varrho_{(n+1)} & = (\rho_{\text{aux}_{t_1}} * \tilde M_1 * R'_1 * \tilde M_2 * \cdots * R'_n * \tilde M_{n+1}) * (A_{1}^{\prime\otimes {n}}\otimes\ldots \otimes A_{N}^{\prime\otimes {n}}).
\end{align}
At the first time step, right after the first internal operation $\tilde \M_1$ has been applied, the simulation output is described by
\begin{align}
    \varrho_{(1)} & = \rho_{\text{aux}_{t_1}}* \tilde M_1 \nonumber \\
    &= \sum_{k_1} \bigotimes_{i=1}^N (\rho_i \otimes \rho'_i) \otimes M_{\emptyset}^{\to k_{1}} \otimes [\![(k_1)]\!]^{C_{1}}.
\end{align}
Applying the first extended channels in parallel along with the swap operations we obtain the Choi operator at the end of the first time step as
\begin{align}
    \varrho_{(1)}' & = \varrho_{(1)} * R'_1 * (A'_{1}\otimes\ldots \otimes A'_{N}) = \varrho_{(1)} * S'^I_1*(A'_1\otimes \ldots \otimes A'_N) * S'^O_1 \nonumber \\
    & = \bigg[\sum_{k_1} \Big[ \bigotimes_{i=1}^N (\rho_i \otimes \rho'_i) \otimes M_{\emptyset}^{\to k_{1}} \Big] * S'^I_{k_1}\otimes [\![(k_1)]\!]^{C^{\circ}_{1}} \bigg] *(A'_1\otimes \ldots \otimes A'_N) * S'^O_1\nonumber \\
    & = \bigg[\sum_{k_1} \Big[ \bigotimes_{\substack{i=1\\ i\neq k_1}}^N \big((\rho_i \otimes \rho'_i) * A'_i\big) \otimes (\rho_{k_1} \otimes \rho'_{k_1}) \otimes \big(M_{\emptyset}^{\to k_{1}} * A'_{k_1}\big) \otimes \bigotimes_{\substack{i=1\\ i\neq k_1}}^N \kketbra{\id}{\id}^{I'_i\,\text{ext}_{i,t_1^{\circ}}} \Big] \otimes [\![(k_1)]\!]^{C^{\circ}_{1}}\bigg] * S'^O_1 \nonumber \\
    & = \sum_{k_1} \Big[ \bigotimes_{\substack{i=1\\ i\neq k_1}}^N \big((\rho_i \otimes \rho'_i) * A'_i\big) \otimes (\rho_{k_1} \otimes \rho'_{k_1}) \otimes \big(M_{\emptyset}^{\to k_{1}} * A'_{k_1}\big) \otimes \bigotimes_{\substack{i=1\\ i\neq k_1}}^N \kketbra{\id}{\id}^{I'_i\,\text{ext}_{i,t_2}} \Big] \otimes [\![(k_1)]\!]^{C'_{1}}.
\end{align}
In the third line, the auxiliary states $\rho_{k_1}$ and $\rho'_{k_1}$ are stored in $\H_{\bar I_{k_1}}$ and $\H_{\text{ext}_{k_1,t_1^{\circ}}}$, respectively, while the copy of $\A'_{k_1}$ acts on the target system output by $M_{\emptyset}^{\to k_1}$ and on the extension system in $\H_{\text{ext}_{k_1,t_1}}=\H_{I'_{k_1}}$, and outputs them in $\H_{\text{aux}_{k_1,t_1^O}}$ and $\H_{\text{aux}'_{k_1,t_1^O}}$, respectively; these systems are then swapped back by $S'^O_1$. In the last line, the target system output by the copy of $\A'_{k_1}$ is thus in $\H_{O_{k_1}}$, its output extension system is stored in the extension wire $\H_{\text{ext}_{k_1,t_2}}$, and the auxiliary states $\rho_{k_1}$ and $\rho'_{k_1}$ are back in $\H_{\text{aux}_{k_1,t_2}}$ and $\H_{\text{aux}'_{k_1,t_2}}$. Moreover, $\kketbra{\id}{\id}^{I'_i\,\text{ext}_{i,t_2}}$ denotes the Choi operator of the identity channel from $\H_{I'_i}=\H_{\text{ext}_{i,t_1}}$ to $\H_{\text{ext}_{i,t_2}}$: it describes the extension wires associated with the extended channels $\A'_i$, $i\neq k_1$, which have not been applied on the target system yet, and which thus transmit the input extension systems unchanged. More generally, at the end of each time step $n$, the extension wire associated with an extended channel $\A'_{k_m}$ that has been applied on the target system at a time step $m\leq n$ carries the output extension system of this copy of $\A'_{k_m}$, transmitted with identity channels from $\H_{\text{ext}_{k_m,t_{m+1}}}$ to $\H_{\text{ext}_{k_m,t_{n+1}}}$. For the sake of conciseness, we absorb these identity channels into the Choi operator $A'_{k_m}$ of the copy of $\A'_{k_m}$ applied on the target system, whose output extension system is thus understood to be in $\H_{\text{ext}_{k_m,t_{n+1}}}$ at the end of time step $n$.
 
The output of the simulation scheme at the second time step before the extended channels are applied is
\begin{align}
     \varrho_{(2)} & = \varrho_{(1)}' * \tilde M_2 \nonumber \\
     &= \sum_{k_1,k_2} \bigg[\bigotimes_{\substack{i=1\\ i\neq k_1}}^N \big((\rho_i \otimes \rho'_i) * A'_i\big) \otimes (\rho_{k_1} \otimes \rho'_{k_1}) \otimes \big(M_{\emptyset}^{\to k_{1}} * A'_{k_1} * M_{(k_1)}^{\to k_{2}}\big) \otimes \bigotimes_{\substack{i=1\\ i\neq k_1}}^N \kketbra{\id}{\id}^{I'_i\,\text{ext}_{i,t_2}} \bigg] \otimes [\![(k_1,k_2)]\!]^{C_{2}}.
\end{align}
At the end of the second time step, once the extended channels and the swap operations have been applied, we note that each extended channel $\mathcal{A}'_k$ is applied twice on the corresponding pair of auxiliary systems in state $\rho_k\otimes\rho'_k$, except for $k=k_1,k_2$. Indeed, $\mathcal{A}'_{k_1},\mathcal{A}'_{k_2}$ are respectively applied only once on the corresponding auxiliary states since the other copy of this operation was applied on the target system state and on the corresponding extension system:
\begin{align}
     \varrho'_{(2)} & = \varrho_{(2)} * R'_2 * (A'_{1}\otimes\ldots \otimes A'_{N})\nonumber \\
     &= \sum_{k_1,k_2} \bigg[\bigotimes_{\substack{i=1\\ i\neq k_1,k_2}}^N \big((\rho_i \otimes \rho'_i) * A_i^{\prime *2}\big) \otimes \bigotimes_{j\in\{k_1,k_2\}} \big((\rho_j \otimes \rho'_j) * A'_j\big) \nonumber \\
     & \ \ \ \ \ \ \ \ \ \ \ \ \ \ \ \otimes \big(M_{\emptyset}^{\to k_{1}} * A'_{k_1} * M_{(k_1)}^{\to k_{2}} * A'_{k_2}\big) \otimes \bigotimes_{\substack{i=1\\ i\neq k_1,k_2}}^N \kketbra{\id}{\id}^{I'_i\,\text{ext}_{i,t_3}} \bigg] \otimes [\![(k_1,k_2)]\!]^{C_{2}'},
\end{align}
where $A_k^{\prime *2}=A'_k*A'_k$ is the Choi operator of the channel $\A_k^{\prime\circ2}=\A'_k\circ \A'_k$ obtained by composing the channel $\A'_k$ with itself, and where the extension wires associated with the extended channels $\A'_i$, $i\neq k_1,k_2$, which have not been applied on the target system yet, now transmit the input extension systems from $\H_{I'_i}$ to $\H_{\text{ext}_{i,t_3}}$.
Deriving iteratively the Choi operator after each time step, we eventually obtain the output of the simulation at the second to last time step $N$:
\begin{align}
     &\varrho'_{(N)} = \sum_{(k_1,\ldots,k_N)} \Big[\bigotimes_{j=1}^N \big((\rho_j \otimes \rho'_j) * A_j^{\prime *(N-1)}\big) \otimes \big(M_\emptyset^{\to k_1} \!*\! A'_{k_1} * \cdots \!*\! M_{(k_1,\ldots,k_{N-1})}^{\to k_N} \!*\! A'_{k_N}\big)\Big] \otimes [\![(k_1,\ldots,k_N)]\!]^{C'_{N}}.
\end{align}
Since all extended channels have now been applied on the target system, no identity channel remains on the extension wires: the input extension systems of the copies of $\A'_{k_1},\ldots,\A'_{k_N}$ applied on the target system are in $\H_{\text{ext}_{t_1}}=\H_{I'_{[N]}}$, and their output extension systems are in $\H_{\text{ext}_{t_{N+1}}}=\H_{O'_{[N]}}$, as in Eq.~\eqref{eq:output_channel_QCCC_2}.
Applying the last internal operation $\tilde {\mathcal{M}}_{N+1}$:
\begin{align}
     &\varrho_{(N+1)} = \varrho_{(N)}' * \tilde M_{N+1} \nonumber \\
     &= \sum_{(k_1,\ldots,k_N)} \bigotimes_{j=1}^N \big((\rho_j \otimes \rho'_j) * A_j^{\prime *(N-1)}\big) \otimes \big(M_\emptyset^{\to k_1} \!*\! A'_{k_1} * \cdots \!*\! M_{(k_1,\ldots,k_{N-1})}^{\to k_N} \!*\! A'_{k_N} \!*\! M_{(k_1,\ldots,k_N)}^{\to F}\big) \nonumber \\
     &= \bigotimes_{j=1}^N \big((\rho_j \otimes \rho'_j) * A_j^{\prime *(N-1)}\big) \otimes \left[\sum_{(k_1,\ldots,k_N)} M_\emptyset^{\to k_1} \!*\! A'_{k_1} * \cdots \!*\! M_{(k_1,\ldots,k_{N-1})}^{\to k_N} \!*\! A'_{k_N} \!*\! M_{(k_1,\ldots,k_N)}^{\to F}\right].
\end{align}
Tracing out the auxiliary systems in $\mathcal{H}_{\text{aux}_{t_{N+1}}}$, which are in the normalized states $(\rho_j\otimes\rho'_j)*A_j^{\prime *(N-1)}$, we indeed retrieve the output $S_{\mathsf{CC}}*(A'_1\otimes \ldots \otimes A'_N)$ of the QC-CC superchannel applied on the extended channels, as defined in Eq.~\eqref{eq:output_channel_QCCC_2}, which concludes the proof.
 
As in Sec.~\ref{subsubsec::proof_simulation_QCCC}, because the correlations with the auxiliary systems are classical, one recovers the intermediate state of the QC-CC applied on the extended channels at each single time step of the simulation by tracing out the auxiliary systems. We also observe that $\A'_k$ is applied exactly $N-1$ times on each pair of auxiliary systems in state $\rho_k\otimes\rho'_k = \ketbra{a_k}{a_k}\otimes\ketbra{a'_k}{a'_k}$, so that this extended simulation is not catalytic either.
\end{proof}

\subsubsection{Catalytic simulation}
\label{subsubsec::proof_catalyticsimulation_QCCC}

We finally show that a catalytic simulation, as defined in Definition~\ref{def:simulation_catalytic}, exists for any QC-CC. For the sake of simplicity, we exhibit such a simulation for input quantum channels $\A_k:\L(\H_{I_k})\to\L(\H_{O_k})$, and we explain at the end of the proof how it can be straightforwardly combined with the extended simulation of Sec.~\ref{subsubsec::proof_extendedsimulation_QCCC}.

In the simulation of Sec.~\ref{subsubsec::proof_simulation_QCCC}, the $N-1$ copies of each input channel $\A_k$ that are not applied on the target system all act on the same auxiliary system, whose final state $\rho_k*A_k^{*(N-1)}$ is eventually traced out. To make the simulation catalytic, we distribute these $N-1$ applications of $\A_k$ over $N-1$ distinct auxiliary systems. The $N-1$ applications of each channel $\A_k$ are returned, unchanged, as part of the output of the simulation, see Fig.~\ref{fig:simu_QCCC_catalytic}.

\begin{proof}
Given a QC-CC superchannel $\boldsymbol{S}_\mathsf{CC}:\bigotimes_{i=1}^N[\L(\H_{I_i})\to\L(\H_{O_i})]\to[\L(\H_{P})\to\L(\H_{F})]$, we construct a fixed-order quantum circuit $\widehat{\boldsymbol{C}}:\bigotimes_{i=1}^N[\L(\H_{I_i})\to\L(\H_{O_i})]^{\otimes N}\to[\L(\H_{P}\otimes\bigotimes_{i=1}^N\H_{I_i}^{\otimes(N-1)})\to\L(\H_{F}\otimes\bigotimes_{i=1}^N\H_{O_i}^{\otimes(N-1)})]$ which uses $N$ queries of each input channel $\A_k:\L(\H_{I_k}) \to \L(\H_{O_k})$ to catalytically simulate the action of the QC-CC. Analogously to the proofs presented in Secs.~\ref{subsubsec::proof_simulation_QCCC} and~\ref{subsubsec::proof_extendedsimulation_QCCC}, we start by formally defining $\widehat{\boldsymbol{C}}$ and showing that it is a valid QC-FO superchannel, before showing that $\widehat{\boldsymbol{C}}$ is indeed a catalytic simulation of $\boldsymbol{S}_\mathsf{CC}$.\\

On top of the $N$ copies of each input channel $\A_k$, the simulation scheme uses, for each $k\in[N]$, $N-1$ auxiliary systems, hence $N(N-1)$ auxiliary systems in total. Contrary to Secs.~\ref{subsubsec::proof_simulation_QCCC} and~\ref{subsubsec::proof_extendedsimulation_QCCC}, these auxiliary systems are not initialized in fixed states: their inputs and outputs are left open, and carry the inputs and outputs of the $N-1$ catalyst copies of each channel $\A_k$. The simulation then proceeds in accordance with the structure of a QC-CC, considering $N+1$ time steps labeled by $n \in [N+1]$, each except the last one corresponding to the execution of an operation $\A_k$ on the target system. At the beginning of each time step $n$, the $N-1$ auxiliary systems associated with $\A_k$ are carried by the wires $\H_{\text{aux}_{k,t_n}^i}$, $i\in[N-1]$, each isomorphic to $\H_{I_k}$, and we say that the auxiliary system carried by $\H_{\text{aux}_{k,t_n}^i}$ is in position $i$. We denote the auxiliary space associated with $\A_k$ at the beginning of time step $n$ as $\H_{\text{aux}_{k,t_n}}\coloneqq\bigotimes_{i=1}^{N-1}\H_{\text{aux}_{k,t_n}^i}$, and the global auxiliary space as $\H_{\text{aux}_{t_n}}\coloneqq\bigotimes_{k=1}^N\H_{\text{aux}_{k,t_n}}$. For consistency with Definition~\ref{def:simulation_catalytic}, the input spaces $\H_{\text{aux}_{k,t_1}^i}$ and output spaces $\H_{\text{aux}_{k,t_{N+1}}^i}$ are isomorphic to $\H_{I_k}$ and $\H_{O_k}$, respectively, i.e., $\H_{\text{aux}_{t_1}}\cong\bigotimes_{k=1}^N\H_{I_k}^{\otimes(N-1)}$ and $\H_{\text{aux}_{t_{N+1}}}\cong\bigotimes_{k=1}^N\H_{O_k}^{\otimes(N-1)}$.

The action of the simulation at each time step $n \in [N]$ corresponds to the sequential composition of the following four steps:
\begin{enumerate}
    \item First, the application of the internal QC-CC operation $\tilde M_{n}$ to the target, auxiliary and control systems, as defined in Eqs.~\eqref{eq:def_tilde_M1_QCCC}--\eqref{eq:def_tilde_MN1_QCCC}.
    \item Then, controlled on the last label $k_n$ of the control state $[\![(k_1,\ldots,k_n)]\!]$, a routing operation $\hat{\mathcal{S}}^I_{k_n}: \mathcal{L}(\mathcal{H}_{\text{aux}_{t_n}} \otimes \mathcal{H}_{I_{k_n}}) \to \mathcal{L}(\mathcal{H}_{\text{aux}_{t_n^I}} \otimes \mathcal{H}_{\bar I_{k_n}})$ swaps the incoming target system in $\H_{I_{k_n}}$ coming from $\M_{(k_1,\ldots,k_{n-1})}^{\to k_n}$ with the auxiliary system in position 1 associated with $\A_{k_n}$, in $\H_{\text{aux}_{k_n,t_n}^1}$, and outputs the target system in $\H_{\text{aux}_{k_n,t_n^I}^1}$ and the auxiliary system in $\H_{\bar I_{k_n}}$ (which is isomorphic to $\H_{I_{k_n}}$), while it transmits with an identity channel all the other auxiliary systems, from $\H_{\text{aux}_{k,t_n}^i}$ to $\H_{\text{aux}_{k,t_n^I}^i}$. The Hilbert space $\mathcal{H}_{\text{aux}_{t_n^I}}\coloneqq\bigotimes_{k=1}^N\bigotimes_{i=1}^{N-1} \H_{\text{aux}_{k,t_n^I}^i}$ describes the collection of all auxiliary Hilbert spaces $\H_{\text{aux}_{k,t_n^I}^i}$ (which are isomorphic to $\H_{I_k}$) right after the application of $\hat{\mathcal{S}}^I_{k_n}$. The overall controlled routing operation $\hat{\map{S}}^I_n$ applied at time step $n \in [N]$ can be described in the Choi picture as
\begin{align}
    \hat S^I_n = \sum_{(k_1,\ldots,k_n)} \hat S^I_{k_n} \otimes [\![(k_1,\ldots,k_n)]\!]^{C_n} \otimes [\![(k_1,\ldots,k_n)]\!]^{C_n^{\circ}},
    \label{eq:routing_in_cat}
\end{align}
    where, as in Eq.~\eqref{eq:routing_in}, $\H_{C_n}$ is the Hilbert space corresponding to the control system stemming from $\tilde{\M}_n$ and input in $\hat{\map{S}}^I_n$, and $\H_{C_n^{\circ}}$ is the Hilbert space corresponding to the control system output by $\hat{\map{S}}^I_n$ and input in $\hat{\map{S}}^O_n$.
    \item A copy $\A_k:\L(\H_{\text{aux}_{k,t_n^I}^1})\to \L(\H_{\text{aux}_{k,t_n^O}^1})$ of each quantum channel (with the same convention as in Sec.~\ref{subsubsec::proof_simulation_QCCC} regarding the input and output Hilbert spaces of the copies) is then applied on the system in position 1 associated with $\A_k$, and outputs it in $\H_{\text{aux}_{k,t_n^O}^1}$, which is isomorphic to $\H_{O_k}$, while the auxiliary systems in positions $i\in\{2,\ldots,N-1\}$ are transmitted with an identity channel from $\H_{\text{aux}_{k,t_n^I}^i}$ to $\H_{\text{aux}_{k,t_n^O}^i}$. Such an identity channel simply relabels the Hilbert space $\H_{\text{aux}_{k,t_n^I}^i}$ as $\H_{\text{aux}_{k,t_n^O}^i}$. The Hilbert space $\mathcal{H}_{\text{aux}_{t_n^O}}\coloneqq\bigotimes_{k=1}^N\bigotimes_{i=1}^{N-1} \H_{\text{aux}_{k,t_n^O}^i}$ describes the collection of all auxiliary Hilbert spaces right after the application of the $N$ external channels. In particular, the copy of $\A_{k_n}$ acts on the target system, while each copy of $\A_k$, $k\neq k_n$, acts on the auxiliary system in position 1 
    associated with $\A_k$.
        \item Finally, the routing is inverted by $\hat{\mathcal{S}}^O_{k_n}: \mathcal{L}(\mathcal{H}_{\text{aux}_{t_n^O}} \otimes \mathcal{H}_{\bar I_{k_n}}) \to \mathcal{L}(\mathcal{H}_{\text{aux}_{t_{n+1}}} \otimes \mathcal{H}_{O_{k_n}})$, which swaps the target state in $\mathcal{H}_{\text{aux}_{k_n,t_n^O}^1}$ with the auxiliary state in $\mathcal{H}_{\bar I_{k_n}}$, the target state being eventually output in $\mathcal{H}_{O_{k_n}}$ and sent to $\M_{(k_1,\ldots,k_{n})}^{\to k_{n+1}}$, and the auxiliary state being eventually output in $\mathcal{H}_{\text{aux}_{k_n,t_{n+1}}^1}$. In addition, for each $k\neq k_n$, $\hat{\mathcal{S}}^O_{k_n}$ cyclically permutes the auxiliary systems associated with $\A_k$: it moves the auxiliary system in $\H_{\text{aux}_{k,t_n^O}^1}$, which has just received the action of $\A_k$, to the last position $\H_{\text{aux}_{k,t_{n+1}}^{N-1}}$, and the auxiliary system in $\H_{\text{aux}_{k,t_n^O}^{i+1}}$ to $\H_{\text{aux}_{k,t_{n+1}}^{i}}$ for $i\in[N-2]$. The other auxiliary systems associated with $\A_{k_n}$, in positions $i\in\{2,\ldots,N-1\}$, keep their positions. The overall controlled routing operation $\hat{\map{S}}^O_n$ applied at time step $n$ can be described in the Choi picture as
    \begin{align}
    \hat S^O_n = \sum_{(k_1,\ldots,k_n)} \hat S^O_{k_n} \otimes [\![(k_1,\ldots,k_n)]\!]^{C_n^{\circ}} \otimes [\![(k_1,\ldots,k_n)]\!]^{C_n'},
     \label{eq:routing_out_cat}
\end{align}
where, as in Eq.~\eqref{eq:routing_out}, $\H_{C_n^{\prime}}$ is the Hilbert space corresponding to the control system output by $\hat{\map{S}}^O_n$ and input in $\tilde{\M}_{n+1}$.

We will denote by $\hat{\mathcal{R}}_n$ the parallel higher-order transformation with $N$ slots obtained by applying $\hat{\map{S}}^I_n$, leaving open one slot $\H_{\text{aux}_{k,t_n^I}^1}\to\H_{\text{aux}_{k,t_n^O}^1}$ for each channel $\A_k$, $k \in [N]$, and then applying $\hat{\map{S}}^O_n$. In the Choi picture, it is described by $\hat R_n = \hat S^I_n * \hat S^O_n$, where the auxiliary systems in positions $i\in\{2,\ldots,N-1\}$, output by $\hat{\map{S}}^I_n$ in $\H_{\text{aux}_{k,t_n^I}^i}$, are input in $\hat{\map{S}}^O_n$ in $\H_{\text{aux}_{k,t_n^O}^i}$ after the relabeling described in the third step.
\end{enumerate}

This routing distributes the $N-1$ copies of each channel $\A_k$ that are not applied on the target system over the $N-1$ auxiliary systems associated with $\A_k$, in such a way that each of them receives the action of $\A_k$ exactly once. Indeed, since $\A_k$ is selected at exactly one time step, the copy of $\A_k$ acts on the auxiliary system in position 1 at exactly $N-1$ time steps, each of which is followed by the cyclic permutation of the auxiliary systems associated with $\A_k$. Hence, the auxiliary system initially in position $j\in[N-1]$ is in position 1 precisely at the $j$-th time step at which $\A_k$ is not selected, and it receives the action of $\A_k$ then, and only then. At the time step at which $\A_k$ is selected, the auxiliary system in position 1 associated with $\A_k$ is stored in $\H_{\bar I_k}$ while the copy of $\A_k$ acts on the target system, and it is put back in position 1 untouched, while the other auxiliary systems associated with $\A_k$ are not permuted. Finally, since the auxiliary systems associated with each channel $\A_k$ undergo exactly $N-1$ times the cyclic permutation of their $N-1$ positions, they are all back in their initial positions at the end of time step $N$, independently of the order $(k_1,\ldots,k_N)$ in which the channels are applied on the target system. Note that the routing operations $\hat{\mathcal{S}}^I_{k_n}$ and $\hat{\mathcal{S}}^O_{k_n}$ only depend on the last label $k_n$ of the control state, as depicted in Fig.~\ref{fig:simu_QCCC_catalytic}.

The last time step $n=N+1$ consists in applying the last QC-CC internal operation $\tilde\M_{N+1}$, which transforms the target, auxiliary and control systems into a state output in the global future Hilbert space $\H_F$. Contrary to Secs.~\ref{subsubsec::proof_simulation_QCCC} and~\ref{subsubsec::proof_extendedsimulation_QCCC}, the auxiliary systems in $\H_{\text{aux}_{t_{N+1}}}$ are not traced out, but are output as the outputs of the catalyst copies of the input channels.

For any QC-CC superchannel $\boldsymbol{S}_\mathsf{CC}:\bigotimes_{i=1}^N[\L(\H_{I_i})\to\L(\H_{O_i})]\to[\L(\H_{P})\to\L(\H_{F})]$, we can thus formally define the superchannel $\widehat{\boldsymbol{C}}:\bigotimes_{i=1}^N[\L(\H_{I_i})\to\L(\H_{O_i})]^{\otimes N}\to[\L(\H_{P}\otimes\H_{\text{aux}_{t_1}})\to\L(\H_{F}\otimes\H_{\text{aux}_{t_{N+1}}})]$ as the alternating composition, step by step, of the procedure described above, except the third step as it corresponds to the application of the quantum channels $\A_k$ which are input in $\widehat{\boldsymbol{C}}$. The Choi operator of the superchannel $\widehat{\boldsymbol{C}}$ is thus described as
\begin{align}
\widehat{C} & = \tilde M_1 * \hat R_1 * \tilde M_2 * \hat R_2  * \cdots * \tilde M_N * \hat R_N * \tilde M_{N+1}.
\label{eq:output_state_simu_QCCC_catalytic}
\end{align}
Note that, contrary to Eqs.~\eqref{eq:output_state_simu_QCCC} and~\eqref{eq:output_state_simu_QCCC_extended}, no auxiliary state is prepared and no auxiliary system is traced out: the auxiliary spaces $\H_{\text{aux}_{t_1}}\cong\bigotimes_{k=1}^N\H_{I_k}^{\otimes(N-1)}$ and $\H_{\text{aux}_{t_{N+1}}}\cong\bigotimes_{k=1}^N\H_{O_k}^{\otimes(N-1)}$ remain open in Eq.~\eqref{eq:output_state_simu_QCCC_catalytic}, and play, together with $\H_P$ and $\H_F$, the role of the global past and global future of $\widehat{\boldsymbol{C}}$.

That $\tilde M_{n}$ is CPTP for $n\in[N]$ follows from the fact that $\{\mathcal{M}_{(k_1,\ldots,k_{n-1})}^{\rightarrow{k_{n}}}\}_{k_{n}\in [N] \setminus \{k_1,\ldots,k_{n-1}\}}$ is a quantum instrument, and that $\tilde M_{N+1}$ is CPTP follows from the fact that the maps $\mathcal{M}_{(k_1,\ldots,k_N)}^{\to F}$ are CPTP. The maps $\hat{\map{S}}^I_n$ and $\hat{\map{S}}^O_n$ are clearly CPTP, and $\hat{\mathcal{R}}_n$ is a valid parallel higher-order superchannel. Thus, at each time step $\tilde M_n * \hat R_n$ defines a valid parallel higher-order process, which can be interpreted as a QC-FO between the $N$ input channels $\A_1,\ldots,\A_N$. Finally, by Eq.~\eqref{eq:output_state_simu_QCCC_catalytic}, the superchannel $\widehat{\boldsymbol{C}}$ is obtained by applying sequentially the blocks $\tilde M_1 * \hat R_1, \ldots, \tilde M_N * \hat R_N$ and then $\tilde M_{N+1}$, the auxiliary systems being carried along as memory systems from the global past to the global future. Since each block $\tilde M_n * \hat R_n$ is a QC-FO, while $\tilde M_{N+1}$ is a CPTP map, this sequential composition is itself a QC-FO, i.e., $\widehat{\boldsymbol{C}}$ is a valid QC-FO superchannel.

\begin{figure*}[h!]
    \centering
    \includegraphics[width=\columnwidth]{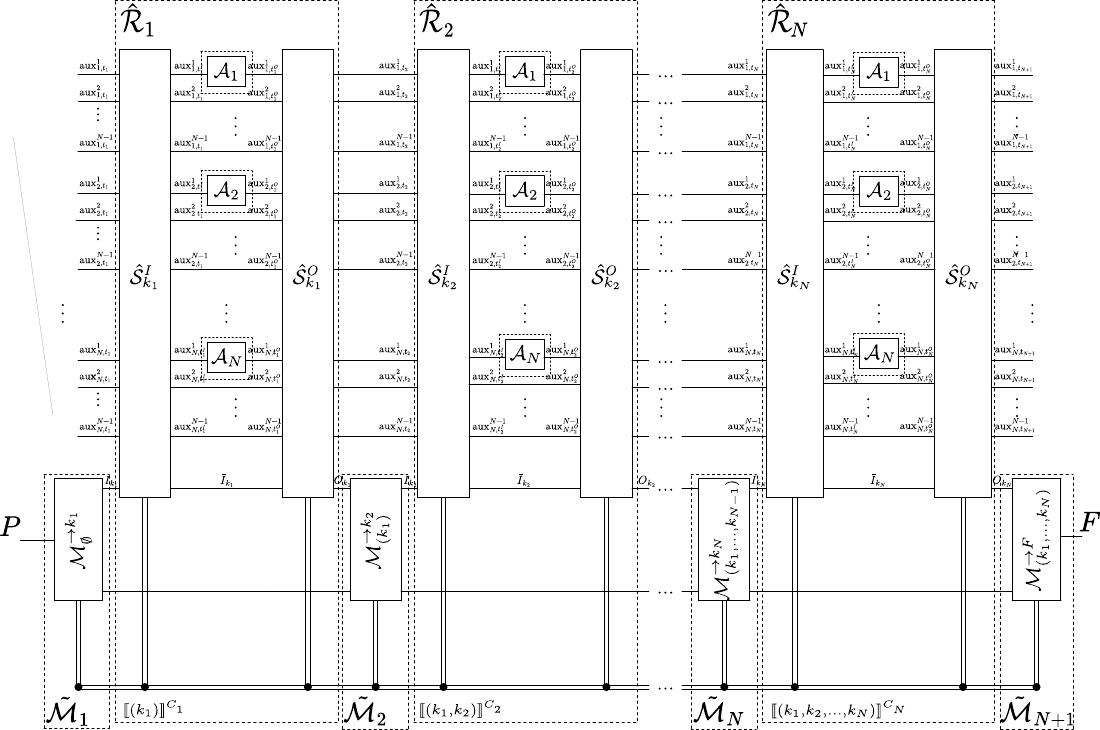}
    \caption{Catalytic fixed order simulation of a QC-CC with input quantum channels $\A_1,\ldots,\A_N$. $N$ queries of each input channel $\A_k$ are made: at each time step, one copy of $\A_k$ acts on the first wire $\text{aux}_k^1$ associated with $\A_k$, which carries either the target system, if $\A_k$ is selected, or an auxiliary system that has not yet received the action of $\A_k$. The inputs and outputs of the $N(N-1)$ auxiliary systems are left open, so that the $N^2-N$ copies of the input channels that are not applied on the target system are returned as part of the output of the simulation.}
    \label{fig:simu_QCCC_catalytic}
\end{figure*}

\vspace{5mm}

We will now show that $\widehat{\boldsymbol{C}}$ applied on $N$ copies of each input channel is a catalytic simulation of $\boldsymbol{S}_\mathsf{CC}$ in the sense of Definition~\ref{def:simulation_catalytic}, for any choice of the input channels. More formally, we want to show that
\begin{align}
    \widehat{\boldsymbol{C}}(\A_{1}^{\otimes N}\otimes\ldots\otimes\A_{N}^{\otimes N}) = \boldsymbol{S}_\mathsf{CC}(\A_1\otimes\ldots\otimes\A_N)\otimes\A_1^{\otimes(N-1)}\otimes\ldots\otimes\A_N^{\otimes(N-1)},
\end{align}
for all possible quantum channels $\A_1,\ldots,\A_N$, where the $N-1$ catalyst copies of each $\A_k$ on the right-hand side act from $\H_{\text{aux}_{k,t_1}^i}$ to $\H_{\text{aux}_{k,t_{N+1}}^i}$, $i\in[N-1]$. Equivalently, we can rephrase the last equation in the Choi picture as
\begin{align}
    \widehat{C}*(A_{1}^{\otimes N}\otimes\ldots \otimes A_{N}^{\otimes N}) = S_{\mathsf{CC}}*(A_1 \otimes \ldots \otimes A_N)\otimes A_1^{\otimes(N-1)}\otimes\ldots\otimes A_N^{\otimes(N-1)},
    \label{eq:catalytic_statement_Choi}
\end{align}
for all possible quantum channels $A_1,\ldots,A_N$.

The proof consists in computing iteratively the output of the simulation scheme presented in Fig.~\ref{fig:simu_QCCC_catalytic} and showing that it factorizes into the output of the QC-CC as defined in Eq.~\eqref{eq:output_channel_QCCC} and $N-1$ copies of each input channel.

Following Definition~\ref{def:QC-CC}, any QC-CC superchannel $\boldsymbol{S}_\mathsf{CC}$ is defined by a collection of internal operations $\{\mathcal{M}_{\emptyset}^{\to k_{1}},\ldots,\allowbreak \,  \mathcal{M}_{(k_1,\ldots,k_n)}^{\to k_{n+1}},\allowbreak\ldots, \, \mathcal{M}_{(k_1,\ldots,k_N)}^{\to F}\}.$

Since the auxiliary inputs of $\widehat{\boldsymbol{C}}$ are left open, we compute the output of the simulation scheme when arbitrary operators $\rho_k^i\in\L(\H_{\text{aux}_{k,t_1}^i})$ are input in them, for $k\in[N]$ and $i\in[N-1]$, and we denote $\rho_{\text{aux}_{t_1}}\coloneqq\bigotimes_{k=1}^N\bigotimes_{i=1}^{N-1}\rho_k^i \in \L(\H_{\text{aux}_{t_1}})$. We refer to the auxiliary system initially in position $i$ associated with $\A_k$ as the auxiliary system carrying $\rho_k^i$. The output of the simulation scheme can then be computed by composing iteratively the operations $\tilde{\M}_n$, $\hat{\mathcal{R}}_n$ and the quantum channels $\A_1 \otimes \ldots \otimes \A_N$. We describe by $\varrho_{(n)}'$ (resp.~$\varrho_{(n+1)}$) the simulation output at time step $n$, before (resp.~after) the operation $\tilde{\M}_{n+1}$ is applied:
\begin{align}
\varrho_{(n)}' & = (\rho_{\text{aux}_{t_1}} * \tilde M_1 * \hat R_1 *  \tilde{M}_2 * \cdots * \tilde M_n * \hat R_n) * (A_{1}^{\otimes {n}}\otimes\ldots \otimes A_{N}^{\otimes {n}}), \notag
\end{align}
and
\begin{align}
\varrho_{(n+1)} & = (\rho_{\text{aux}_{t_1}} * \tilde M_1 * \hat R_1 * \tilde M_2 * \cdots * \hat R_n * \tilde M_{n+1}) * (A_{1}^{\otimes {n}}\otimes\ldots \otimes A_{N}^{\otimes {n}}).
\end{align}
At the first time step, right after the first internal operation $\tilde \M_1$ has been applied, the simulation output is described by
\begin{align}
    \varrho_{(1)} & = \rho_{\text{aux}_{t_1}}* \tilde M_1 \nonumber \\
    &= \sum_{k_1} \bigotimes_{k=1}^N\bigotimes_{i=1}^{N-1} \rho_k^i \otimes M_{\emptyset}^{\to k_{1}} \otimes [\![(k_1)]\!]^{C_{1}}.
\end{align}
Applying the first external channels in parallel along with the routing operations, we obtain the Choi operator at the end of the first time step as
\begin{align}
    \varrho_{(1)}' & = \varrho_{(1)} * \hat R_1 * (A_{1}\otimes\ldots \otimes A_{N}) = \varrho_{(1)} * \hat S^I_1*(A_1\otimes \ldots \otimes A_N) * \hat S^O_1 \nonumber \\
    & = \bigg[\sum_{k_1} \Big[ \bigotimes_{k=1}^N\bigotimes_{i=1}^{N-1} \rho_k^i \otimes M_{\emptyset}^{\to k_{1}} \Big] * \hat S^I_{k_1}\otimes [\![(k_1)]\!]^{C^{\circ}_{1}} \bigg] *(A_1\otimes \ldots \otimes A_N) * \hat S^O_1\nonumber \\
    & = \bigg[\sum_{k_1} \Big[ \bigotimes_{\substack{k=1\\ k\neq k_1}}^N \Big(\big(\rho_k^1 * A_k\big) \otimes \bigotimes_{i=2}^{N-1}\rho_k^i\Big) \otimes \bigotimes_{i=1}^{N-1}\rho_{k_1}^i \otimes \big(M_{\emptyset}^{\to k_{1}} * A_{k_1}\big) \Big] \otimes [\![(k_1)]\!]^{C^{\circ}_{1}}\bigg] * \hat S^O_1 \nonumber \\
    & = \sum_{k_1} \Big[ \bigotimes_{\substack{k=1\\ k\neq k_1}}^N \Big(\big(\rho_k^1 * A_k\big) \otimes \bigotimes_{i=2}^{N-1}\rho_k^i\Big) \otimes \bigotimes_{i=1}^{N-1}\rho_{k_1}^i \otimes \big(M_{\emptyset}^{\to k_{1}} * A_{k_1}\big) \Big] \otimes [\![(k_1)]\!]^{C'_{1}}.
\end{align}
In the third line, the auxiliary system carrying $\rho_{k_1}^1$ is stored in $\H_{\bar I_{k_1}}$, while the copy of $\A_{k_1}$ acts on the target system output by $M_{\emptyset}^{\to k_1}$, and each copy of $\A_k$, $k\neq k_1$, acts on the auxiliary system in position 1 associated with $\A_k$, which carries $\rho_k^1$. In the last line, the target system output by the copy of $\A_{k_1}$ is in $\H_{O_{k_1}}$ and the auxiliary system carrying $\rho_{k_1}^1$ is back in $\H_{\text{aux}_{k_1,t_2}^1}$, while, for each $k\neq k_1$, the auxiliary systems associated with $\A_k$ have been cyclically permuted: $\rho_k^1*A_k$ is now in $\H_{\text{aux}_{k,t_2}^{N-1}}$, and $\rho_k^i$ is in $\H_{\text{aux}_{k,t_2}^{i-1}}$ for $i\in\{2,\ldots,N-1\}$. In particular, for $N\geq3$, the auxiliary system in position 1 associated with $\A_k$, $k\neq k_1$, now carries $\rho_k^2$, on which $\A_k$ has not acted yet.

The output of the simulation scheme at the second time step before the external channels are applied is
\begin{align}
     \varrho_{(2)} & = \varrho_{(1)}' * \tilde M_2 \nonumber \\
     &= \sum_{k_1,k_2} \bigg[\bigotimes_{\substack{k=1\\ k\neq k_1}}^N \Big(\big(\rho_k^1 * A_k\big) \otimes \bigotimes_{i=2}^{N-1}\rho_k^i\Big) \otimes \bigotimes_{i=1}^{N-1}\rho_{k_1}^i \otimes \big(M_{\emptyset}^{\to k_{1}} * A_{k_1} * M_{(k_1)}^{\to k_{2}}\big) \bigg] \otimes [\![(k_1,k_2)]\!]^{C_{2}}.
\end{align}
At the second time step, the copy of $\A_{k_2}$ acts on the target system, while, for each $k\neq k_2$, the copy of $\A_k$ acts on the auxiliary system in position 1 associated with $\A_k$, which carries $\rho_k^2$ if $k\neq k_1$, and $\rho_{k_1}^1$ if $k=k_1$ (since the auxiliary systems associated with $\A_{k_1}$ have not been permuted at the first time step). In both cases, $\A_k$ acts on an auxiliary system on which it has not acted before. Hence, once the external channels and the routing operations have been applied, each channel $\A_k$ has acted on the auxiliary systems carrying $\rho_k^1$ and $\rho_k^2$ if $k\notin\{k_1,k_2\}$, and only on the auxiliary system carrying $\rho_k^1$ if $k\in\{k_1,k_2\}$, the other copy of $\A_k$ having been applied on the target system:
\begin{align}
     \varrho'_{(2)} & = \varrho_{(2)} * \hat R_2 * (A_{1}\otimes\ldots \otimes A_{N})\nonumber \\
     &= \sum_{k_1,k_2} \bigg[\bigotimes_{\substack{k=1\\ k\notin\{k_1,k_2\}}}^N \Big(\bigotimes_{i=1}^{2}\big(\rho_k^i * A_k\big) \otimes \bigotimes_{i=3}^{N-1}\rho_k^i\Big) \otimes \bigotimes_{j\in\{k_1,k_2\}} \Big(\big(\rho_j^1 * A_j\big) \otimes \bigotimes_{i=2}^{N-1}\rho_j^i\Big) \nonumber \\
     & \ \ \ \ \ \ \ \ \ \ \ \ \ \ \ \otimes \big(M_{\emptyset}^{\to k_{1}} * A_{k_1} * M_{(k_1)}^{\to k_{2}} * A_{k_2}\big) \bigg] \otimes [\![(k_1,k_2)]\!]^{C_{2}'}.
\end{align}
More generally, deriving iteratively the Choi operator after each time step, we obtain at the end of each time step $n\in[N]$
\begin{align}
     \varrho'_{(n)} = \sum_{(k_1,\ldots,k_n)} \bigg[\bigotimes_{k=1}^N \Big(\bigotimes_{i=1}^{s_k}\big(\rho_k^i * A_k\big) \otimes \bigotimes_{i=s_k+1}^{N-1}\rho_k^i\Big) \otimes \big(M_\emptyset^{\to k_1} \!*\! A_{k_1} * \cdots \!*\! M_{(k_1,\ldots,k_{n-1})}^{\to k_n} \!*\! A_{k_n}\big)\bigg] \otimes [\![(k_1,\ldots,k_n)]\!]^{C'_{n}},
     \label{eq:rho_n_catalytic}
\end{align}
where $s_k\in\{n-1,n\}$, which depends on $(k_1,\ldots,k_n)$, is the number of time steps among the first $n$ at which $\A_k$ has not been selected, i.e., $s_k=n-1$ if $k\in\{k_1,\ldots,k_n\}$ and $s_k=n$ otherwise, and where the auxiliary system carrying $\rho_k^i$ is in $\H_{\text{aux}_{k,t_{n+1}}^{i-s_k}}$ if $i>s_k$, and in $\H_{\text{aux}_{k,t_{n+1}}^{i-s_k+N-1}}$ if $i\leq s_k$. Indeed, Eq.~\eqref{eq:rho_n_catalytic} holds for $n=1,2$, and if it holds at time step $n-1$, then at time step $n$ the copy of each $\A_k$, $k\neq k_n$, acts on the auxiliary system in position 1 associated with $\A_k$, which carries $\rho_k^{s_k}$ and has not yet received the action of $\A_k$, before these auxiliary systems are cyclically permuted, while the auxiliary systems associated with $\A_{k_n}$ are left untouched. At the second to last time step $N$, every channel has been selected once, so that $s_k=N-1$ for all $k\in[N]$:
\begin{align}
     &\varrho'_{(N)} = \sum_{(k_1,\ldots,k_N)} \Big[\bigotimes_{k=1}^N\bigotimes_{i=1}^{N-1} \big(\rho_k^i * A_k\big) \otimes \big(M_\emptyset^{\to k_1} \!*\! A_{k_1} * \cdots \!*\! M_{(k_1,\ldots,k_{N-1})}^{\to k_N} \!*\! A_{k_N}\big)\Big] \otimes [\![(k_1,\ldots,k_N)]\!]^{C'_{N}}.
\end{align}
Each auxiliary system has thus received the action of the corresponding channel exactly once, and is back in its initial position, i.e., $\rho_k^i * A_k\in\L(\H_{\text{aux}_{k,t_{N+1}}^i})$, independently of the order $(k_1,\ldots,k_N)$.
Applying the last internal operation $\tilde {\mathcal{M}}_{N+1}$:
\begin{align}
     &\varrho_{(N+1)} = \varrho_{(N)}' * \tilde M_{N+1} \nonumber \\
     &= \sum_{(k_1,\ldots,k_N)} \bigotimes_{k=1}^N\bigotimes_{i=1}^{N-1} \big(\rho_k^i * A_k\big) \otimes \big(M_\emptyset^{\to k_1} \!*\! A_{k_1} * \cdots \!*\! M_{(k_1,\ldots,k_{N-1})}^{\to k_N} \!*\! A_{k_N} \!*\! M_{(k_1,\ldots,k_N)}^{\to F}\big) \nonumber \\
     &= \bigotimes_{k=1}^N\bigotimes_{i=1}^{N-1} \big(\rho_k^i * A_k\big) \otimes \left[\sum_{(k_1,\ldots,k_N)} M_\emptyset^{\to k_1} \!*\! A_{k_1} * \cdots \!*\! M_{(k_1,\ldots,k_{N-1})}^{\to k_N} \!*\! A_{k_N} \!*\! M_{(k_1,\ldots,k_N)}^{\to F}\right] \nonumber \\
     &= \bigotimes_{k=1}^N\bigotimes_{i=1}^{N-1} \big(\rho_k^i * A_k\big) \otimes S_{\mathsf{CC}}*(A_1\otimes \ldots \otimes A_N).
\end{align}
Since the final state of the auxiliary systems does not depend on the order $(k_1,\ldots,k_N)$, the output factorizes into the output $S_{\mathsf{CC}}*(A_1\otimes \ldots \otimes A_N)$ of the QC-CC superchannel, as defined in Eq.~\eqref{eq:output_channel_QCCC}, and the operators $\rho_k^i * A_k$, where $A_k$ is the Choi operator of a copy of $\A_k$ acting from $\H_{\text{aux}_{k,t_1}^i}$ to $\H_{\text{aux}_{k,t_{N+1}}^i}$. We have thus shown that
\begin{align}
    \rho_{\text{aux}_{t_1}} * \Big[\widehat{C}*(A_{1}^{\otimes N}\otimes\ldots \otimes A_{N}^{\otimes N})\Big] = \rho_{\text{aux}_{t_1}} * \Big[S_{\mathsf{CC}}*(A_1 \otimes \ldots \otimes A_N)\otimes A_1^{\otimes(N-1)}\otimes\ldots\otimes A_N^{\otimes(N-1)}\Big]
\end{align}
for all product operators $\rho_{\text{aux}_{t_1}}=\bigotimes_{k=1}^N\bigotimes_{i=1}^{N-1}\rho_k^i$. Since these operators span $\L(\H_{\text{aux}_{t_1}})$, and since an operator acting on $\H_{\text{aux}_{t_1}}$ and other Hilbert spaces is fully determined by its link products with all operators in $\L(\H_{\text{aux}_{t_1}})$, Eq.~\eqref{eq:catalytic_statement_Choi} follows, which concludes the proof.

Finally, as stated in Theorem~\ref{thm::QCCC_app}, this catalytic simulation can be straightforwardly combined with the extended simulation of Sec.~\ref{subsubsec::proof_extendedsimulation_QCCC}, so as to simultaneously perform an extended and a catalytic simulation of $\boldsymbol{S}_\mathsf{CC}$, still with $N$ queries of each extended channel $\A'_k:\L(\H_{I_k}\otimes\H_{I'_k})\to\L(\H_{O_k}\otimes\H_{O'_k})$. It suffices to replace each auxiliary system in position $i$ associated with $\A'_k$ by a pair of auxiliary systems in $\H_{\text{aux}_{k,t_n}^i}\otimes\H_{\text{aux}'^{i}_{k,t_n}}$, isomorphic to $\H_{I_k}\otimes\H_{I'_k}$, whose inputs and outputs are left open, to add the extension wires of Sec.~\ref{subsubsec::proof_extendedsimulation_QCCC}, and to let the routing operations permute these pairs as above, while the target system and the extension wire associated with $\A'_{k_n}$ are swapped with the pair in position 1 associated with $\A'_{k_n}$ as in Sec.~\ref{subsubsec::proof_extendedsimulation_QCCC}. The same computation as above then shows that the resulting fixed-order quantum circuit outputs $(\boldsymbol{S}_\mathsf{CC}\otimes\boldsymbol{I})(\A'_1\otimes\ldots\otimes\A'_N)\otimes\A_1^{\prime\otimes(N-1)}\otimes\ldots\otimes\A_N^{\prime\otimes(N-1)}$, each of the $N^2-N$ catalyst copies of the extended channels acting on its own pair of auxiliary inputs and outputs.
\end{proof}

\section{Simulating quantum control of computational order}
\label{sec::simulatingQCQC}

\subsection{Quantum circuits with quantum control of computational order (QC-QC)}
\label{subsec::QCQC}

\begin{figure*}[h]
    \centering
    \includegraphics[width=\columnwidth]{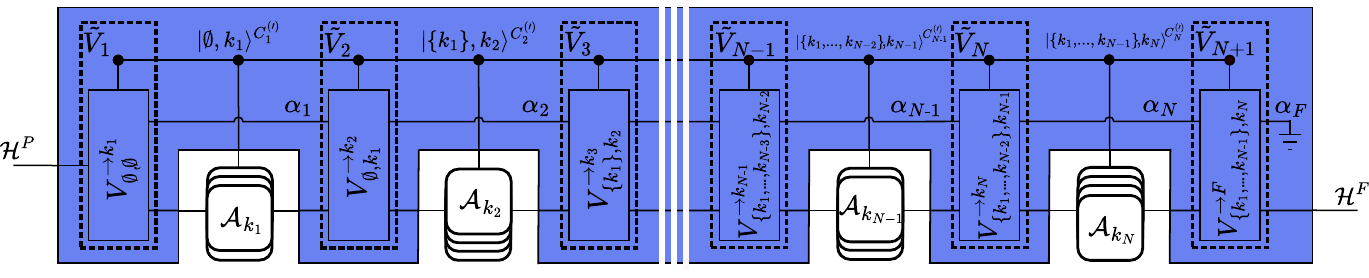}
    \caption{\textbf{Quantum circuits with quantum control of computational order (QC-QC).} Graphical representation of a quantum circuit with quantum control of computational order. Adapted from Ref.~\cite{wechs21}.}
    \label{fig:QCQC}
\end{figure*}

We now define quantum circuits with quantum control (QC-QCs) of computational order. QC-QCs generalize QC-CCs by allowing the control of the computational order to be a quantum system, allowing the choice of the next operation to be controlled coherently. They were introduced in Ref.~\cite{wechs21} as a broader class of superchannels that include all QC-CCs as well as some superchannels with indefinite computational order. 

Consider a QC-QC superchannel that acts on $N$ quantum channels, and hence, has $N$ slots, labeled $n\in\{1,\ldots,N\}$. Such a superchannel
\begin{align}
    \boldsymbol{S}_\mathsf{QC}:\bigotimes_{i=1}^N[\L(\H_{I_i})\to\L(\H_{O_i})]\to[\L(\H_{P})\to\L(\H_{F})]
\end{align}
is defined by a collection of internal operators 
$\{\{V^{\to k_1}_{\emptyset,\emptyset}\}_{k_1}, 
\{V^{\to k_2}_{\emptyset,k_1}\}_{k_2}, \ldots, 
\{V^{\to k_{n+1}}_{\{k_1,\ldots,k_{n-1}\},k_{n}}\}_{k_{n+1}}, \ldots, 
V^{\to F}_{\{k_1,\ldots,k_{N-1}\},k_{N}}\}$ acting on 
\begin{align}
    V^{\to k_1}_{\emptyset,\emptyset}&: \mathcal H_P \to \mathcal H_{I_{k_1}}\otimes\mathcal H_{\alpha_1}, \\
    V^{\to k_{n+1}}_{\{k_1,\ldots,k_{n-1}\},k_{n}}&: \mathcal H_{O_{k_n}}\otimes\mathcal H_{\alpha_n}\to \mathcal H_{I_{k_{n+1}}}\otimes\mathcal H_{\alpha_{n+1}} \ \ \ \forall\ n\in\{1,\ldots,N-1\}, \\
    V^{\to F}_{\{k_1,\ldots,k_{N-1}\},k_{N}}&: \mathcal H_{O_{k_N}}\otimes\mathcal H_{\alpha_N} \to \mathcal H_F\otimes\mathcal H_{\alpha_F}.
\end{align}
where $\mathcal H_{\alpha_1},\ldots,\mathcal H_{\alpha_N}, \mathcal H_{\alpha_F}$ are auxiliary systems. We recall that throughout this construction, $(k_1,\ldots,k_n)$ denotes an ordered sequence of $n$ distinct elements of $[N]=\{1,\ldots,N\}$; for $n=N$, it is therefore a permutation of $[N]$. The symbol $\emptyset$ denotes the empty set. In $V_{\emptyset,k_1}^{\to k_2}$, it indicates that no channel preceded $\mathcal A_{k_1}$, while in $V_{\emptyset,\emptyset}^{\to k_1}$, the two empty labels indicate that no channel has yet been applied and there is no preceding channel whose output is being processed. Here, we allow the auxiliary systems to be enlarged and the internal operations to be extended beyond the subspaces reached during the computation, without changing the implemented superchannel. The normalization and orthogonality conditions below are imposed on the full input spaces of these extended operations. These operations satisfy normalization constraints
\begin{align}
    \sum_{k_1\in[N]} (V^{\to k_1}_{\emptyset,\emptyset})^\dagger V^{\to k_1}_{\emptyset,\emptyset} &= \id_P \\
    \sum_{k_2\in[N]\setminus\{k_1\}}\left(V_{\emptyset,k_1}^{\to k_2}\right)^\dagger V_{\emptyset,k_1}^{\to k_2} &= \id_{O_{k_1}}\otimes\id_{\alpha_1} \quad \forall \ k_1\in[N] \\
    \sum_{k_{n+1}\in[N]\setminus\{k_1,\ldots,k_n\}} \left( V_{\{k_1,\ldots,k_{n-1}\},k_n}^{\to k_{n+1}}\right)^\dagger V_{\{k_1,\ldots,k_{n-1}\},k_n}^{\to k_{n+1}} &= \id_{O_{k_n}}\otimes\id_{\alpha_n} \quad \forall \ n\in\{2,\ldots,N-1\}, \ \forall \ (k_1,\ldots,k_n) \\
    \left(V_{\{k_1,\ldots,k_{N-1}\},k_N}^{\to F} \right)^\dagger V_{\{k_1,\ldots,k_{N-1}\},k_N}^{\to F} &= \id_{O_{k_N}}\otimes\id_{\alpha_N} \quad \forall \ (k_1,\ldots,k_N)
\end{align}
and orthogonality constraints
\begin{align}
    \sum_{k_{n+1}\in[N]\setminus\{k_1,\ldots,k_n\}} \left( V_{\{k_1,\ldots,k_{n-1}\},k_n}^{\to k_{n+1}} \right)^\dagger V_{\{k_1,\ldots,k_n\}\setminus\{\ell\},\ell}^{\to k_{n+1}} &=0 
\end{align}
$\forall\ n\in\{2,\ldots,N-1\}$, $
\forall\ (k_1,\ldots,k_n)$ with distinct entries in $[N]$, and $\forall\ \ell\in\{k_1,\ldots,k_{n-1}\}$, and
 \begin{align}   
    \left( V_{\{k_1,\ldots,k_{N-1}\},k_N}^{\to F} \right)^\dagger V_{[N]\setminus\{\ell\},\ell}^{\to F} &=0,
\end{align}
$\forall\ (k_1,\ldots,k_N)$     permutations of $[N]$, and $\forall\ 
\ell\in[N]\setminus\{k_N\}$.

Note that each operator $V_{\{k_1,\ldots,k_{n-1}\},k_n}^{\to k_{n+1}}$ need not be itself an isometry. As we will see later, these orthogonality constraints ensure that the overall description of a QC-QC in terms of quantum control of orders is well defined.

In $V_{\{k_1,\ldots,k_{n-1}\},k_n}^{\to k_{n+1}}$, the subscript set $\{k_1,\ldots,k_{n-1}\}$ denotes a set of previously applied channels, without specifying their order, and $k_n$ labels the channel whose output is being processed, while the superscript $\to k_{n+1}$ labels the channel to whose input this component of the internal operation sends the system. See an illustration of a generic QC-QC in Fig.~\ref{fig:QCQC}. 

Finally, let their respective Choi vectors be denoted 
$\kket{V^{\to k_1}_{\emptyset,\emptyset}}, \ldots, 
\kket{V^{\to k_{n+1}}_{\{k_1,\ldots,k_{n-1}\},k_{n}}}, \ldots, 
\kket{V^{\to F}_{\{k_1,\ldots,k_{N-1}\},k_{N}}}$. We can then define a QC-QC superchannel.

\begin{definition}[Quantum circuit with quantum control of computational order (QC-QC)~\cite{wechs21}]
    A higher-order linear map $\boldsymbol{S}_{\mathsf{QC}}$, with associated Choi operator $S_{\mathsf{QC}}$, is a quantum circuit with quantum control of computational order iff there exists a collection of operators $V^{\to k_1}_{\emptyset,\emptyset},\ldots, V^{\to k_{n+1}}_{\{k_1,\ldots,k_{n-1}\},k_{n}}, \ldots, V^{\to F}_{\{k_1,\ldots,k_{N-1}\},k_{N}}$, with Choi vectors $\kket{V^{\to k_1}_{\emptyset,\emptyset}}, \ldots, \kket{V^{\to k_{n+1}}_{\{k_1,\ldots,k_{n-1}\},k_{n}}}, \ldots, \kket{V^{\to F}_{\{k_1,\ldots,k_{N-1}\},k_{N}}}$, satisfying the normalization and orthogonality constraints defined above, such that 
    \begin{align}
        S_{\mathsf{QC}} = \tr_{\alpha_F} \left( \ketbra{s_{\mathsf{QC}}}{s_{\mathsf{QC}}} \right),
    \end{align}
    where 
    \begin{align}\label{eq::ket_s}
        \ket{s_{\mathsf{QC}}}\coloneqq \sum_{(k_1,\ldots,k_N)} \kket{V^{\to k_1}_{\emptyset,\emptyset}} * \kket{V^{\to k_2}_{\emptyset,k_1}}* \ldots* \kket{V^{\to k_{n+1}}_{\{k_1,\ldots,k_{n-1}\},k_{n}}}* \ldots* \kket{V^{\to F}_{\{k_1,\ldots,k_{N-1}\},k_{N}}},
    \end{align}
    where the sum runs over all permutations $(k_1,\ldots,k_N)$ of $[N]$.
    \label{def:QC-QC}
\end{definition}

The action of the QC-QC superchannel on the input quantum channels $\mathcal{A}_k$ is then expressed in the Choi picture as:
\begin{align}
    S_{\mathsf{QC}}*(A_1\otimes \ldots \otimes A_N) & = \tr_{\alpha_F} \left( \ketbra{s_{\mathsf{QC}}}{s_{\mathsf{QC}}} \right) * (A_1 \otimes \ldots \otimes A_N).
    \label{eq:output_channel_QCQC}
\end{align}
In the particular case where the input quantum channels are unitary channels $\map{U}_i$, with associated Choi vectors $\kket{U_i}$, we can equivalently write 
\begin{align}
    S_{\mathsf{QC}}*(\kketbra{U_1}{U_1}\otimes \ldots \otimes \kketbra{U_N}{U_N}) & = \tr_{\alpha_F} \Big( \ketbra{s_{\boldsymbol{U}}}{s_{\boldsymbol{U}}} \Big) \label{eq:output_channel_QCQC(unitary}
\end{align}
where 
\begin{align}
    \ket{s_{\boldsymbol{U}}}
    \coloneqq \sum_{(k_1,\ldots,k_N)} &\kket{V^{\to k_1}_{\emptyset,\emptyset}} * \kket{U_{k_1}} * \kket{V^{\to k_2}_{\emptyset,k_1}} * \kket{U_{k_2}} * \ldots * \kket{V^{\to k_{n+1}}_{\{k_1,\ldots,k_{n-1}\},k_{n}}} * \kket{U_{k_{n+1}}} * \ldots * \kket{U_{k_{N}}} * \kket{V^{\to F}_{\{k_1,\ldots,k_{N-1}\},k_{N}}}.
\end{align}

The quantum control of computational order can be made explicit by introducing a control register alongside the target and auxiliary systems. At the beginning of time step \(n\), its orthonormal basis states are labeled
\begin{align}
    \ket{\{k_1,\ldots,k_{n-1}\},k_n}_{C_n} \in \H_{C_n},
\end{align}
where $\{k_1,\ldots,k_{n-1}\}$ specifies the set of channels already applied and $k_n$ identifies the channel to be applied next. Unlike the classical control description of QC-CCs, the joint state of the target, auxiliary, and control systems may contain coherence between different control labels. Moreover, the control label records only the set of previously applied channels, rather than their complete ordered history. Any further information about this history may be carried by the auxiliary system.

The internal operators introduced above describe the transitions between these control states. Denoting by $K_n$ the (unordered) set of length $n$ identified with $\{k_1,\ldots,k_n\}$, with $K_0\coloneqq\emptyset$, the component $V_{K_{n-1},k_n}^{\to k_{n+1}}$ processes the output of $\mathcal A_{k_n}$, prepares the input to $\mathcal A_{k_{n+1}}$, and accompanies the control update
\begin{align}
    \ket{K_{n-1},k_n} \longmapsto \ket{K_n,k_{n+1}},
\end{align}
where $K_n=K_{n-1}\cup\{k_n\}$. These transitions are combined coherently into a single internal operation at each time step. As in the QC-CC description, we identify the target input and output spaces with common spaces at each time step, leaving these identifications implicit.

More precisely, let $\mathcal H_{C_n}$ and $\mathcal H_{C_n'}\cong\mathcal H_{C_n}$ denote the control spaces immediately before and after the application of the input channel at time step $n$. The combined internal operators have Choi vectors
\begin{align}
    \kket{\tilde V_1}
    ={}&
    \sum_{k_1\in[N]}
    \kket{V_{\emptyset,\emptyset}^{\to k_1}}
    \otimes\ket{\emptyset,k_1}_{C_1},\label{eq:def_tilde_V1_QCQC}      
    \\
    \kket{\tilde V_{n+1}}
    ={}&
    \sum_{\substack{K_{n-1}\subseteq[N]}}
    \ \sum_{k_n\in[N]\setminus K_{n-1}}
    \ \sum_{k_{n+1}\in[N]\setminus K_n}
    \kket{V_{K_{n-1},k_n}^{\to k_{n+1}}} \otimes\ket{K_{n-1},k_n}_{C_n'}
    \otimes\ket{K_n,k_{n+1}}_{C_{n+1}},\label{eq:Choi_Vn1_QCQC}        
    \\
    \kket{\tilde V_{N+1}}
    ={}&
    \sum_{k_N\in[N]}
    \kket{V_{[N]\setminus\{k_N\},k_N}^{\to F}}
    \otimes\ket{[N]\setminus\{k_N\},k_N}_{C_N'}.\label{eq:def_tilde_VN1_QCQC}   
\end{align}
The intermediate expression applies for $n\in\{1,\ldots,N-1\}$. Here, the sum over $K_{n-1}$ runs over the subsets of $[N]$ of size $n-1$, and $K_n\coloneqq K_{n-1}\cup\{k_n\}$, so that each control label is included exactly once. The normalization and orthogonality conditions imposed above ensure that the combined operators $\tilde V_n$ are isometries on their respective input spaces.

The first internal operation prepares the target, auxiliary, and control systems jointly, with the control selecting the first channel to be applied. Each subsequent internal operation updates this selection while retaining the coherence allowed between different alternatives. During the application of an input channel, the control label is unchanged; $C_n$ and $C_n'$ distinguish its input and output occurrences. Finally, $\tilde V_{N+1}$ maps the target, auxiliary, and control systems into $\mathcal H_F\otimes\mathcal H_{\alpha_F}$, after which the final auxiliary system is traced out.

This representation gives an operational interpretation of the sum over permutations in Eq.~\eqref{eq::ket_s}: each term describes one sequence of channel applications, and these contributions are added at the level of probability amplitudes. Their relative phases can therefore affect the resulting channel through interference terms. Whether such interference survives the final partial trace depends on the information retained in the discarded auxiliary system.

\subsection{Proof of Theorem~\ref{thm::qcqc} (Non-universal simulation of quantum control of computational order)}
\label{subsec::theoremQCQC}

We are now ready to prove our second theorem.

\begin{theorem}[Non-universal simulation of quantum control of computational order]\label{thm::qcqc_app}
    For every quantum circuit with quantum control of computational order $\boldsymbol{S}_{\mathsf{QC}}\in\textup{\textsf{QC-QC}}$ that acts on $N$ quantum channels, there exists a fixed-order quantum circuit $\boldsymbol{C}\in\textup{\textsf{QC-FO}}$ using at most $N$ queries of each input channel, hence at most $N^2$ total queries, that simulates the action of $\boldsymbol{S}_{\mathsf{QC}}$ on unitary channels deterministically and exactly.

    Moreover, these fixed-order simulations can be adapted to perform extended and catalytic simulations of the action of any given $\boldsymbol{S}_{\mathsf{QC}}$ on unitary channels at no additional overhead in query complexity, and in the catalytic case, recovering all $N^2-N$ extra queries.
\end{theorem}

The proof is broken down into three parts: simulation, extended simulation, and catalytic simulation.

\subsubsection{Simulation}
\label{subsubsec::proof_simulation_QCQC}

\begin{proof}
Given a QC-QC superchannel $\boldsymbol{S}_\mathsf{QC}:\bigotimes_{i=1}^N[\L(\H_{I_i})\to\L(\H_{O_i})]\to[\L(\H_{P})\to\L(\H_{F})]$, we construct a fixed-order quantum circuit $\boldsymbol{C}:\bigotimes_{i=1}^N[\L(\H_{I_i})\to\L(\H_{O_i})]^{\otimes N}\to[\L(\H_{P})\to\L(\H_{F})]$ which uses $N$ queries of each input channel to simulate the action of the QC-QC on unitary channels $\map{U}_k:\L(\H_{I_k}) \to \L(\H_{O_k})$, with Choi vectors $\kket{U_k}$. We start by formally defining $\boldsymbol{C}$ and showing that it is a valid QC-FO superchannel, before showing that $\boldsymbol{C}$ is indeed simulating the action of $\boldsymbol{S}_\mathsf{QC}$ on unitary channels.\\

The fixed-order quantum circuit $\boldsymbol{C}$ is an adaptation of the simulation scheme of Sec.~\ref{subsubsec::proof_simulation_QCCC}, see Fig.~\ref{fig:simu_QCCC}, which is itself based on the simulation scheme proposed to simulate the quantum $N$-switch~\cite{Colnaghi12,araujo14,facchini15}. It incorporates into this simulation scheme the QC-QC architecture, i.e., the internal operators $V_{K_{n-1},k_n}^{\to k_{n+1}}$ whose order of application is coherently controlled, see Fig.~\ref{fig:simu_QCQC}. In particular, the routing operations of Sec.~\ref{subsubsec::proof_simulation_QCCC} are now coherently controlled by the quantum control system of the QC-QC.

On top of the $N$ copies of each input channel, the simulation scheme uses $N$ auxiliary registers initialized in state $\ket{a_k} \in \H_{\text{aux}_{k,t_1}}$, where $\H_{\text{aux}_{k,t_1}}$ is isomorphic to the input Hilbert space $\H_{I_k}$. The simulation then proceeds in accordance with the structure of a QC-QC, considering $N+1$ time steps labeled by $n \in [N+1]$, each except the last one corresponding to the execution of an input channel on the target system. We will denote the global auxiliary space at the beginning of each time step $n$ as $\H_{\text{aux}_{t_n}}\coloneqq\bigotimes_{k=1}^N \H_{\text{aux}_{k,t_n}}$, where each individual Hilbert space $\H_{\text{aux}_{k,t_n}}$ is isomorphic to $\H_{I_k}$.

The action of the simulation at each time step $n \in [N]$ corresponds to the sequential composition of the following four steps:
\begin{enumerate}
    \item First, the application of the combined internal QC-QC operator $\tilde V_{n}$ to the target, auxiliary and control systems, with Choi vector $\kket{\tilde V_n}$ as defined in Eqs.~\eqref{eq:def_tilde_V1_QCQC}--\eqref{eq:def_tilde_VN1_QCQC}.
    \item Then, controlled on the last label $k_n$ of the control state $\ket{K_{n-1},k_n}$, a routing operator $S^I_{k_n}: \H_{\text{aux}_{t_n}} \otimes \H_{I_{k_n}} \to \H_{\text{aux}_{t_n^I}} \otimes \H_{\bar I_{k_n}}$ swaps the incoming target system in $\H_{I_{k_n}}$ coming from $\tilde V_n$ with the auxiliary system in $\H_{\text{aux}_{k_n,t_n}}$, and outputs the target system in $\H_{\text{aux}_{k_n,t_n^I}}$ and the auxiliary system in $\H_{\bar I_{k_n}}$ (which is isomorphic to $\H_{I_{k_n}}$), while it leaves all the other systems unchanged. The Hilbert space $\H_{\text{aux}_{t_n^I}}\coloneqq\bigotimes_{k=1}^N \H_{\text{aux}_{k,t_n^I}}$ describes the collection of all auxiliary Hilbert spaces $\H_{\text{aux}_{k,t_n^I}}$ (which are isomorphic to $\H_{I_k}$) right after the application of $S^I_{k_n}$. Since it only permutes subsystems, $S^I_{k_n}$ is a unitary operator, which implements the same routing as the operation $\map{S}^I_{k_n}$ of Sec.~\ref{subsubsec::proof_simulation_QCCC}. The overall controlled routing operator $S^I_n$ applied at time step $n \in [N]$ can be described by its Choi vector
\begin{align}
    \kket{S^I_n} = \sum_{K_{n-1}\subseteq[N]}\ \sum_{k_n\in[N]\setminus K_{n-1}} \kket{S^I_{k_n}} \otimes \ket{K_{n-1},k_n}_{C_n} \otimes \ket{K_{n-1},k_n}_{C_n^{\circ}},
    \label{eq:routing_in_pure}
\end{align}
    where, as in Eq.~\eqref{eq:Choi_Vn1_QCQC}, the sum over $K_{n-1}$ runs over the subsets of $[N]$ of size $n-1$, so that each control state is included exactly once, where $\H_{C_n}$ is the Hilbert space corresponding to the control system stemming from $\tilde V_n$ and input in $S^I_n$, and where $\H_{C_n^{\circ}}$ is the Hilbert space corresponding to the control system output by $S^I_n$ and input in $S^O_n$.
    \item A copy $\map{U}_k:\L(\H_{\text{aux}_{k,t_n^I}})\to \L(\H_{\text{aux}_{k,t_n^O}})$ of each unitary channel, with Choi vector $\kket{U_k}\in\H_{\text{aux}_{k,t_n^I}}\otimes\H_{\text{aux}_{k,t_n^O}}$ (with the same convention as in Sec.~\ref{subsubsec::proof_simulation_QCCC} regarding the input and output Hilbert spaces of the copies), is then applied and transforms the corresponding auxiliary state in $\H_{\text{aux}_{k,t_n^I}}$ into an output state in $\H_{\text{aux}_{k,t_n^O}}$, the latter Hilbert space being isomorphic to $\H_{O_k}$, and where $\H_{\text{aux}_{t_n^O}}\coloneqq\bigotimes_{k=1}^N \H_{\text{aux}_{k,t_n^O}}$ describes the collection of all auxiliary Hilbert spaces right after the application of the $N$ unitary channels. In particular, the copy of $\map{U}_{k_n}$ acts on the target system, while each copy of $\map{U}_k$, $k\neq k_n$, acts on the auxiliary system associated with $\map{U}_k$.
    \item Finally, the routing is inverted by $S^O_{k_n}: \H_{\text{aux}_{t_n^O}} \otimes \H_{\bar I_{k_n}} \to \H_{\text{aux}_{t_{n+1}}} \otimes \H_{O_{k_n}}$, which swaps the target state in $\H_{\text{aux}_{k_n,t_n^O}}$ with the auxiliary state in $\H_{\bar I_{k_n}}$, the target state being eventually output in $\H_{O_{k_n}}$ and sent to $\tilde V_{n+1}$, and the auxiliary state being eventually output in $\H_{\text{aux}_{k_n,t_{n+1}}}$. The overall controlled routing operator $S^O_n$ applied at time step $n$ can be described by its Choi vector
    \begin{align}
    \kket{S^O_n} = \sum_{K_{n-1}\subseteq[N]}\ \sum_{k_n\in[N]\setminus K_{n-1}} \kket{S^O_{k_n}} \otimes \ket{K_{n-1},k_n}_{C_n^{\circ}} \otimes \ket{K_{n-1},k_n}_{C_n'},
     \label{eq:routing_out_pure}
\end{align}
where $\H_{C_n'}$ is the Hilbert space corresponding to the control system output by $S^O_n$ and input in $\tilde V_{n+1}$.

We will denote by $\mathcal{Q}_n$ the parallel higher-order transformation with $N$ slots obtained by applying $S^I_n$, leaving open one slot $\H_{\text{aux}_{k,t_n^I}}\to\H_{\text{aux}_{k,t_n^O}}$ for each input channel, $k \in [N]$, and then applying $S^O_n$. In the Choi picture, it is described by the Choi vector $\kket{Q_n} = \kket{S^I_n} * \kket{S^O_n}$.
\end{enumerate}

The last time step $n=N+1$ consists in applying the last combined internal QC-QC operator $\tilde V_{N+1}$, which transforms the target, auxiliary and control systems into a state in $\H_F\otimes\H_{\alpha_F}$, and in tracing out the final auxiliary system of the QC-QC in $\H_{\alpha_F}$, as well as all the auxiliary registers in $\H_{\text{aux}_{t_{N+1}}}$.

For any QC-QC superchannel $\boldsymbol{S}_\mathsf{QC}:\bigotimes_{i=1}^N[\L(\H_{I_i})\to\L(\H_{O_i})]\to[\L(\H_{P})\to\L(\H_{F})]$, we can thus formally define the superchannel $\boldsymbol{C}:\bigotimes_{i=1}^N[\L(\H_{I_i})\to\L(\H_{O_i})]^{\otimes N}\to[\L(\H_{P})\to\L(\H_{F})]$ as the alternating composition, step by step, of the procedure described above, except the third step as it corresponds to the application of the quantum channels which are input in $\boldsymbol{C}$. Since the auxiliary registers are initialized in a pure state and since the operators $\tilde V_n$, $S^I_n$ and $S^O_n$ are isometries, the Choi operator of the superchannel $\boldsymbol{C}$ is described, analogously to the Choi operator $S_{\mathsf{QC}}$ in Definition~\ref{def:QC-QC}, as
\begin{align}
C & = \tr_{\alpha_F\,\text{aux}_{t_{N+1}}}\big(\ketbra{c}{c}\big), \label{eq:output_state_simu_QCQC} \\
\ket{c} & \coloneqq \ket{\psi_{\text{aux}_{t_1}}} * \kket{\tilde V_1} * \kket{Q_1} * \kket{\tilde V_2} * \kket{Q_2}  * \cdots * \kket{\tilde V_N} * \kket{Q_N} * \kket{\tilde V_{N+1}}, \label{eq:ket_c_simu_QCQC}
\end{align}
where $\ket{\psi_{\text{aux}_{t_1}}}=\bigotimes_{i=1}^N \ket{a_i} \in \H_{\text{aux}_{t_1}}$ is the initial auxiliary state.

That the combined operators $\tilde V_n$ are isometries follows from the normalization and orthogonality conditions satisfied by the operators $V_{K_{n-1},k_n}^{\to k_{n+1}}$, so that the corresponding maps $\rho\mapsto\tilde V_n\rho\tilde V_n^\dagger$ are CPTP. The routing operators $S^I_n$ and $S^O_n$ are controlled permutations of subsystems, hence unitary, and $\mathcal{Q}_n$ is a valid parallel higher-order superchannel. Thus, at each time step, the composition of $\tilde V_n$ and $\mathcal{Q}_n$, with Choi vector $\kket{\tilde V_n} * \kket{Q_n}$, defines a valid parallel higher-order process, which can be interpreted as a QC-FO between the $N$ input channels. Finally, by Eqs.~\eqref{eq:output_state_simu_QCQC} and~\eqref{eq:ket_c_simu_QCQC}, the superchannel $\boldsymbol{C}$ is obtained by preparing the auxiliary registers in the state $\ket{\psi_{\text{aux}_{t_1}}}$, applying sequentially the blocks $\kket{\tilde V_1} * \kket{Q_1}, \ldots, \kket{\tilde V_N} * \kket{Q_N}$ and then $\kket{\tilde V_{N+1}}$, and tracing out $\H_{\alpha_F}$ and the auxiliary registers in $\H_{\text{aux}_{t_{N+1}}}$. Since each block $\kket{\tilde V_n} * \kket{Q_n}$ describes a QC-FO, while $\tilde V_{N+1}$, the state preparation and the partial traces define CPTP maps, this sequential composition is itself a QC-FO, i.e., $\boldsymbol{C}$ is a valid QC-FO superchannel.

\begin{figure*}[h]
    \centering
    \includegraphics[width=\columnwidth]{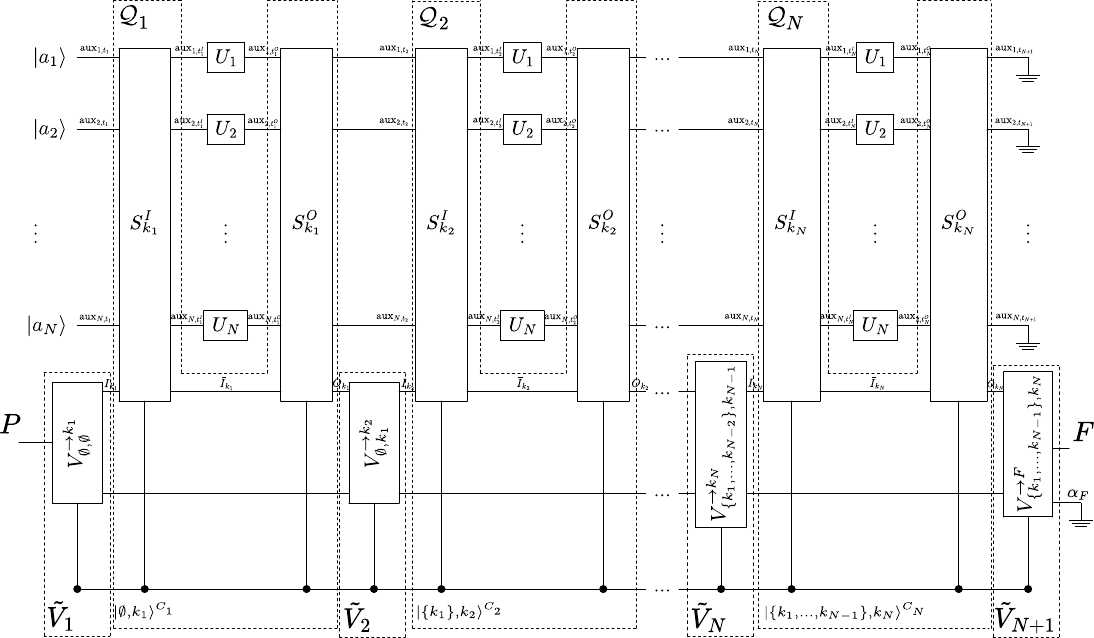}
    \caption{Fixed order simulation of a QC-QC with input unitary channels $\map{U}_1,\ldots,\map{U}_N$ as represented in Fig.~\ref{fig:QCQC}. $N$ queries of each unitary channel $\map{U}_k$ are made, so that the simulation makes $N^2$ queries to the input channels.}
    \label{fig:simu_QCQC}
\end{figure*}

\vspace{5mm}

We will now show that $\boldsymbol{C}$ applied on $N$ copies of each unitary channel is simulating the action of $\boldsymbol{S}_\mathsf{QC}$ on these unitary channels, for any choice of the input unitary channels. More formally, we want to show that
\begin{align}
    \boldsymbol{C}(\map{U}_{1}^{\otimes N}\otimes\ldots\otimes\map{U}_{N}^{\otimes N}) = \boldsymbol{S}_\mathsf{QC}(\map{U}_1\otimes\ldots\otimes\map{U}_N),
\end{align}
for all possible unitary channels $\map{U}_1,\ldots,\map{U}_N$. Equivalently, we can rephrase the last equation in the Choi picture as
\begin{align}
    C*(\kketbra{U_1}{U_1}^{\otimes N}\otimes\ldots \otimes \kketbra{U_N}{U_N}^{\otimes N}) = S_{\mathsf{QC}}*(\kketbra{U_1}{U_1} \otimes \ldots \otimes \kketbra{U_N}{U_N}),
\end{align}
for all possible unitary operators $U_1,\ldots,U_N$.

The proof consists in computing iteratively the output of the simulation scheme presented in Fig.~\ref{fig:simu_QCQC} and showing that it matches the output of the QC-QC on unitary channels, as defined in Eq.~\eqref{eq:output_channel_QCQC(unitary}.

Following Definition~\ref{def:QC-QC}, any QC-QC superchannel $\boldsymbol{S}_\mathsf{QC}$ is defined by a collection of internal operators $\{V_{\emptyset,\emptyset}^{\to k_{1}},\ldots,\allowbreak \, V_{K_{n-1},k_n}^{\to k_{n+1}},\allowbreak\ldots, \, V_{K_{N-1},k_N}^{\to F}\}$.

Since the input channels are unitary, with rank-one Choi operators $\kketbra{U_k}{U_k}$, and since the simulation scheme only involves a pure initial state and isometries, the output of the simulation scheme can be computed at the level of Choi vectors, by composing iteratively the operations $\tilde V_n$, $\mathcal{Q}_n$ and the unitary channels $\map{U}_1 \otimes \ldots \otimes \map{U}_N$. The auxiliary registers are initialized in the pure state $\ket{\psi_{\text{aux}_{t_1}}}=\bigotimes_{i=1}^N \ket{a_i}$. We describe by $\ket{\varphi_{(n)}'}$ (resp.~$\ket{\varphi_{(n+1)}}$) the simulation output at time step $n$, before (resp.~after) the operation $\tilde V_{n+1}$ is applied:
\begin{align}
\ket{\varphi_{(n)}'} & = \big(\ket{\psi_{\text{aux}_{t_1}}} * \kket{\tilde V_1} * \kket{Q_1} *  \kket{\tilde V_2} * \cdots * \kket{\tilde V_n} * \kket{Q_n}\big) * (\kket{U_{1}}^{\otimes {n}}\otimes\ldots \otimes \kket{U_{N}}^{\otimes {n}}), \notag
\end{align}
and
\begin{align}
\ket{\varphi_{(n+1)}} & = \big(\ket{\psi_{\text{aux}_{t_1}}} * \kket{\tilde V_1} * \kket{Q_1} * \kket{\tilde V_2} * \cdots * \kket{Q_n} * \kket{\tilde V_{n+1}}\big) * (\kket{U_{1}}^{\otimes {n}}\otimes\ldots \otimes \kket{U_{N}}^{\otimes {n}}).
\end{align}
At the first time step, right after the first internal operation $\tilde V_1$ has been applied, the simulation output is described by
\begin{align}
    \ket{\varphi_{(1)}} & = \ket{\psi_{\text{aux}_{t_1}}} * \kket{\tilde V_1} \nonumber \\
    &= \sum_{k_1} \big(\ket{a_1} \otimes \ldots \otimes \ket{a_N}\big) \otimes \kket{V_{\emptyset,\emptyset}^{\to k_{1}}} \otimes \ket{\emptyset,k_1}_{C_{1}},
\end{align}
i.e., $\tilde V_1$ prepares a coherent superposition of the different choices $k_1$ of the first channel, recorded in the control system.
Applying the first unitary channels in parallel along with the swap operations, we obtain the Choi vector at the end of the first time step as
\begin{align}
    \ket{\varphi_{(1)}'} & = \ket{\varphi_{(1)}} * \kket{Q_1} * (\kket{U_{1}}\otimes\ldots \otimes \kket{U_{N}}) = \ket{\varphi_{(1)}} * \kket{S^I_1}*(\kket{U_1}\otimes \ldots \otimes \kket{U_N}) * \kket{S^O_1} \nonumber \\
    & = \bigg[\sum_{k_1} \Big[ \big(\ket{a_1} \otimes \ldots \otimes \ket{a_N}\big) \otimes \kket{V_{\emptyset,\emptyset}^{\to k_{1}}} \Big] * \kket{S^I_{k_1}}\otimes \ket{\emptyset,k_1}_{C^{\circ}_{1}} \bigg] *(\kket{U_1}\otimes \ldots \otimes \kket{U_N}) * \kket{S^O_1}\nonumber \\
    & = \bigg[\sum_{k_1} \Big[ \bigotimes_{\substack{i=1\\ i\neq k_1}}^N \big(\ket{a_i} * \kket{U_i}\big) \otimes \ket{a_{k_1}} \otimes \big(\kket{V_{\emptyset,\emptyset}^{\to k_{1}}} * \kket{U_{k_1}}\big) \Big] \otimes \ket{\emptyset,k_1}_{C^{\circ}_{1}}\bigg] * \kket{S^O_1} \nonumber \\
    & = \sum_{k_1} \Big[ \bigotimes_{\substack{i=1\\ i\neq k_1}}^N \big(\ket{a_i} * \kket{U_i}\big) \otimes \ket{a_{k_1}} \otimes \big(\kket{V_{\emptyset,\emptyset}^{\to k_{1}}} * \kket{U_{k_1}}\big) \Big] \otimes \ket{\emptyset,k_1}_{C'_{1}},
\end{align}
where $\ket{a_i} * \kket{U_i}=U_i\ket{a_i}$. In the third line, the auxiliary state $\ket{a_{k_1}}$ is stored in $\H_{\bar I_{k_1}}$, while the copy of $\map{U}_{k_1}$ acts on the target system output by $V_{\emptyset,\emptyset}^{\to k_1}$ and outputs it in $\H_{\text{aux}_{k_1,t_1^O}}$, and each copy of $\map{U}_i$, $i\neq k_1$, acts on the auxiliary state $\ket{a_i}$; the target system and the auxiliary state $\ket{a_{k_1}}$ are then swapped back by $S^O_{k_1}$, so that, in the last line, the target system is in $\H_{O_{k_1}}$ and the auxiliary state $\ket{a_{k_1}}$ is back in $\H_{\text{aux}_{k_1,t_2}}$.

The output of the simulation scheme at the second time step before the unitary channels are applied is
\begin{align}
     \ket{\varphi_{(2)}} & = \ket{\varphi_{(1)}'} * \kket{\tilde V_2} \nonumber \\
     &= \sum_{k_1,k_2} \Big[\bigotimes_{\substack{i=1\\ i\neq k_1}}^N \big(\ket{a_i} * \kket{U_i}\big) \otimes \ket{a_{k_1}} \otimes \big(\kket{V_{\emptyset,\emptyset}^{\to k_{1}}} * \kket{U_{k_1}} * \kket{V_{\emptyset,k_1}^{\to k_{2}}}\big) \Big] \otimes \ket{\{k_1\},k_2}_{C_{2}},
\end{align}
where, for each $k_1$, the link product over $\H_{C_1'}$ selects the components of $\kket{\tilde V_2}$ associated with the control state $\ket{\emptyset,k_1}$, which update it into $\ket{\{k_1\},k_2}$ for all $k_2\neq k_1$.
At the end of the second time step, once the unitary channels and the swap operations have been applied, we note that each unitary channel $\map{U}_k$ is applied twice on the corresponding auxiliary system in state $\ket{a_k}$, except for $k=k_1,k_2$. Indeed, $\map{U}_{k_1},\map{U}_{k_2}$ are respectively applied only once on the corresponding auxiliary state since the other copy of this operation was applied on the target system:
\begin{align}
     \ket{\varphi_{(2)}'} & = \ket{\varphi_{(2)}} * \kket{Q_2} * (\kket{U_{1}}\otimes\ldots \otimes \kket{U_{N}})\nonumber \\
     &= \sum_{k_1,k_2} \bigg[\bigotimes_{\substack{i=1\\ i\notin\{k_1,k_2\}}}^N \big(\ket{a_i} * \kket{U_i}^{*2}\big) \otimes \bigotimes_{j\in\{k_1,k_2\}} \big(\ket{a_j} * \kket{U_j}\big) \otimes \big(\kket{V_{\emptyset,\emptyset}^{\to k_{1}}} * \kket{U_{k_1}} * \kket{V_{\emptyset,k_1}^{\to k_{2}}} * \kket{U_{k_2}}\big) \bigg] \otimes \ket{\{k_1\},k_2}_{C_{2}'},
\end{align}
where $\kket{U_k}^{*2}=\kket{U_k}*\kket{U_k}$ is the Choi vector of the unitary operator $U_k^2$, so that $\ket{a_k}*\kket{U_k}^{*2}=U_k^2\ket{a_k}$. We note that the auxiliary state depends on the sequence $(k_1,k_2)$ only through the set $\{k_1,k_2\}$.
More generally, deriving iteratively the Choi vector after each time step, we obtain at the end of each time step $n\in[N]$
\begin{align}
     \ket{\varphi_{(n)}'} = \sum_{(k_1,\ldots,k_n)} \bigg[&\bigotimes_{i\in K_n} \big(\ket{a_i} * \kket{U_i}^{*(n-1)}\big) \otimes \bigotimes_{i\in[N]\setminus K_n} \big(\ket{a_i} * \kket{U_i}^{*n}\big) \nonumber \\
     &\otimes \big(\kket{V_{\emptyset,\emptyset}^{\to k_1}} \!*\! \kket{U_{k_1}} * \cdots \!*\! \kket{V_{K_{n-2},k_{n-1}}^{\to k_n}} \!*\! \kket{U_{k_n}}\big)\bigg] \otimes \ket{K_{n-1},k_n}_{C'_{n}},
     \label{eq:varphi_n_QCQC}
\end{align}
where the sum runs over all ordered sequences $(k_1,\ldots,k_n)$ of distinct elements of $[N]$, where, for each such sequence, $K_m=\{k_1,\ldots,k_m\}$ for $m\in\{0,\ldots,n\}$, as in Sec.~\ref{subsec::QCQC}, and where $\kket{U_i}^{*m}$ is the Choi vector of $U_i^m$, with $\ket{a_i}*\kket{U_i}^{*0}=\ket{a_i}$. Indeed, Eq.~\eqref{eq:varphi_n_QCQC} holds for $n=1,2$, and if it holds at time step $n-1$, then at time step $n$, $\tilde V_n$ appends $\kket{V_{K_{n-2},k_{n-1}}^{\to k_n}}$ to each term and updates its control state into $\ket{K_{n-1},k_n}$, for all $k_n\in[N]\setminus K_{n-1}$; the copy of $\map{U}_{k_n}$ then acts on the target system, so that the auxiliary system associated with $\map{U}_{k_n}$, which has received the action of $\map{U}_{k_n}$ $n-1$ times since $k_n\notin K_{n-1}$, is left untouched, while each copy of $\map{U}_i$, $i\neq k_n$, acts once more on the auxiliary system associated with $\map{U}_i$, which has thus received the action of $\map{U}_i$ $n-1$ times if $i\in K_{n-1}$, and $n$ times if $i\notin K_n$.

The auxiliary state in Eq.~\eqref{eq:varphi_n_QCQC} depends on the sequence $(k_1,\ldots,k_n)$ only through the set $K_n=K_{n-1}\cup\{k_n\}$, which is determined by the control state $\ket{K_{n-1},k_n}$. Hence, the auxiliary registers do not record the order in which the channels in $K_{n-1}$ have been applied: the terms of Eq.~\eqref{eq:varphi_n_QCQC} that share the same control state $\ket{K_{n-1},k_n}$, which correspond to the different orders of the channels in $K_{n-1}$, come with the same auxiliary state. These terms are thus processed coherently by the subsequent internal operators, and their contributions add up at the level of probability amplitudes, as in Eq.~\eqref{eq::ket_s}. This is crucial in the QC-QC case: if the auxiliary registers recorded the order of the channels, tracing them out at the end of the simulation would suppress the interference between the different orders.

At the second to last time step $N$, all the channels have been applied on the target system, i.e., $K_N=[N]$ for all sequences, so that each auxiliary system has received the action of the corresponding unitary channel exactly $N-1$ times, independently of the order $(k_1,\ldots,k_N)$:
\begin{align}
     &\ket{\varphi_{(N)}'} = \sum_{(k_1,\ldots,k_N)} \Big[\bigotimes_{j=1}^N \big(\ket{a_j} * \kket{U_j}^{*(N-1)}\big) \otimes \big(\kket{V_{\emptyset,\emptyset}^{\to k_1}} \!*\! \kket{U_{k_1}} * \cdots \!*\! \kket{V_{K_{N-2},k_{N-1}}^{\to k_N}} \!*\! \kket{U_{k_N}}\big)\Big] \otimes \ket{K_{N-1},k_N}_{C'_{N}}.
\end{align}
Applying the last internal operation $\tilde V_{N+1}$:
\begin{align}
     &\ket{\varphi_{(N+1)}} = \ket{\varphi_{(N)}'} * \kket{\tilde V_{N+1}} \nonumber \\
     &= \sum_{(k_1,\ldots,k_N)} \bigotimes_{j=1}^N \big(\ket{a_j} * \kket{U_j}^{*(N-1)}\big) \otimes \big(\kket{V_{\emptyset,\emptyset}^{\to k_1}} \!*\! \kket{U_{k_1}} * \cdots \!*\! \kket{V_{K_{N-2},k_{N-1}}^{\to k_N}} \!*\! \kket{U_{k_N}} \!*\! \kket{V_{K_{N-1},k_N}^{\to F}}\big) \nonumber \\
     &= \bigotimes_{j=1}^N \big(\ket{a_j} * \kket{U_j}^{*(N-1)}\big) \otimes \left[\sum_{(k_1,\ldots,k_N)} \kket{V_{\emptyset,\emptyset}^{\to k_1}} \!*\! \kket{U_{k_1}} * \cdots \!*\! \kket{V_{K_{N-2},k_{N-1}}^{\to k_N}} \!*\! \kket{U_{k_N}} \!*\! \kket{V_{K_{N-1},k_N}^{\to F}}\right] \nonumber \\
     &= \bigotimes_{j=1}^N \big(\ket{a_j} * \kket{U_j}^{*(N-1)}\big) \otimes \ket{s_{\boldsymbol{U}}},
\end{align}
where we used $K_{N-1}=[N]\setminus\{k_N\}$ and the definition of $\ket{s_{\boldsymbol{U}}}$ given below Eq.~\eqref{eq:output_channel_QCQC(unitary}. Since the auxiliary state does not depend on the order $(k_1,\ldots,k_N)$, it factorizes from the rest of the output and it is normalized. Hence, Eq.~\eqref{eq:output_state_simu_QCQC} gives
\begin{align}
    C*(\kketbra{U_1}{U_1}^{\otimes N}\otimes\ldots \otimes \kketbra{U_N}{U_N}^{\otimes N}) &= \tr_{\alpha_F\,\text{aux}_{t_{N+1}}}\big(\ketbra{\varphi_{(N+1)}}{\varphi_{(N+1)}}\big) \nonumber \\
    &= \tr_{\alpha_F}\big(\ketbra{s_{\boldsymbol{U}}}{s_{\boldsymbol{U}}}\big),
\end{align}
which is the output $S_{\mathsf{QC}}*(\kketbra{U_1}{U_1}\otimes\ldots\otimes\kketbra{U_N}{U_N})$ of the QC-QC superchannel on unitary channels, as defined in Eq.~\eqref{eq:output_channel_QCQC(unitary}, which concludes the proof.

We note that, contrary to the QC-CC case, one does not in general recover the intermediate QC-QC state at each single time step of the simulation by tracing out the auxiliary registers. Indeed, at the end of a time step $n<N$, the auxiliary state depends on the set $K_n$, so that tracing it out would reduce the coherences between control states associated with different sets $K_n$. These coherences are nevertheless preserved in the global state of the simulation, and they are restored at the end of time step $N$, once the auxiliary state no longer depends on the order. The restriction to unitary input channels is essential in the present construction: for a non-unitary channel $\A_k$, the query of $\A_k$ that acts on the target system depends on the time step at which $\A_k$ is selected, so that the environments of the different queries of $\A_k$ would in general carry information about the order of the channels, and thereby decohere the superposition of different orders. Finally, as in the QC-CC case, we observe that $\map{U}_k$ is applied exactly $N-1$ times on each auxiliary register in state $\ket{a_k}$, thus the simulation is not catalytic.
\end{proof}

\subsubsection{Extended simulation}
\label{subsubsec::proof_extendedsimulation_QCQC}

In full generality, the quantum channels input in a QC-QC can be extended quantum channels, which act jointly on the target system and on extension systems that do not enter the QC-QC, as in Sec.~\ref{subsubsec::proof_extendedsimulation_QCCC}. Here, we consider extended unitary channels $\map{U}'_k:\L(\H_{I_k}\otimes\H_{I'_k})\to\L(\H_{O_k}\otimes\H_{O'_k})$, described by unitary operators $U'_k:\H_{I_k}\otimes\H_{I'_k}\to\H_{O_k}\otimes\H_{O'_k}$ with Choi vectors $\kket{U'_k}$, where the primed Hilbert spaces $\H_{I'_k}$ and $\H_{O'_k}$ describe the extension systems. Extending the QC-QC superchannel by the higher-order identity map $\boldsymbol{I}$ acting on these extension systems defines the higher-order linear map
\begin{align}
    \boldsymbol{S}_\mathsf{QC}\otimes\boldsymbol{I}:\bigotimes_{i=1}^N[\L(\H_{I_i}\otimes\H_{I'_i})\to\L(\H_{O_i}\otimes\H_{O'_i})]\to[\L(\H_{P}\otimes\H_{I'_{[N]}})\to\L(\H_{F}\otimes\H_{O'_{[N]}})],
\end{align}
where $\H_{I'_{[N]}}\coloneqq\bigotimes_{k=1}^N\H_{I'_k}$ and $\H_{O'_{[N]}}\coloneqq\bigotimes_{k=1}^N\H_{O'_k}$. Analogously to Eq.~\eqref{eq:output_channel_QCQC(unitary}, the action of the QC-QC superchannel on extended unitary channels is expressed in the Choi picture as
\begin{align}
    S_{\mathsf{QC}}*(\kketbra{U'_1}{U'_1}\otimes \ldots \otimes \kketbra{U'_N}{U'_N}) & = \tr_{\alpha_F} \Big( \ketbra{s_{\boldsymbol{U}'}}{s_{\boldsymbol{U}'}} \Big),
    \label{eq:output_channel_QCQC_2}
\end{align}
where
\begin{align}
    \ket{s_{\boldsymbol{U}'}}
    \coloneqq \sum_{(k_1,\ldots,k_N)} \kket{V^{\to k_1}_{\emptyset,\emptyset}} * \kket{U'_{k_1}} * \kket{V^{\to k_2}_{\emptyset,k_1}} * \kket{U'_{k_2}} * \cdots * \kket{V^{\to k_N}_{K_{N-2},k_{N-1}}} * \kket{U'_{k_N}} * \kket{V^{\to F}_{K_{N-1},k_{N}}}.
    \label{eq:ket_s_U_ext}
\end{align}
Contrary to $\ket{s_{\boldsymbol{U}}}$, the vector $\ket{s_{\boldsymbol{U}'}}\in\H_P\otimes\H_{I'_{[N]}}\otimes\H_F\otimes\H_{O'_{[N]}}\otimes\H_{\alpha_F}$ also involves the extension spaces, which are left open.

We highlight that a fixed-order simulation as described in Sec.~\ref{subsubsec::proof_simulation_QCQC}, which reproduces the output of a QC-QC for any choice of the input unitary channels, does not necessarily reproduce it when extended unitary channels are applied. We thus update the simulation scheme of Fig.~\ref{fig:simu_QCQC} by routing the extension systems together with the target system, as in Sec.~\ref{subsubsec::proof_extendedsimulation_QCCC}, and provide a fixed-order simulation of the action of any QC-QC on unitary channels that remains valid when extended unitary channels are applied, see Fig.~\ref{fig:simu_QCQC_bipartite_channels}.

\begin{proof}
Given a QC-QC superchannel $\boldsymbol{S}_\mathsf{QC}:\bigotimes_{i=1}^N[\L(\H_{I_i})\to\L(\H_{O_i})]\to[\L(\H_{P})\to\L(\H_{F})]$ and extension spaces $\H_{I'_k}$ and $\H_{O'_k}$ for $k\in[N]$, we construct a fixed-order quantum circuit $\boldsymbol{C}':\bigotimes_{i=1}^N[\L(\H_{I_i}\otimes\H_{I'_i})\to\L(\H_{O_i}\otimes\H_{O'_i})]^{\otimes N}\to[\L(\H_{P}\otimes\H_{I'_{[N]}})\to\L(\H_{F}\otimes\H_{O'_{[N]}})]$ which uses $N$ queries of each extended input channel to simulate the action of $\boldsymbol{S}_\mathsf{QC}\otimes\boldsymbol{I}$ on extended unitary channels $\map{U}'_k:\L(\H_{I_k}\otimes\H_{I'_k}) \to \L(\H_{O_k}\otimes\H_{O'_k})$. Analogously to the proof for the non-extended unitary channels presented in Sec.~\ref{subsubsec::proof_simulation_QCQC}, we start by formally defining $\boldsymbol{C}'$ and showing that it is a valid QC-FO superchannel, before showing that $\boldsymbol{C}'$ is indeed simulating the action of $\boldsymbol{S}_\mathsf{QC}\otimes\boldsymbol{I}$ on extended unitary channels.\\

On top of the $N$ copies of each extended input channel, the simulation scheme uses, for each $k\in[N]$, two auxiliary registers initialized in the states $\ket{a_k} \in \H_{\text{aux}_{k,t_1}}$ and $\ket{a'_k} \in \H_{\text{aux}'_{k,t_1}}$, where $\H_{\text{aux}_{k,t_1}}$ and $\H_{\text{aux}'_{k,t_1}}$ are isomorphic to the input Hilbert spaces $\H_{I_k}$ and $\H_{I'_k}$, respectively. It also uses, for each $k\in[N]$, an extension wire which carries the input extension system of $\map{U}'_k$ until $\map{U}'_k$ is applied on the target system, and its output extension system afterwards. The simulation then proceeds in accordance with the structure of a QC-QC, considering $N+1$ time steps labeled by $n \in [N+1]$, each except the last one corresponding to the execution of an extended input channel on the target system. As in Sec.~\ref{subsubsec::proof_extendedsimulation_QCCC}, we denote the global auxiliary space at the beginning of each time step $n$ as $\H_{\text{aux}_{t_n}}\coloneqq\bigotimes_{k=1}^N (\H_{\text{aux}_{k,t_n}}\otimes\H_{\text{aux}'_{k,t_n}})$, where each individual Hilbert space $\H_{\text{aux}_{k,t_n}}$ (resp.~$\H_{\text{aux}'_{k,t_n}}$) is isomorphic to $\H_{I_k}$ (resp.~$\H_{I'_k}$), and the global extension space at the beginning of each time step $n$ as $\H_{\text{ext}_{t_n}}\coloneqq\bigotimes_{k=1}^N \H_{\text{ext}_{k,t_n}}$, where each individual Hilbert space $\H_{\text{ext}_{k,t_n}}$, describing the extension wire associated with $\map{U}'_k$, is isomorphic to $\H_{I'_k}$. For consistency with $\boldsymbol{S}_\mathsf{QC}\otimes\boldsymbol{I}$ being a transformation into quantum channels from $\L(\H_{P}\otimes\H_{I'_{[N]}})$ to $\L(\H_{F}\otimes\H_{O'_{[N]}})$, we impose $\H_{\text{ext}_{k,t_1}}\coloneqq\H_{I'_k}$ and $\H_{\text{ext}_{k,t_{N+1}}}\coloneqq\H_{O'_k}$, i.e., $\H_{\text{ext}_{t_1}}=\H_{I'_{[N]}}$ and $\H_{\text{ext}_{t_{N+1}}}=\H_{O'_{[N]}}$.

The action of the simulation at each time step $n \in [N]$ corresponds to the sequential composition of the following four steps:
\begin{enumerate}
    \item First, the application of the combined internal QC-QC operator $\tilde V_{n}$ to the target, auxiliary and control systems, with Choi vector $\kket{\tilde V_n}$ as defined in Eqs.~\eqref{eq:def_tilde_V1_QCQC}--\eqref{eq:def_tilde_VN1_QCQC}.
    \item Then, controlled on the last label $k_n$ of the control state $\ket{K_{n-1},k_n}$, a routing operator $S'^I_{k_n}: \H_{\text{aux}_{t_n}} \otimes \H_{\text{ext}_{t_n}} \otimes \H_{I_{k_n}} \to \H_{\text{aux}_{t_n^I}} \otimes \H_{\text{ext}_{t_n^{\circ}}} \otimes \H_{\bar I_{k_n}}$ performs a double swap. First, it swaps the incoming target system in $\H_{I_{k_n}}$ coming from $\tilde V_n$ with the auxiliary system in $\H_{\text{aux}_{k_n,t_n}}$, and outputs the target system in $\H_{\text{aux}_{k_n,t_n^I}}$ and the auxiliary system in $\H_{\bar I_{k_n}}$ (which is isomorphic to $\H_{I_{k_n}}$). Second, it swaps the incoming extension system in $\H_{\text{ext}_{k_n,t_n}}$ with the auxiliary system in $\H_{\text{aux}'_{k_n,t_n}}$, and outputs the extension system in $\H_{\text{aux}'_{k_n,t_n^I}}$ and the auxiliary system in $\H_{\text{ext}_{k_n,t_n^{\circ}}}$. It leaves all the other systems unchanged. As in Sec.~\ref{subsubsec::proof_extendedsimulation_QCCC}, the Hilbert spaces $\H_{\text{aux}_{t_n^I}}\coloneqq\bigotimes_{k=1}^N (\H_{\text{aux}_{k,t_n^I}}\otimes\H_{\text{aux}'_{k,t_n^I}})$ and $\H_{\text{ext}_{t_n^{\circ}}}\coloneqq\bigotimes_{k=1}^N \H_{\text{ext}_{k,t_n^{\circ}}}$ describe the collection of all auxiliary Hilbert spaces and the collection of all extension wires, respectively, right after the application of $S'^I_{k_n}$. Since it only permutes subsystems, $S'^I_{k_n}$ is a unitary operator, which implements the same routing as the operation $\map{S}'^I_{k_n}$ of Sec.~\ref{subsubsec::proof_extendedsimulation_QCCC}. The overall controlled routing operator $S'^I_n$ applied at time step $n \in [N]$ can be described by its Choi vector
\begin{align}
    \kket{S'^I_n} = \sum_{K_{n-1}\subseteq[N]}\ \sum_{k_n\in[N]\setminus K_{n-1}} \kket{S'^I_{k_n}} \otimes \ket{K_{n-1},k_n}_{C_n} \otimes \ket{K_{n-1},k_n}_{C_n^{\circ}},
    \label{eq:routing_in_pure_ext}
\end{align}
    where the sum over $K_{n-1}$, as well as the control spaces $\H_{C_n}$ and $\H_{C_n^{\circ}}$, are defined as in Eq.~\eqref{eq:routing_in_pure}.
    \item A copy $\map{U}'_k:\L(\H_{\text{aux}_{k,t_n^I}}\otimes\H_{\text{aux}'_{k,t_n^I}})\to \L(\H_{\text{aux}_{k,t_n^O}}\otimes\H_{\text{aux}'_{k,t_n^O}})$ of each extended unitary channel, with Choi vector $\kket{U'_k}$ (with the same convention as in Sec.~\ref{subsubsec::proof_simulation_QCCC} regarding the input and output Hilbert spaces of the copies), is then applied and transforms the corresponding joint state in $\H_{\text{aux}_{k,t_n^I}}\otimes\H_{\text{aux}'_{k,t_n^I}}$ into an output state in $\H_{\text{aux}_{k,t_n^O}}\otimes\H_{\text{aux}'_{k,t_n^O}}$, the latter Hilbert spaces being isomorphic to $\H_{O_k}$ and $\H_{O'_k}$, respectively, and where $\H_{\text{aux}_{t_n^O}}\coloneqq\bigotimes_{k=1}^N (\H_{\text{aux}_{k,t_n^O}}\otimes\H_{\text{aux}'_{k,t_n^O}})$ describes the collection of all auxiliary Hilbert spaces right after the application of the $N$ extended unitary channels. In particular, the copy of $\map{U}'_{k_n}$ acts jointly on the target system and on the extension system associated with $\map{U}'_{k_n}$, while each copy of $\map{U}'_k$, $k\neq k_n$, acts on the pair of auxiliary systems associated with $\map{U}'_k$.
    \item Finally, the routing is inverted by $S'^O_{k_n}: \H_{\text{aux}_{t_n^O}} \otimes \H_{\text{ext}_{t_n^{\circ}}} \otimes \H_{\bar I_{k_n}} \to \H_{\text{aux}_{t_{n+1}}} \otimes \H_{\text{ext}_{t_{n+1}}} \otimes \H_{O_{k_n}}$, which performs a double swap. First, it swaps the target state in $\H_{\text{aux}_{k_n,t_n^O}}$ with the auxiliary state in $\H_{\bar I_{k_n}}$, the target state being eventually output in $\H_{O_{k_n}}$ and sent to $\tilde V_{n+1}$, and the auxiliary state being eventually output in $\H_{\text{aux}_{k_n,t_{n+1}}}$. Second, it swaps the extension state in $\H_{\text{aux}'_{k_n,t_n^O}}$ with the auxiliary state in $\H_{\text{ext}_{k_n,t_n^{\circ}}}$, the extension state being eventually output in the extension wire $\H_{\text{ext}_{k_n,t_{n+1}}}$, and the auxiliary state being eventually output in $\H_{\text{aux}'_{k_n,t_{n+1}}}$. It leaves all the other systems unchanged. The overall controlled routing operator $S'^O_n$ applied at time step $n$ can be described by its Choi vector
    \begin{align}
    \kket{S'^O_n} = \sum_{K_{n-1}\subseteq[N]}\ \sum_{k_n\in[N]\setminus K_{n-1}} \kket{S'^O_{k_n}} \otimes \ket{K_{n-1},k_n}_{C_n^{\circ}} \otimes \ket{K_{n-1},k_n}_{C_n'},
     \label{eq:routing_out_pure_ext}
\end{align}
where, as in Eq.~\eqref{eq:routing_out_pure}, $\H_{C_n'}$ is the Hilbert space corresponding to the control system output by $S'^O_n$ and input in $\tilde V_{n+1}$.

We will denote by $\mathcal{Q}'_n$ the parallel higher-order transformation with $N$ slots obtained by applying $S'^I_n$, leaving open one slot $\H_{\text{aux}_{k,t_n^I}}\otimes\H_{\text{aux}'_{k,t_n^I}}\to\H_{\text{aux}_{k,t_n^O}}\otimes\H_{\text{aux}'_{k,t_n^O}}$ for each extended input channel, $k \in [N]$, and then applying $S'^O_n$. In the Choi picture, it is described by the Choi vector $\kket{Q'_n} = \kket{S'^I_n} * \kket{S'^O_n}$.
\end{enumerate}

The last time step $n=N+1$ consists in applying the last combined internal QC-QC operator $\tilde V_{N+1}$, which transforms the target, auxiliary and control systems into a state in $\H_F\otimes\H_{\alpha_F}$, and in tracing out the final auxiliary system of the QC-QC in $\H_{\alpha_F}$, as well as all the auxiliary registers in $\H_{\text{aux}_{t_{N+1}}}$. The extension wires in $\H_{\text{ext}_{t_{N+1}}}=\H_{O'_{[N]}}$ are output as the output extension systems.

For any QC-QC superchannel $\boldsymbol{S}_\mathsf{QC}:\bigotimes_{i=1}^N[\L(\H_{I_i})\to\L(\H_{O_i})]\to[\L(\H_{P})\to\L(\H_{F})]$ and extension spaces $\H_{I'_k}$ and $\H_{O'_k}$, we can thus formally define the superchannel $\boldsymbol{C}':\bigotimes_{i=1}^N[\L(\H_{I_i}\otimes\H_{I'_i})\to\L(\H_{O_i}\otimes\H_{O'_i})]^{\otimes N}\to[\L(\H_{P}\otimes\H_{I'_{[N]}})\to\L(\H_{F}\otimes\H_{O'_{[N]}})]$ as the alternating composition, step by step, of the procedure described above, except the third step as it corresponds to the application of the extended channels which are input in $\boldsymbol{C}'$. As in Sec.~\ref{subsubsec::proof_simulation_QCQC}, since the auxiliary registers are initialized in a pure state and since the operators $\tilde V_n$, $S'^I_n$ and $S'^O_n$ are isometries, the Choi operator of the superchannel $\boldsymbol{C}'$ is described as
\begin{align}
C' & = \tr_{\alpha_F\,\text{aux}_{t_{N+1}}}\big(\ketbra{c'}{c'}\big), \label{eq:output_state_simu_QCQC_extended} \\
\ket{c'} & \coloneqq \ket{\psi_{\text{aux}_{t_1}}} * \kket{\tilde V_1} * \kket{Q'_1} * \kket{\tilde V_2} * \kket{Q'_2}  * \cdots * \kket{\tilde V_N} * \kket{Q'_N} * \kket{\tilde V_{N+1}}, \label{eq:ket_c_simu_QCQC_extended}
\end{align}
where $\ket{\psi_{\text{aux}_{t_1}}}=\bigotimes_{i=1}^N \ket{a_i}\otimes\ket{a'_i} \in \H_{\text{aux}_{t_1}}$ is the initial auxiliary state. Note that the extension spaces $\H_{\text{ext}_{t_1}}=\H_{I'_{[N]}}$ and $\H_{\text{ext}_{t_{N+1}}}=\H_{O'_{[N]}}$ remain open in Eq.~\eqref{eq:ket_c_simu_QCQC_extended}, and play, together with $\H_P$ and $\H_F$, the role of the global past and global future of $\boldsymbol{C}'$.

That the combined operators $\tilde V_n$ are isometries follows from the normalization and orthogonality conditions satisfied by the operators $V_{K_{n-1},k_n}^{\to k_{n+1}}$, so that the corresponding maps $\rho\mapsto\tilde V_n\rho\tilde V_n^\dagger$ are CPTP. The routing operators $S'^I_n$ and $S'^O_n$ are controlled permutations of subsystems, hence unitary, and $\mathcal{Q}'_n$ is a valid parallel higher-order superchannel. Thus, at each time step, the composition of $\tilde V_n$ and $\mathcal{Q}'_n$, with Choi vector $\kket{\tilde V_n} * \kket{Q'_n}$, defines a valid parallel higher-order process, which can be interpreted as a QC-FO between the $N$ extended input channels, in which the extension wires are carried along as memory systems. Finally, by Eqs.~\eqref{eq:output_state_simu_QCQC_extended} and~\eqref{eq:ket_c_simu_QCQC_extended}, the superchannel $\boldsymbol{C}'$ is obtained by preparing the auxiliary registers in the state $\ket{\psi_{\text{aux}_{t_1}}}$, applying sequentially the blocks $\kket{\tilde V_1} * \kket{Q'_1}, \ldots, \kket{\tilde V_N} * \kket{Q'_N}$ and then $\kket{\tilde V_{N+1}}$, and tracing out $\H_{\alpha_F}$ and the auxiliary registers in $\H_{\text{aux}_{t_{N+1}}}$. Since each block $\kket{\tilde V_n} * \kket{Q'_n}$ describes a QC-FO, while $\tilde V_{N+1}$, the state preparation and the partial traces define CPTP maps, this sequential composition is itself a QC-FO, i.e., $\boldsymbol{C}'$ is a valid QC-FO superchannel.

\begin{figure*}[h]
    \centering
    \includegraphics[width=\columnwidth]{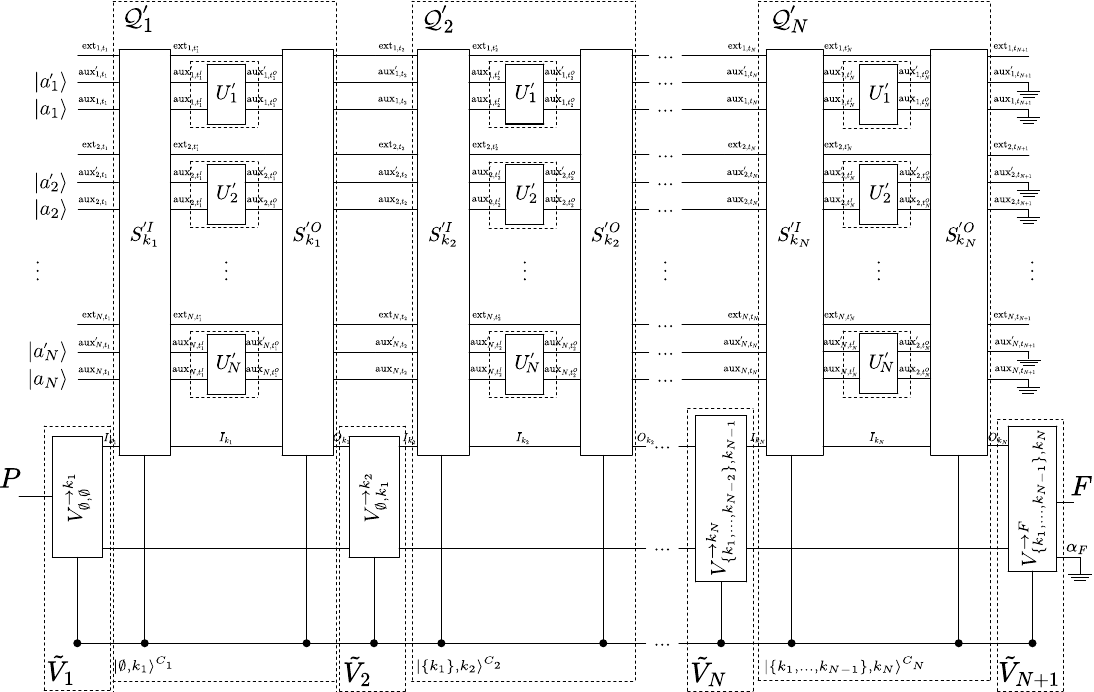}
    \caption{Fixed order simulation of a QC-QC with input extended unitary channels $\map{U}'_1,\ldots,\map{U}'_N$. $N$ queries of each extended unitary channel $\map{U}'_k$ are made, so that the simulation makes $N^2$ queries to the input channels. Compared to Fig.~\ref{fig:simu_QCQC}, the routing operations also swap the extension wires with additional auxiliary registers, so that the copy of $\map{U}'_{k_n}$ applied on the target system at time step $n$ also acts on the extension system associated with $\map{U}'_{k_n}$.}
    \label{fig:simu_QCQC_bipartite_channels}
\end{figure*}

\vspace{5mm}

We will now show that $\boldsymbol{C}'$ applied on $N$ copies of each extended unitary channel is simulating the action of $\boldsymbol{S}_\mathsf{QC}\otimes\boldsymbol{I}$ on these extended unitary channels, for any choice of the extended unitary channels, i.e., that $\boldsymbol{C}'$ is an extended simulation of the action of $\boldsymbol{S}_\mathsf{QC}$ on unitary channels, in the sense of Definition~\ref{def:simulation_extended}. More formally, we want to show that
\begin{align}
    \boldsymbol{C}'(\map{U}_{1}^{\prime\otimes N}\otimes\ldots\otimes\map{U}_{N}^{\prime\otimes N}) = (\boldsymbol{S}_\mathsf{QC}\otimes\boldsymbol{I})(\map{U}'_1\otimes\ldots\otimes\map{U}'_N),
\end{align}
for all possible extended unitary channels $\map{U}'_1,\ldots,\map{U}'_N$. Equivalently, we can rephrase the last equation in the Choi picture as
\begin{align}
    C'*(\kketbra{U'_1}{U'_1}^{\otimes N}\otimes\ldots \otimes \kketbra{U'_N}{U'_N}^{\otimes N}) = S_{\mathsf{QC}}*(\kketbra{U'_1}{U'_1} \otimes \ldots \otimes \kketbra{U'_N}{U'_N}),
\end{align}
for all possible unitary operators $U'_1,\ldots,U'_N$.

The proof consists in computing iteratively the output of the simulation scheme presented in Fig.~\ref{fig:simu_QCQC_bipartite_channels} and showing that it matches the output of the QC-QC applied on the extended unitary channels, as defined in Eq.~\eqref{eq:output_channel_QCQC_2}.

Following Definition~\ref{def:QC-QC}, any QC-QC superchannel $\boldsymbol{S}_\mathsf{QC}$ is defined by a collection of internal operators $\{V_{\emptyset,\emptyset}^{\to k_{1}},\ldots,\allowbreak \, V_{K_{n-1},k_n}^{\to k_{n+1}},\allowbreak\ldots, \, V_{K_{N-1},k_N}^{\to F}\}$.

As in Sec.~\ref{subsubsec::proof_simulation_QCQC}, since the input channels are unitary and the simulation scheme only involves a pure initial state and isometries, the output of the simulation scheme can be computed at the level of Choi vectors, by composing iteratively the operations $\tilde V_n$, $\mathcal{Q}'_n$ and the extended unitary channels $\map{U}'_1 \otimes \ldots \otimes \map{U}'_N$. The auxiliary registers are initialized in the pure state $\ket{\psi_{\text{aux}_{t_1}}}=\bigotimes_{i=1}^N \ket{a_i}\otimes\ket{a'_i}$. We describe by $\ket{\varphi_{(n)}'}$ (resp.~$\ket{\varphi_{(n+1)}}$) the simulation output at time step $n$, before (resp.~after) the operation $\tilde V_{n+1}$ is applied:
\begin{align}
\ket{\varphi_{(n)}'} & = \big(\ket{\psi_{\text{aux}_{t_1}}} * \kket{\tilde V_1} * \kket{Q'_1} *  \kket{\tilde V_2} * \cdots * \kket{\tilde V_n} * \kket{Q'_n}\big) * (\kket{U'_{1}}^{\otimes {n}}\otimes\ldots \otimes \kket{U'_{N}}^{\otimes {n}}), \notag
\end{align}
and
\begin{align}
\ket{\varphi_{(n+1)}} & = \big(\ket{\psi_{\text{aux}_{t_1}}} * \kket{\tilde V_1} * \kket{Q'_1} * \kket{\tilde V_2} * \cdots * \kket{Q'_n} * \kket{\tilde V_{n+1}}\big) * (\kket{U'_{1}}^{\otimes {n}}\otimes\ldots \otimes \kket{U'_{N}}^{\otimes {n}}).
\end{align}
At the first time step, right after the first internal operation $\tilde V_1$ has been applied, the simulation output is described by
\begin{align}
    \ket{\varphi_{(1)}} & = \ket{\psi_{\text{aux}_{t_1}}} * \kket{\tilde V_1} \nonumber \\
    &= \sum_{k_1} \bigotimes_{i=1}^N \big(\ket{a_i}\otimes\ket{a'_i}\big) \otimes \kket{V_{\emptyset,\emptyset}^{\to k_{1}}} \otimes \ket{\emptyset,k_1}_{C_{1}}.
\end{align}
Applying the first extended unitary channels in parallel along with the swap operations, we obtain the Choi vector at the end of the first time step as
\begin{align}
    \ket{\varphi_{(1)}'} & = \ket{\varphi_{(1)}} * \kket{Q'_1} * (\kket{U'_{1}}\otimes\ldots \otimes \kket{U'_{N}}) = \ket{\varphi_{(1)}} * \kket{S'^I_1}*(\kket{U'_1}\otimes \ldots \otimes \kket{U'_N}) * \kket{S'^O_1} \nonumber \\
    & = \bigg[\sum_{k_1} \Big[ \bigotimes_{i=1}^N \big(\ket{a_i}\otimes\ket{a'_i}\big) \otimes \kket{V_{\emptyset,\emptyset}^{\to k_{1}}} \Big] * \kket{S'^I_{k_1}}\otimes \ket{\emptyset,k_1}_{C^{\circ}_{1}} \bigg] *(\kket{U'_1}\otimes \ldots \otimes \kket{U'_N}) * \kket{S'^O_1}\nonumber \\
    & = \bigg[\sum_{k_1} \Big[ \bigotimes_{\substack{i=1\\ i\neq k_1}}^N \Big(\big(\ket{a_i}\otimes\ket{a'_i}\big) * \kket{U'_i}\Big) \otimes \big(\ket{a_{k_1}}\otimes\ket{a'_{k_1}}\big) \otimes \big(\kket{V_{\emptyset,\emptyset}^{\to k_{1}}} * \kket{U'_{k_1}}\big) \nonumber \\
    & \ \ \ \ \ \ \ \ \ \ \ \ \ \ \ \otimes \bigotimes_{\substack{i=1\\ i\neq k_1}}^N \kket{\id}^{I'_i\,\text{ext}_{i,t_1^{\circ}}} \Big] \otimes \ket{\emptyset,k_1}_{C^{\circ}_{1}}\bigg] * \kket{S'^O_1} \nonumber \\
    & = \sum_{k_1} \Big[ \bigotimes_{\substack{i=1\\ i\neq k_1}}^N \Big(\big(\ket{a_i}\otimes\ket{a'_i}\big) * \kket{U'_i}\Big) \otimes \big(\ket{a_{k_1}}\otimes\ket{a'_{k_1}}\big) \otimes \big(\kket{V_{\emptyset,\emptyset}^{\to k_{1}}} * \kket{U'_{k_1}}\big) \otimes \bigotimes_{\substack{i=1\\ i\neq k_1}}^N \kket{\id}^{I'_i\,\text{ext}_{i,t_2}} \Big] \otimes \ket{\emptyset,k_1}_{C'_{1}},
\end{align}
where $(\ket{a_i}\otimes\ket{a'_i}) * \kket{U'_i}=U'_i(\ket{a_i}\otimes\ket{a'_i})$. In the third line, the auxiliary states $\ket{a_{k_1}}$ and $\ket{a'_{k_1}}$ are stored in $\H_{\bar I_{k_1}}$ and $\H_{\text{ext}_{k_1,t_1^{\circ}}}$, respectively, while the copy of $\map{U}'_{k_1}$ acts on the target system output by $V_{\emptyset,\emptyset}^{\to k_1}$ and on the extension system in $\H_{\text{ext}_{k_1,t_1}}=\H_{I'_{k_1}}$, and outputs them in $\H_{\text{aux}_{k_1,t_1^O}}$ and $\H_{\text{aux}'_{k_1,t_1^O}}$, respectively; these systems are then swapped back by $S'^O_{k_1}$. In the last line, the target system output by the copy of $\map{U}'_{k_1}$ is thus in $\H_{O_{k_1}}$, its output extension system is stored in the extension wire $\H_{\text{ext}_{k_1,t_2}}$, and the auxiliary states $\ket{a_{k_1}}$ and $\ket{a'_{k_1}}$ are back in $\H_{\text{aux}_{k_1,t_2}}$ and $\H_{\text{aux}'_{k_1,t_2}}$. Moreover, $\kket{\id}^{I'_i\,\text{ext}_{i,t_2}}$ denotes the Choi vector of the identity operator from $\H_{I'_i}=\H_{\text{ext}_{i,t_1}}$ to $\H_{\text{ext}_{i,t_2}}$: it describes the extension wires associated with the extended unitary channels $\map{U}'_i$, $i\neq k_1$, which have not been applied on the target system yet, and which thus transmit the input extension systems unchanged. More generally, at the end of each time step $n$, the extension wire associated with an extended unitary channel $\map{U}'_{k_m}$ that has been applied on the target system at a time step $m\leq n$ carries the output extension system of this copy of $\map{U}'_{k_m}$, transmitted with identity operators from $\H_{\text{ext}_{k_m,t_{m+1}}}$ to $\H_{\text{ext}_{k_m,t_{n+1}}}$. For the sake of conciseness, we absorb these identity operators into the Choi vector $\kket{U'_{k_m}}$ of the copy of $\map{U}'_{k_m}$ applied on the target system, whose output extension system is thus understood to be in $\H_{\text{ext}_{k_m,t_{n+1}}}$ at the end of time step $n$.

The output of the simulation scheme at the second time step before the extended unitary channels are applied is
\begin{align}
     \ket{\varphi_{(2)}} & = \ket{\varphi_{(1)}'} * \kket{\tilde V_2} \nonumber \\
     &= \sum_{k_1,k_2} \bigg[\bigotimes_{\substack{i=1\\ i\neq k_1}}^N \Big(\big(\ket{a_i}\otimes\ket{a'_i}\big) * \kket{U'_i}\Big) \otimes \big(\ket{a_{k_1}}\otimes\ket{a'_{k_1}}\big) \nonumber \\
     & \ \ \ \ \ \ \ \ \ \ \ \ \ \ \  \ \ \ \ \ \ \ \ \ \ \ \ \ \ \  \ \ \ \ \otimes \big(\kket{V_{\emptyset,\emptyset}^{\to k_{1}}} * \kket{U'_{k_1}} * \kket{V_{\emptyset,k_1}^{\to k_{2}}}\big)  \otimes \bigotimes_{\substack{i=1\\ i\neq k_1}}^N \kket{\id}^{I'_i\,\text{ext}_{i,t_2}} \bigg] \otimes \ket{\{k_1\},k_2}_{C_{2}}.
\end{align}
At the end of the second time step, once the extended unitary channels and the swap operations have been applied, we note that each extended unitary channel $\map{U}'_k$ is applied twice on the corresponding pair of auxiliary systems in state $\ket{a_k}\otimes\ket{a'_k}$, except for $k=k_1,k_2$. Indeed, $\map{U}'_{k_1},\map{U}'_{k_2}$ are respectively applied only once on the corresponding auxiliary states since the other copy of this operation was applied on the target system and on the corresponding extension system:
\begin{align}
     \ket{\varphi_{(2)}'} & = \ket{\varphi_{(2)}} * \kket{Q'_2} * (\kket{U'_{1}}\otimes\ldots \otimes \kket{U'_{N}})\nonumber \\
     &= \sum_{k_1,k_2} \bigg[\bigotimes_{\substack{i=1\\ i\notin\{k_1,k_2\}}}^N \Big(\big(\ket{a_i}\otimes\ket{a'_i}\big) * \kket{U'_i}^{*2}\Big) \otimes \bigotimes_{j\in\{k_1,k_2\}} \Big(\big(\ket{a_j}\otimes\ket{a'_j}\big) * \kket{U'_j}\Big) \nonumber \\
     & \ \ \ \ \ \ \ \ \ \ \ \ \ \ \ \otimes \big(\kket{V_{\emptyset,\emptyset}^{\to k_{1}}} * \kket{U'_{k_1}} * \kket{V_{\emptyset,k_1}^{\to k_{2}}} * \kket{U'_{k_2}}\big) \otimes \bigotimes_{\substack{i=1\\ i\notin\{k_1,k_2\}}}^N \kket{\id}^{I'_i\,\text{ext}_{i,t_3}} \bigg] \otimes \ket{\{k_1\},k_2}_{C_{2}'},
\end{align}
where $\kket{U'_k}^{*2}=\kket{U'_k}*\kket{U'_k}$ is the Choi vector of the unitary operator $U'_kU'_k$, which describes the channel $\map{U}'_k\circ\map{U}'_k$, and where the extension wires associated with the extended unitary channels $\map{U}'_i$, $i\notin\{k_1,k_2\}$, which have not been applied on the target system yet, now transmit the input extension systems from $\H_{I'_i}$ to $\H_{\text{ext}_{i,t_3}}$.
More generally, deriving iteratively the Choi vector after each time step, we obtain at the end of each time step $n\in[N]$
\begin{align}
     \ket{\varphi_{(n)}'} = \sum_{(k_1,\ldots,k_n)} \bigg[&\bigotimes_{i\in K_n} \Big(\big(\ket{a_i}\otimes\ket{a'_i}\big) * \kket{U'_i}^{*(n-1)}\Big) \otimes \bigotimes_{i\in[N]\setminus K_n} \Big(\big(\ket{a_i}\otimes\ket{a'_i}\big) * \kket{U'_i}^{*n}\Big) \nonumber \\
     &\otimes \big(\kket{V_{\emptyset,\emptyset}^{\to k_1}} \!*\! \kket{U'_{k_1}} * \cdots \!*\! \kket{V_{K_{n-2},k_{n-1}}^{\to k_n}} \!*\! \kket{U'_{k_n}}\big) \otimes \bigotimes_{i\in[N]\setminus K_n} \kket{\id}^{I'_i\,\text{ext}_{i,t_{n+1}}}\bigg] \otimes \ket{K_{n-1},k_n}_{C'_{n}},
     \label{eq:varphi_n_QCQC_ext}
\end{align}
where the notations are the same as in Eq.~\eqref{eq:varphi_n_QCQC}, with $\kket{U'_i}^{*m}$ the Choi vector of $U_i^{\prime m}$, and where, following the convention above, the input and output extension systems of the copy of $\map{U}'_{k_m}$ applied on the target system, $m\in[n]$, are in $\H_{I'_{k_m}}$ and $\H_{\text{ext}_{k_m,t_{n+1}}}$, respectively. Indeed, as in Sec.~\ref{subsubsec::proof_simulation_QCQC}, Eq.~\eqref{eq:varphi_n_QCQC_ext} holds for $n=1,2$, and if it holds at time step $n-1$, then at time step $n$, $\tilde V_n$ appends $\kket{V_{K_{n-2},k_{n-1}}^{\to k_n}}$ to each term and updates its control state into $\ket{K_{n-1},k_n}$, for all $k_n\in[N]\setminus K_{n-1}$; the copy of $\map{U}'_{k_n}$ then acts on the target system and on the input extension system of $\map{U}'_{k_n}$, which has been transmitted unchanged from $\H_{I'_{k_n}}$ since $k_n\notin K_{n-1}$, and its output extension system is stored in $\H_{\text{ext}_{k_n,t_{n+1}}}$, while the pair of auxiliary systems associated with $\map{U}'_{k_n}$ is left untouched; each copy of $\map{U}'_i$, $i\neq k_n$, acts once more on the pair of auxiliary systems associated with $\map{U}'_i$, and the extension wires associated with $\map{U}'_i$, $i\notin K_n$, keep transmitting the input extension systems unchanged.

As in Sec.~\ref{subsubsec::proof_simulation_QCQC}, the auxiliary state in Eq.~\eqref{eq:varphi_n_QCQC_ext} depends on the sequence $(k_1,\ldots,k_n)$ only through the set $K_n$, which is determined by the control state $\ket{K_{n-1},k_n}$, and so do the identity operators describing the extension wires associated with the extended unitary channels that have not been applied on the target system yet. Moreover, since all the copies of an extended unitary channel $\map{U}'_k$ are described by the same Choi vector $\kket{U'_k}$, the term of Eq.~\eqref{eq:varphi_n_QCQC_ext} associated with a given sequence does not depend on which copy of $\map{U}'_k$ has been applied on the target system and on the extension system associated with $\map{U}'_k$, i.e., on the time step at which $\map{U}'_k$ has been selected. Hence, the terms that share the same control state $\ket{K_{n-1},k_n}$, which correspond to the different orders of the channels in $K_{n-1}$, come with the same auxiliary state and the same extension wires, and their contributions add up at the level of probability amplitudes, as in Eq.~\eqref{eq:ket_s_U_ext}.

At the second to last time step $N$, all the extended unitary channels have been applied on the target system, i.e., $K_N=[N]$ for all sequences, so that each pair of auxiliary systems has received the action of the corresponding extended unitary channel exactly $N-1$ times, independently of the order $(k_1,\ldots,k_N)$:
\begin{align}
     \ket{\varphi_{(N)}'} = \sum_{(k_1,\ldots,k_N)} \Big[&\bigotimes_{j=1}^N \Big(\big(\ket{a_j}\otimes\ket{a'_j}\big) * \kket{U'_j}^{*(N-1)}\Big) \nonumber \\
     &\otimes \big(\kket{V_{\emptyset,\emptyset}^{\to k_1}} \!*\! \kket{U'_{k_1}} * \cdots \!*\! \kket{V_{K_{N-2},k_{N-1}}^{\to k_N}} \!*\! \kket{U'_{k_N}}\big)\Big] \otimes \ket{K_{N-1},k_N}_{C'_{N}}.
\end{align}
Since all extended unitary channels have now been applied on the target system, no identity operator remains on the extension wires: the input extension systems of the copies of $\map{U}'_{k_1},\ldots,\map{U}'_{k_N}$ applied on the target system are in $\H_{\text{ext}_{t_1}}=\H_{I'_{[N]}}$, and their output extension systems are in $\H_{\text{ext}_{t_{N+1}}}=\H_{O'_{[N]}}$, as in Eq.~\eqref{eq:ket_s_U_ext}.
Applying the last internal operation $\tilde V_{N+1}$:
\begin{align}
     \ket{\varphi_{(N+1)}} &= \ket{\varphi_{(N)}'} * \kket{\tilde V_{N+1}} \nonumber \\
     &= \sum_{(k_1,\ldots,k_N)} \bigotimes_{j=1}^N \Big(\big(\ket{a_j}\otimes\ket{a'_j}\big) * \kket{U'_j}^{*(N-1)}\Big) \nonumber \\
     & \ \ \ \ \ \ \ \ \ \ \ \ \ \ \ \otimes \big(\kket{V_{\emptyset,\emptyset}^{\to k_1}} \!*\! \kket{U'_{k_1}} * \cdots \!*\! \kket{V_{K_{N-2},k_{N-1}}^{\to k_N}} \!*\! \kket{U'_{k_N}} \!*\! \kket{V_{K_{N-1},k_N}^{\to F}}\big) \nonumber \\
     &= \bigotimes_{j=1}^N \Big(\big(\ket{a_j}\otimes\ket{a'_j}\big) * \kket{U'_j}^{*(N-1)}\Big) \otimes \left[\sum_{(k_1,\ldots,k_N)} \kket{V_{\emptyset,\emptyset}^{\to k_1}} \!*\! \kket{U'_{k_1}} * \cdots \!*\! \kket{V_{K_{N-2},k_{N-1}}^{\to k_N}} \!*\! \kket{U'_{k_N}} \!*\! \kket{V_{K_{N-1},k_N}^{\to F}}\right] \nonumber \\
     &= \bigotimes_{j=1}^N \Big(\big(\ket{a_j}\otimes\ket{a'_j}\big) * \kket{U'_j}^{*(N-1)}\Big) \otimes \ket{s_{\boldsymbol{U}'}},
\end{align}
where we used $K_{N-1}=[N]\setminus\{k_N\}$ and the definition of $\ket{s_{\boldsymbol{U}'}}$ in Eq.~\eqref{eq:ket_s_U_ext}. Since the auxiliary state does not depend on the order $(k_1,\ldots,k_N)$, it factorizes from the rest of the output and it is normalized. Hence, Eq.~\eqref{eq:output_state_simu_QCQC_extended} gives
\begin{align}
    C'*(\kketbra{U'_1}{U'_1}^{\otimes N}\otimes\ldots \otimes \kketbra{U'_N}{U'_N}^{\otimes N}) &= \tr_{\alpha_F\,\text{aux}_{t_{N+1}}}\big(\ketbra{\varphi_{(N+1)}}{\varphi_{(N+1)}}\big) \nonumber \\
    &= \tr_{\alpha_F}\big(\ketbra{s_{\boldsymbol{U}'}}{s_{\boldsymbol{U}'}}\big),
\end{align}
which is the output $S_{\mathsf{QC}}*(\kketbra{U'_1}{U'_1}\otimes\ldots\otimes\kketbra{U'_N}{U'_N})$ of the QC-QC superchannel applied on the extended unitary channels, as defined in Eq.~\eqref{eq:output_channel_QCQC_2}, which concludes the proof.

As in Sec.~\ref{subsubsec::proof_simulation_QCQC}, one does not in general recover the intermediate state of the QC-QC applied on the extended unitary channels at each single time step of the simulation by tracing out the auxiliary registers, while its final output is exactly recovered. We also observe that $\map{U}'_k$ is applied exactly $N-1$ times on each pair of auxiliary registers in state $\ket{a_k}\otimes\ket{a'_k}$, so that this extended simulation is not catalytic either.
\end{proof}

\subsubsection{Catalytic simulation}
\label{subsubsec::proof_catalyticsimulation_QCQC}

We finally show that a catalytic simulation, as defined in Definition~\ref{def:simulation_catalytic}, exists for the action of any QC-QC on unitary channels. For the sake of simplicity, we exhibit such a simulation for input unitary channels $\map{U}_k:\L(\H_{I_k})\to\L(\H_{O_k})$, and we explain at the end of the proof how it can be straightforwardly combined with the extended simulation of Sec.~\ref{subsubsec::proof_extendedsimulation_QCQC}.

In the simulation of Sec.~\ref{subsubsec::proof_simulation_QCQC}, the $N-1$ copies of each unitary channel $\map{U}_k$ that are not applied on the target system all act on the same auxiliary register, whose final state $\ket{a_k}*\kket{U_k}^{*(N-1)}$ is eventually traced out. As in Sec.~\ref{subsubsec::proof_catalyticsimulation_QCCC}, to make the simulation catalytic, we distribute these $N-1$ applications of $\map{U}_k$ over $N-1$ distinct auxiliary systems. The $N-1$ applications of each unitary channel $\map{U}_k$ are returned, unchanged, as part of the output of the simulation, see Fig.~\ref{fig:simu_QCQC_catalytic}.

\begin{proof}
Given a QC-QC superchannel $\boldsymbol{S}_\mathsf{QC}:\bigotimes_{i=1}^N[\L(\H_{I_i})\to\L(\H_{O_i})]\to[\L(\H_{P})\to\L(\H_{F})]$, we construct a fixed-order quantum circuit $\widehat{\boldsymbol{C}}:\bigotimes_{i=1}^N[\L(\H_{I_i})\to\L(\H_{O_i})]^{\otimes N}\to[\L(\H_{P}\otimes\bigotimes_{i=1}^N\H_{I_i}^{\otimes(N-1)})\to\L(\H_{F}\otimes\bigotimes_{i=1}^N\H_{O_i}^{\otimes(N-1)})]$ which uses $N$ queries of each input channel to catalytically simulate the action of the QC-QC on unitary channels $\map{U}_k:\L(\H_{I_k}) \to \L(\H_{O_k})$, with Choi vectors $\kket{U_k}$. Analogously to the proofs presented in Secs.~\ref{subsubsec::proof_simulation_QCQC} and~\ref{subsubsec::proof_extendedsimulation_QCQC}, we start by formally defining $\widehat{\boldsymbol{C}}$ and showing that it is a valid QC-FO superchannel, before showing that $\widehat{\boldsymbol{C}}$ is indeed a catalytic simulation of the action of $\boldsymbol{S}_\mathsf{QC}$ on unitary channels.\\

On top of the $N$ copies of each input channel, the simulation scheme uses, for each $k\in[N]$, $N-1$ auxiliary systems, hence $N(N-1)$ auxiliary systems in total. As in Sec.~\ref{subsubsec::proof_catalyticsimulation_QCCC}, and contrary to Secs.~\ref{subsubsec::proof_simulation_QCQC} and~\ref{subsubsec::proof_extendedsimulation_QCQC}, these auxiliary systems are not initialized in fixed states: their inputs and outputs are left open, and carry the inputs and outputs of the $N-1$ catalyst copies of each unitary channel $\map{U}_k$. The simulation then proceeds in accordance with the structure of a QC-QC, considering $N+1$ time steps labeled by $n \in [N+1]$, each except the last one corresponding to the execution of an input channel on the target system. At the beginning of each time step $n$, the $N-1$ auxiliary systems associated with $\map{U}_k$ are carried by the wires $\H_{\text{aux}_{k,t_n}^i}$, $i\in[N-1]$, each isomorphic to $\H_{I_k}$, and we say that the auxiliary system carried by $\H_{\text{aux}_{k,t_n}^i}$ is in position $i$. We denote the auxiliary space associated with $\map{U}_k$ at the beginning of time step $n$ as $\H_{\text{aux}_{k,t_n}}\coloneqq\bigotimes_{i=1}^{N-1}\H_{\text{aux}_{k,t_n}^i}$, and the global auxiliary space as $\H_{\text{aux}_{t_n}}\coloneqq\bigotimes_{k=1}^N\H_{\text{aux}_{k,t_n}}$. For consistency with Definition~\ref{def:simulation_catalytic}, the input spaces $\H_{\text{aux}_{k,t_1}^i}$ and output spaces $\H_{\text{aux}_{k,t_{N+1}}^i}$ are isomorphic to $\H_{I_k}$ and $\H_{O_k}$, respectively, i.e., $\H_{\text{aux}_{t_1}}\cong\bigotimes_{k=1}^N\H_{I_k}^{\otimes(N-1)}$ and $\H_{\text{aux}_{t_{N+1}}}\cong\bigotimes_{k=1}^N\H_{O_k}^{\otimes(N-1)}$.

The action of the simulation at each time step $n \in [N]$ corresponds to the sequential composition of the following four steps:
\begin{enumerate}
    \item First, the application of the combined internal QC-QC operator $\tilde V_{n}$ to the target, auxiliary and control systems, with Choi vector $\kket{\tilde V_n}$ as defined in Eqs.~\eqref{eq:def_tilde_V1_QCQC}--\eqref{eq:def_tilde_VN1_QCQC}.
    \item Then, controlled on the last label $k_n$ of the control state $\ket{K_{n-1},k_n}$, a routing operator $\hat S^I_{k_n}: \H_{\text{aux}_{t_n}} \otimes \H_{I_{k_n}} \to \H_{\text{aux}_{t_n^I}} \otimes \H_{\bar I_{k_n}}$ swaps the incoming target system in $\H_{I_{k_n}}$ coming from $\tilde V_n$ with the auxiliary system in position 1 associated with $\map{U}_{k_n}$, in $\H_{\text{aux}_{k_n,t_n}^1}$, and outputs the target system in $\H_{\text{aux}_{k_n,t_n^I}^1}$ and the auxiliary system in $\H_{\bar I_{k_n}}$ (which is isomorphic to $\H_{I_{k_n}}$), while it leaves all the other auxiliary systems unchanged, transmitting them from $\H_{\text{aux}_{k,t_n}^i}$ to $\H_{\text{aux}_{k,t_n^I}^i}$. The Hilbert space $\H_{\text{aux}_{t_n^I}}\coloneqq\bigotimes_{k=1}^N\bigotimes_{i=1}^{N-1} \H_{\text{aux}_{k,t_n^I}^i}$ describes the collection of all auxiliary Hilbert spaces $\H_{\text{aux}_{k,t_n^I}^i}$ (which are isomorphic to $\H_{I_k}$) right after the application of $\hat S^I_{k_n}$. Since it only permutes subsystems, $\hat S^I_{k_n}$ is a unitary operator, which implements the same routing as the operation $\hat{\map{S}}^I_{k_n}$ of Sec.~\ref{subsubsec::proof_catalyticsimulation_QCCC}. The overall controlled routing operator $\hat S^I_n$ applied at time step $n \in [N]$ can be described by its Choi vector
\begin{align}
    \kket{\hat S^I_n} = \sum_{K_{n-1}\subseteq[N]}\ \sum_{k_n\in[N]\setminus K_{n-1}} \kket{\hat S^I_{k_n}} \otimes \ket{K_{n-1},k_n}_{C_n} \otimes \ket{K_{n-1},k_n}_{C_n^{\circ}},
    \label{eq:routing_in_pure_cat}
\end{align}
    where the sum over $K_{n-1}$, as well as the control spaces $\H_{C_n}$ and $\H_{C_n^{\circ}}$, are defined as in Eq.~\eqref{eq:routing_in_pure}.
    \item A copy $\map{U}_k:\L(\H_{\text{aux}_{k,t_n^I}^1})\to \L(\H_{\text{aux}_{k,t_n^O}^1})$ of each unitary channel, with Choi vector $\kket{U_k}\in\H_{\text{aux}_{k,t_n^I}^1}\otimes\H_{\text{aux}_{k,t_n^O}^1}$ (with the same convention as in Sec.~\ref{subsubsec::proof_simulation_QCCC} regarding the input and output Hilbert spaces of the copies), is then applied on the system in position 1 associated with $\map{U}_k$, and outputs it in $\H_{\text{aux}_{k,t_n^O}^1}$, which is isomorphic to $\H_{O_k}$, while the auxiliary systems in positions $i\in\{2,\ldots,N-1\}$ are transmitted with an identity operator from $\H_{\text{aux}_{k,t_n^I}^i}$ to $\H_{\text{aux}_{k,t_n^O}^i}$, which simply relabels the Hilbert space $\H_{\text{aux}_{k,t_n^I}^i}$ as $\H_{\text{aux}_{k,t_n^O}^i}$. The Hilbert space $\H_{\text{aux}_{t_n^O}}\coloneqq\bigotimes_{k=1}^N\bigotimes_{i=1}^{N-1} \H_{\text{aux}_{k,t_n^O}^i}$ describes the collection of all auxiliary Hilbert spaces right after the application of the $N$ unitary channels. In particular, the copy of $\map{U}_{k_n}$ acts on the target system, while each copy of $\map{U}_k$, $k\neq k_n$, acts on the auxiliary system in position 1 associated with $\map{U}_k$.
    \item Finally, the routing is inverted by $\hat S^O_{k_n}: \H_{\text{aux}_{t_n^O}} \otimes \H_{\bar I_{k_n}} \to \H_{\text{aux}_{t_{n+1}}} \otimes \H_{O_{k_n}}$, which swaps the target state in $\H_{\text{aux}_{k_n,t_n^O}^1}$ with the auxiliary state in $\H_{\bar I_{k_n}}$, the target state being eventually output in $\H_{O_{k_n}}$ and sent to $\tilde V_{n+1}$, and the auxiliary state being eventually output in $\H_{\text{aux}_{k_n,t_{n+1}}^1}$. In addition, for each $k\neq k_n$, $\hat S^O_{k_n}$ cyclically permutes the auxiliary systems associated with $\map{U}_k$: it moves the auxiliary system in $\H_{\text{aux}_{k,t_n^O}^1}$, which has just received the action of $\map{U}_k$, to the last position $\H_{\text{aux}_{k,t_{n+1}}^{N-1}}$, and the auxiliary system in $\H_{\text{aux}_{k,t_n^O}^{i+1}}$ to $\H_{\text{aux}_{k,t_{n+1}}^{i}}$ for $i\in[N-2]$. The other auxiliary systems associated with $\map{U}_{k_n}$, in positions $i\in\{2,\ldots,N-1\}$, keep their positions. Like $\hat S^I_{k_n}$, the routing operator $\hat S^O_{k_n}$ is unitary, and it implements the same routing as the operation $\hat{\map{S}}^O_{k_n}$ of Sec.~\ref{subsubsec::proof_catalyticsimulation_QCCC}. The overall controlled routing operator $\hat S^O_n$ applied at time step $n$ can be described by its Choi vector
    \begin{align}
    \kket{\hat S^O_n} = \sum_{K_{n-1}\subseteq[N]}\ \sum_{k_n\in[N]\setminus K_{n-1}} \kket{\hat S^O_{k_n}} \otimes \ket{K_{n-1},k_n}_{C_n^{\circ}} \otimes \ket{K_{n-1},k_n}_{C_n'},
     \label{eq:routing_out_pure_cat}
\end{align}
where, as in Eq.~\eqref{eq:routing_out_pure}, $\H_{C_n'}$ is the Hilbert space corresponding to the control system output by $\hat S^O_n$ and input in $\tilde V_{n+1}$.

We will denote by $\hat{\mathcal{Q}}_n$ the parallel higher-order transformation with $N$ slots obtained by applying $\hat S^I_n$, leaving open one slot $\H_{\text{aux}_{k,t_n^I}^1}\to\H_{\text{aux}_{k,t_n^O}^1}$ for each input channel, $k \in [N]$, and then applying $\hat S^O_n$. In the Choi picture, it is described by the Choi vector $\kket{\hat Q_n} = \kket{\hat S^I_n} * \kket{\hat S^O_n}$, where the auxiliary systems in positions $i\in\{2,\ldots,N-1\}$, output by $\hat S^I_n$ in $\H_{\text{aux}_{k,t_n^I}^i}$, are input in $\hat S^O_n$ in $\H_{\text{aux}_{k,t_n^O}^i}$ after the relabeling described in the third step.
\end{enumerate}

As in Sec.~\ref{subsubsec::proof_catalyticsimulation_QCCC}, this routing distributes the $N-1$ copies of each unitary channel $\map{U}_k$ that are not applied on the target system over the $N-1$ auxiliary systems associated with $\map{U}_k$, in such a way that each of them receives the action of $\map{U}_k$ exactly once, and these auxiliary systems are all back in their initial positions at the end of time step $N$, independently of the order $(k_1,\ldots,k_N)$ in which the channels are applied on the target system. Note that the routing operators $\hat S^I_{k_n}$ and $\hat S^O_{k_n}$ only depend on the last label $k_n$ of the control state, as depicted in Fig.~\ref{fig:simu_QCQC_catalytic}. As shown below, this property is crucial in the QC-QC case, since it ensures that the auxiliary systems do not record the order in which the channels are applied.

The last time step $n=N+1$ consists in applying the last combined internal QC-QC operator $\tilde V_{N+1}$, which transforms the target, auxiliary and control systems into a state in $\H_F\otimes\H_{\alpha_F}$, and in tracing out the final auxiliary system of the QC-QC in $\H_{\alpha_F}$. Contrary to Secs.~\ref{subsubsec::proof_simulation_QCQC} and~\ref{subsubsec::proof_extendedsimulation_QCQC}, the auxiliary systems in $\H_{\text{aux}_{t_{N+1}}}$ are not traced out, but are output as the outputs of the catalyst copies of the input channels.

For any QC-QC superchannel $\boldsymbol{S}_\mathsf{QC}:\bigotimes_{i=1}^N[\L(\H_{I_i})\to\L(\H_{O_i})]\to[\L(\H_{P})\to\L(\H_{F})]$, we can thus formally define the superchannel $\widehat{\boldsymbol{C}}:\bigotimes_{i=1}^N[\L(\H_{I_i})\to\L(\H_{O_i})]^{\otimes N}\to[\L(\H_{P}\otimes\H_{\text{aux}_{t_1}})\to\L(\H_{F}\otimes\H_{\text{aux}_{t_{N+1}}})]$ as the alternating composition, step by step, of the procedure described above, except the third step as it corresponds to the application of the quantum channels which are input in $\widehat{\boldsymbol{C}}$. As in Sec.~\ref{subsubsec::proof_simulation_QCQC}, since the operators $\tilde V_n$, $\hat S^I_n$ and $\hat S^O_n$ are isometries, the Choi operator of the superchannel $\widehat{\boldsymbol{C}}$ is described as
\begin{align}
\widehat{C} & = \tr_{\alpha_F}\big(\ketbra{\hat c}{\hat c}\big), \label{eq:output_state_simu_QCQC_catalytic} \\
\ket{\hat c} & \coloneqq \kket{\tilde V_1} * \kket{\hat Q_1} * \kket{\tilde V_2} * \kket{\hat Q_2}  * \cdots * \kket{\tilde V_N} * \kket{\hat Q_N} * \kket{\tilde V_{N+1}}. \label{eq:ket_c_simu_QCQC_catalytic}
\end{align}
Note that, contrary to Eqs.~\eqref{eq:ket_c_simu_QCQC} and~\eqref{eq:ket_c_simu_QCQC_extended}, no auxiliary state is prepared and no auxiliary register is traced out: the auxiliary spaces $\H_{\text{aux}_{t_1}}\cong\bigotimes_{k=1}^N\H_{I_k}^{\otimes(N-1)}$ and $\H_{\text{aux}_{t_{N+1}}}\cong\bigotimes_{k=1}^N\H_{O_k}^{\otimes(N-1)}$ remain open in Eq.~\eqref{eq:ket_c_simu_QCQC_catalytic}, and play, together with $\H_P$ and $\H_F$, the role of the global past and global future of $\widehat{\boldsymbol{C}}$.

That the combined operators $\tilde V_n$ are isometries follows from the normalization and orthogonality conditions satisfied by the operators $V_{K_{n-1},k_n}^{\to k_{n+1}}$, so that the corresponding maps $\rho\mapsto\tilde V_n\rho\tilde V_n^\dagger$ are CPTP. The routing operators $\hat S^I_n$ and $\hat S^O_n$ are controlled permutations of subsystems, hence unitary, and $\hat{\mathcal{Q}}_n$ is a valid parallel higher-order superchannel. Thus, at each time step, the composition of $\tilde V_n$ and $\hat{\mathcal{Q}}_n$, with Choi vector $\kket{\tilde V_n} * \kket{\hat Q_n}$, defines a valid parallel higher-order process, which can be interpreted as a QC-FO between the $N$ input channels. Finally, by Eqs.~\eqref{eq:output_state_simu_QCQC_catalytic} and~\eqref{eq:ket_c_simu_QCQC_catalytic}, the superchannel $\widehat{\boldsymbol{C}}$ is obtained by applying sequentially the blocks $\kket{\tilde V_1} * \kket{\hat Q_1}, \ldots, \kket{\tilde V_N} * \kket{\hat Q_N}$ and then $\kket{\tilde V_{N+1}}$, and tracing out $\H_{\alpha_F}$, the auxiliary systems being carried along as memory systems from the global past to the global future. Since each block $\kket{\tilde V_n} * \kket{\hat Q_n}$ describes a QC-FO, while $\tilde V_{N+1}$ and the partial trace define CPTP maps, this sequential composition is itself a QC-FO, i.e., $\widehat{\boldsymbol{C}}$ is a valid QC-FO superchannel.

\begin{figure*}[h!]
    \centering
    \includegraphics[width=\columnwidth]{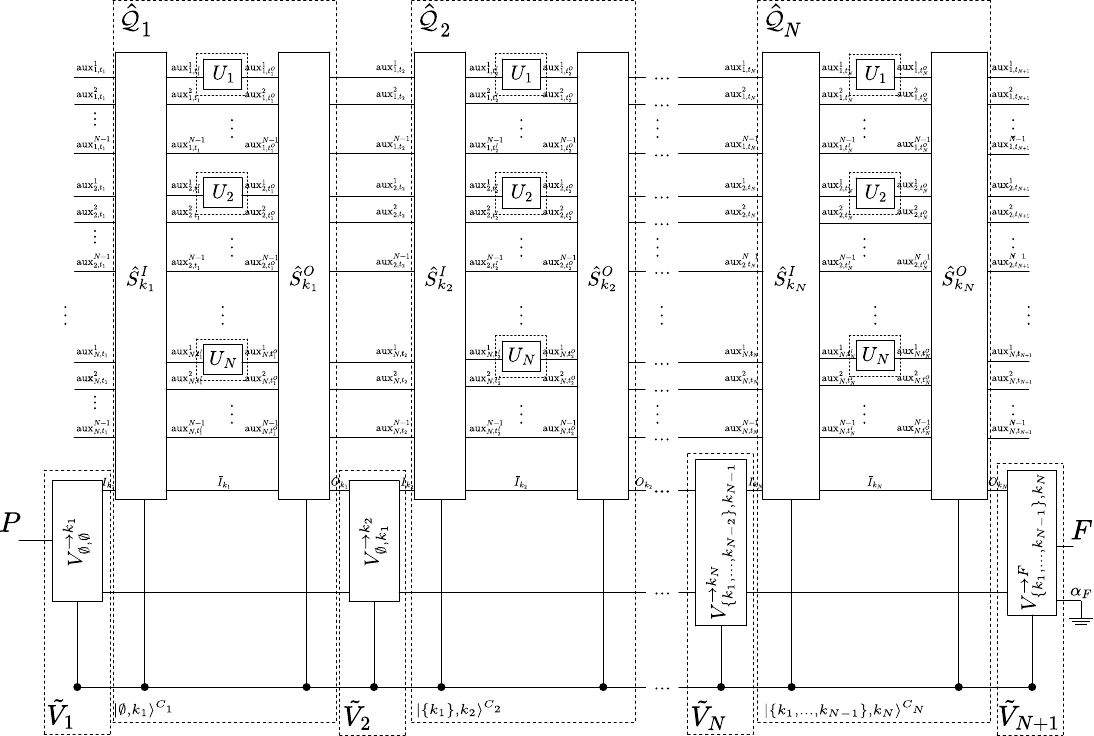}
    \caption{Catalytic fixed order simulation of a QC-QC with input unitary channels $\map{U}_1,\ldots,\map{U}_N$. $N$ queries of each unitary channel $\map{U}_k$ are made: at each time step, one copy of $\map{U}_k$ acts on the first wire $\text{aux}_k^1$ associated with $\map{U}_k$, which carries either the target system, if $\map{U}_k$ is selected, or an auxiliary system that has not yet received the action of $\map{U}_k$. The inputs and outputs of the $N(N-1)$ auxiliary systems are left open, so that the $N^2-N$ copies of the input channels that are not applied on the target system are returned as part of the output of the simulation.}
    \label{fig:simu_QCQC_catalytic}
\end{figure*}

\vspace{5mm}

We will now show that $\widehat{\boldsymbol{C}}$ applied on $N$ copies of each unitary channel is a catalytic simulation of the action of $\boldsymbol{S}_\mathsf{QC}$ on unitary channels, in the sense of Definition~\ref{def:simulation_catalytic}, for any choice of the input unitary channels. More formally, we want to show that
\begin{align}
    \widehat{\boldsymbol{C}}(\map{U}_{1}^{\otimes N}\otimes\ldots\otimes\map{U}_{N}^{\otimes N}) = \boldsymbol{S}_\mathsf{QC}(\map{U}_1\otimes\ldots\otimes\map{U}_N)\otimes\map{U}_1^{\otimes(N-1)}\otimes\ldots\otimes\map{U}_N^{\otimes(N-1)},
\end{align}
for all possible unitary channels $\map{U}_1,\ldots,\map{U}_N$, where the $N-1$ catalyst copies of each $\map{U}_k$ on the right-hand side act from $\H_{\text{aux}_{k,t_1}^i}$ to $\H_{\text{aux}_{k,t_{N+1}}^i}$, $i\in[N-1]$. Equivalently, we can rephrase the last equation in the Choi picture as
\begin{align}
    \widehat{C}*(\kketbra{U_1}{U_1}^{\otimes N}\otimes\ldots \otimes \kketbra{U_N}{U_N}^{\otimes N}) ={}& S_{\mathsf{QC}}*(\kketbra{U_1}{U_1} \otimes \ldots \otimes \kketbra{U_N}{U_N}) \nonumber \\
    &\otimes \kketbra{U_1}{U_1}^{\otimes(N-1)}\otimes\ldots\otimes \kketbra{U_N}{U_N}^{\otimes(N-1)},
    \label{eq:catalytic_statement_Choi_QCQC}
\end{align}
for all possible unitary operators $U_1,\ldots,U_N$.

The proof consists in computing iteratively the output of the simulation scheme presented in Fig.~\ref{fig:simu_QCQC_catalytic} and showing that it factorizes into the output of the QC-QC on unitary channels, as defined in Eq.~\eqref{eq:output_channel_QCQC(unitary}, and $N-1$ copies of each input unitary channel.

Following Definition~\ref{def:QC-QC}, any QC-QC superchannel $\boldsymbol{S}_\mathsf{QC}$ is defined by a collection of internal operators $\{V_{\emptyset,\emptyset}^{\to k_{1}},\ldots,\allowbreak \, V_{K_{n-1},k_n}^{\to k_{n+1}},\allowbreak\ldots, \, V_{K_{N-1},k_N}^{\to F}\}$.

As in Sec.~\ref{subsubsec::proof_simulation_QCQC}, since the input channels are unitary and the simulation scheme only involves isometries, the output of the simulation scheme can be computed at the level of Choi vectors. Since the auxiliary inputs of $\widehat{\boldsymbol{C}}$ are left open, we compute the output of the simulation scheme when arbitrary vectors $\ket{a_k^i}\in\H_{\text{aux}_{k,t_1}^i}$ are input in them, for $k\in[N]$ and $i\in[N-1]$, and we denote $\ket{\psi_{\text{aux}_{t_1}}}\coloneqq\bigotimes_{k=1}^N\bigotimes_{i=1}^{N-1}\ket{a_k^i} \in \H_{\text{aux}_{t_1}}$. We refer to the auxiliary system initially in position $i$ associated with $\map{U}_k$ as the auxiliary system carrying $\ket{a_k^i}$. The output of the simulation scheme can then be computed by composing iteratively the operations $\tilde V_n$, $\hat{\mathcal{Q}}_n$ and the unitary channels $\map{U}_1 \otimes \ldots \otimes \map{U}_N$. We describe by $\ket{\varphi_{(n)}'}$ (resp.~$\ket{\varphi_{(n+1)}}$) the simulation output at time step $n$, before (resp.~after) the operation $\tilde V_{n+1}$ is applied:
\begin{align}
\ket{\varphi_{(n)}'} & = \big(\ket{\psi_{\text{aux}_{t_1}}} * \kket{\tilde V_1} * \kket{\hat Q_1} *  \kket{\tilde V_2} * \cdots * \kket{\tilde V_n} * \kket{\hat Q_n}\big) * (\kket{U_{1}}^{\otimes {n}}\otimes\ldots \otimes \kket{U_{N}}^{\otimes {n}}), \notag
\end{align}
and
\begin{align}
\ket{\varphi_{(n+1)}} & = \big(\ket{\psi_{\text{aux}_{t_1}}} * \kket{\tilde V_1} * \kket{\hat Q_1} * \kket{\tilde V_2} * \cdots * \kket{\hat Q_n} * \kket{\tilde V_{n+1}}\big) * (\kket{U_{1}}^{\otimes {n}}\otimes\ldots \otimes \kket{U_{N}}^{\otimes {n}}).
\end{align}
At the first time step, right after the first internal operation $\tilde V_1$ has been applied, the simulation output is described by
\begin{align}
    \ket{\varphi_{(1)}} & = \ket{\psi_{\text{aux}_{t_1}}} * \kket{\tilde V_1} \nonumber \\
    &= \sum_{k_1} \bigotimes_{k=1}^N\bigotimes_{i=1}^{N-1} \ket{a_k^i} \otimes \kket{V_{\emptyset,\emptyset}^{\to k_{1}}} \otimes \ket{\emptyset,k_1}_{C_{1}}.
\end{align}
Applying the first unitary channels in parallel along with the routing operations, we obtain the Choi vector at the end of the first time step as
\begin{align}
    \ket{\varphi_{(1)}'} & = \ket{\varphi_{(1)}} * \kket{\hat Q_1} * (\kket{U_{1}}\otimes\ldots \otimes \kket{U_{N}}) = \ket{\varphi_{(1)}} * \kket{\hat S^I_1}*(\kket{U_1}\otimes \ldots \otimes \kket{U_N}) * \kket{\hat S^O_1} \nonumber \\
    & = \bigg[\sum_{k_1} \Big[ \bigotimes_{k=1}^N\bigotimes_{i=1}^{N-1} \ket{a_k^i} \otimes \kket{V_{\emptyset,\emptyset}^{\to k_{1}}} \Big] * \kket{\hat S^I_{k_1}}\otimes \ket{\emptyset,k_1}_{C^{\circ}_{1}} \bigg] *(\kket{U_1}\otimes \ldots \otimes \kket{U_N}) * \kket{\hat S^O_1}\nonumber \\
    & = \bigg[\sum_{k_1} \Big[ \bigotimes_{\substack{k=1\\ k\neq k_1}}^N \Big(\big(\ket{a_k^1} * \kket{U_k}\big) \otimes \bigotimes_{i=2}^{N-1}\ket{a_k^i}\Big) \otimes \bigotimes_{i=1}^{N-1}\ket{a_{k_1}^i} \otimes \big(\kket{V_{\emptyset,\emptyset}^{\to k_{1}}} * \kket{U_{k_1}}\big) \Big] \otimes \ket{\emptyset,k_1}_{C^{\circ}_{1}}\bigg] * \kket{\hat S^O_1} \nonumber \\
    & = \sum_{k_1} \Big[ \bigotimes_{\substack{k=1\\ k\neq k_1}}^N \Big(\big(\ket{a_k^1} * \kket{U_k}\big) \otimes \bigotimes_{i=2}^{N-1}\ket{a_k^i}\Big) \otimes \bigotimes_{i=1}^{N-1}\ket{a_{k_1}^i} \otimes \big(\kket{V_{\emptyset,\emptyset}^{\to k_{1}}} * \kket{U_{k_1}}\big) \Big] \otimes \ket{\emptyset,k_1}_{C'_{1}},
\end{align}
where $\ket{a_k^i}*\kket{U_k}=U_k\ket{a_k^i}$. In the third line, the auxiliary system carrying $\ket{a_{k_1}^1}$ is stored in $\H_{\bar I_{k_1}}$, while the copy of $\map{U}_{k_1}$ acts on the target system output by $V_{\emptyset,\emptyset}^{\to k_1}$, and each copy of $\map{U}_k$, $k\neq k_1$, acts on the auxiliary system in position 1 associated with $\map{U}_k$, which carries $\ket{a_k^1}$. In the last line, the target system output by the copy of $\map{U}_{k_1}$ is in $\H_{O_{k_1}}$ and the auxiliary system carrying $\ket{a_{k_1}^1}$ is back in $\H_{\text{aux}_{k_1,t_2}^1}$, while, for each $k\neq k_1$, the auxiliary systems associated with $\map{U}_k$ have been cyclically permuted: $\ket{a_k^1}*\kket{U_k}$ is now in $\H_{\text{aux}_{k,t_2}^{N-1}}$, and $\ket{a_k^i}$ is in $\H_{\text{aux}_{k,t_2}^{i-1}}$ for $i\in\{2,\ldots,N-1\}$. In particular, for $N\geq3$, the auxiliary system in position 1 associated with $\map{U}_k$, $k\neq k_1$, now carries $\ket{a_k^2}$, on which $\map{U}_k$ has not acted yet.

The output of the simulation scheme at the second time step before the unitary channels are applied is
\begin{align}
     \ket{\varphi_{(2)}} & = \ket{\varphi_{(1)}'} * \kket{\tilde V_2} \nonumber \\
     &= \sum_{k_1,k_2} \bigg[\bigotimes_{\substack{k=1\\ k\neq k_1}}^N \Big(\big(\ket{a_k^1} * \kket{U_k}\big) \otimes \bigotimes_{i=2}^{N-1}\ket{a_k^i}\Big) \otimes \bigotimes_{i=1}^{N-1}\ket{a_{k_1}^i} \otimes \big(\kket{V_{\emptyset,\emptyset}^{\to k_{1}}} * \kket{U_{k_1}} * \kket{V_{\emptyset,k_1}^{\to k_{2}}}\big) \bigg] \otimes \ket{\{k_1\},k_2}_{C_{2}}.
\end{align}
At the second time step, the copy of $\map{U}_{k_2}$ acts on the target system, while, for each $k\neq k_2$, the copy of $\map{U}_k$ acts on the auxiliary system in position 1 associated with $\map{U}_k$, which carries $\ket{a_k^2}$ if $k\neq k_1$, and $\ket{a_{k_1}^1}$ if $k=k_1$ (since the auxiliary systems associated with $\map{U}_{k_1}$ have not been permuted at the first time step). In both cases, $\map{U}_k$ acts on an auxiliary system on which it has not acted before. Hence, once the unitary channels and the routing operations have been applied, each unitary channel $\map{U}_k$ has acted on the auxiliary systems carrying $\ket{a_k^1}$ and $\ket{a_k^2}$ if $k\notin\{k_1,k_2\}$, and only on the auxiliary system carrying $\ket{a_k^1}$ if $k\in\{k_1,k_2\}$, the other copy of $\map{U}_k$ having been applied on the target system:
\begin{align}
     \ket{\varphi_{(2)}'} & = \ket{\varphi_{(2)}} * \kket{\hat Q_2} * (\kket{U_{1}}\otimes\ldots \otimes \kket{U_{N}})\nonumber \\
     &= \sum_{k_1,k_2} \bigg[\bigotimes_{\substack{k=1\\ k\notin\{k_1,k_2\}}}^N \Big(\bigotimes_{i=1}^{2}\big(\ket{a_k^i} * \kket{U_k}\big) \otimes \bigotimes_{i=3}^{N-1}\ket{a_k^i}\Big) \otimes \bigotimes_{j\in\{k_1,k_2\}} \Big(\big(\ket{a_j^1} * \kket{U_j}\big) \otimes \bigotimes_{i=2}^{N-1}\ket{a_j^i}\Big) \nonumber \\
     & \ \ \ \ \ \ \ \ \ \ \ \ \ \ \ \otimes \big(\kket{V_{\emptyset,\emptyset}^{\to k_{1}}} * \kket{U_{k_1}} * \kket{V_{\emptyset,k_1}^{\to k_{2}}} * \kket{U_{k_2}}\big) \bigg] \otimes \ket{\{k_1\},k_2}_{C_{2}'}.
\end{align}
We note that the auxiliary state, including the positions of the auxiliary systems, depends on the sequence $(k_1,k_2)$ only through the set $\{k_1,k_2\}$.
More generally, deriving iteratively the Choi vector after each time step, we obtain at the end of each time step $n\in[N]$
\begin{align}
     \ket{\varphi_{(n)}'} = \sum_{(k_1,\ldots,k_n)} \bigg[&\bigotimes_{k=1}^N \Big(\bigotimes_{i=1}^{s_k}\big(\ket{a_k^i} * \kket{U_k}\big) \otimes \bigotimes_{i=s_k+1}^{N-1}\ket{a_k^i}\Big) \nonumber \\
     &\otimes \big(\kket{V_{\emptyset,\emptyset}^{\to k_1}} \!*\! \kket{U_{k_1}} * \cdots \!*\! \kket{V_{K_{n-2},k_{n-1}}^{\to k_n}} \!*\! \kket{U_{k_n}}\big)\bigg] \otimes \ket{K_{n-1},k_n}_{C'_{n}},
     \label{eq:varphi_n_QCQC_cat}
\end{align}
where the notations are the same as in Eq.~\eqref{eq:varphi_n_QCQC}, and where, as in Eq.~\eqref{eq:rho_n_catalytic}, $s_k\in\{n-1,n\}$ is the number of time steps among the first $n$ at which $\map{U}_k$ has not been selected, i.e., $s_k=n-1$ if $k\in K_n$ and $s_k=n$ otherwise, and the auxiliary system carrying $\ket{a_k^i}$ is in $\H_{\text{aux}_{k,t_{n+1}}^{i-s_k}}$ if $i>s_k$, and in $\H_{\text{aux}_{k,t_{n+1}}^{i-s_k+N-1}}$ if $i\leq s_k$. Indeed, Eq.~\eqref{eq:varphi_n_QCQC_cat} holds for $n=1,2$, and if it holds at time step $n-1$, then at time step $n$, $\tilde V_n$ appends $\kket{V_{K_{n-2},k_{n-1}}^{\to k_n}}$ to each term and updates its control state into $\ket{K_{n-1},k_n}$, for all $k_n\in[N]\setminus K_{n-1}$; the copy of $\map{U}_{k_n}$ then acts on the target system, while the copy of each $\map{U}_k$, $k\neq k_n$, acts on the auxiliary system in position 1 associated with $\map{U}_k$, which carries $\ket{a_k^{s_k}}$ and has not yet received the action of $\map{U}_k$, before these auxiliary systems are cyclically permuted, and the auxiliary systems associated with $\map{U}_{k_n}$ are left untouched.

As in Sec.~\ref{subsubsec::proof_simulation_QCQC}, the auxiliary state in Eq.~\eqref{eq:varphi_n_QCQC_cat} depends on the sequence $(k_1,\ldots,k_n)$ only through the set $K_n$, which is determined by the control state $\ket{K_{n-1},k_n}$. Indeed, for each $k\in[N]$, both the number $s_k$ of auxiliary systems associated with $\map{U}_k$ that have received the action of $\map{U}_k$ and their positions only depend on whether $k\in K_n$. This follows from the fact that the routing operators only depend on the last label $k_n$ of the control state: the auxiliary systems associated with $\map{U}_k$ are cyclically permuted exactly at the time steps at which $\map{U}_k$ is not selected, whatever the order in which the other channels are applied. Hence, the terms of Eq.~\eqref{eq:varphi_n_QCQC_cat} that share the same control state $\ket{K_{n-1},k_n}$, which correspond to the different orders of the channels in $K_{n-1}$, come with the same auxiliary state, and their contributions add up at the level of probability amplitudes, as in Eq.~\eqref{eq::ket_s}.

At the second to last time step $N$, all the channels have been applied on the target system, i.e., $K_N=[N]$ and $s_k=N-1$ for all $k\in[N]$:
\begin{align}
     \ket{\varphi_{(N)}'} = \sum_{(k_1,\ldots,k_N)} \Big[&\bigotimes_{k=1}^N\bigotimes_{i=1}^{N-1} \big(\ket{a_k^i} * \kket{U_k}\big) \nonumber \\
     &\otimes \big(\kket{V_{\emptyset,\emptyset}^{\to k_1}} \!*\! \kket{U_{k_1}} * \cdots \!*\! \kket{V_{K_{N-2},k_{N-1}}^{\to k_N}} \!*\! \kket{U_{k_N}}\big)\Big] \otimes \ket{K_{N-1},k_N}_{C'_{N}}.
\end{align}
Each auxiliary system has thus received the action of the corresponding unitary channel exactly once, and is back in its initial position, i.e., $\ket{a_k^i} * \kket{U_k}\in\H_{\text{aux}_{k,t_{N+1}}^i}$, independently of the order $(k_1,\ldots,k_N)$.
Applying the last internal operation $\tilde V_{N+1}$:
\begin{align}
     &\ket{\varphi_{(N+1)}} = \ket{\varphi_{(N)}'} * \kket{\tilde V_{N+1}} \nonumber \\
     &= \sum_{(k_1,\ldots,k_N)} \bigotimes_{k=1}^N\bigotimes_{i=1}^{N-1} \big(\ket{a_k^i} * \kket{U_k}\big) \otimes \big(\kket{V_{\emptyset,\emptyset}^{\to k_1}} \!*\! \kket{U_{k_1}} * \cdots \!*\! \kket{V_{K_{N-2},k_{N-1}}^{\to k_N}} \!*\! \kket{U_{k_N}} \!*\! \kket{V_{K_{N-1},k_N}^{\to F}}\big) \nonumber \\
     &= \bigotimes_{k=1}^N\bigotimes_{i=1}^{N-1} \big(\ket{a_k^i} * \kket{U_k}\big) \otimes \left[\sum_{(k_1,\ldots,k_N)} \kket{V_{\emptyset,\emptyset}^{\to k_1}} \!*\! \kket{U_{k_1}} * \cdots \!*\! \kket{V_{K_{N-2},k_{N-1}}^{\to k_N}} \!*\! \kket{U_{k_N}} \!*\! \kket{V_{K_{N-1},k_N}^{\to F}}\right] \nonumber \\
     &= \bigotimes_{k=1}^N\bigotimes_{i=1}^{N-1} \big(\ket{a_k^i} * \kket{U_k}\big) \otimes \ket{s_{\boldsymbol{U}}},
\end{align}
where we used $K_{N-1}=[N]\setminus\{k_N\}$ and the definition of $\ket{s_{\boldsymbol{U}}}$ given below Eq.~\eqref{eq:output_channel_QCQC(unitary}. Since the final state of the auxiliary systems does not depend on the order $(k_1,\ldots,k_N)$, the output factorizes into $\ket{s_{\boldsymbol{U}}}$ and the vectors $\ket{a_k^i} * \kket{U_k}$, where $\kket{U_k}$ is the Choi vector of a copy of $\map{U}_k$ acting from $\H_{\text{aux}_{k,t_1}^i}$ to $\H_{\text{aux}_{k,t_{N+1}}^i}$. We have thus shown that
\begin{align}
    \ket{\psi_{\text{aux}_{t_1}}} * \Big[\ket{\hat c}*(\kket{U_{1}}^{\otimes N}\otimes\ldots \otimes \kket{U_{N}}^{\otimes N})\Big] = \ket{\psi_{\text{aux}_{t_1}}} * \Big[\ket{s_{\boldsymbol{U}}}\otimes \kket{U_1}^{\otimes(N-1)}\otimes\ldots\otimes \kket{U_N}^{\otimes(N-1)}\Big]
\end{align}
for all product vectors $\ket{\psi_{\text{aux}_{t_1}}}=\bigotimes_{k=1}^N\bigotimes_{i=1}^{N-1}\ket{a_k^i}$. Since these vectors span $\H_{\text{aux}_{t_1}}$, and since a vector of $\H_{\text{aux}_{t_1}}$ tensored with other Hilbert spaces is fully determined by its link products with all vectors of $\H_{\text{aux}_{t_1}}$, it follows that
\begin{align}
    \ket{\hat c}*(\kket{U_{1}}^{\otimes N}\otimes\ldots \otimes \kket{U_{N}}^{\otimes N}) = \ket{s_{\boldsymbol{U}}}\otimes \kket{U_1}^{\otimes(N-1)}\otimes\ldots\otimes \kket{U_N}^{\otimes(N-1)}.
\end{align}
Hence, as in Sec.~\ref{subsubsec::proof_simulation_QCQC}, Eq.~\eqref{eq:output_state_simu_QCQC_catalytic} gives
\begin{align}
    \widehat{C}*(\kketbra{U_1}{U_1}^{\otimes N}\otimes\ldots \otimes \kketbra{U_N}{U_N}^{\otimes N}) = \tr_{\alpha_F}\big(\ketbra{s_{\boldsymbol{U}}}{s_{\boldsymbol{U}}}\big)\otimes \kketbra{U_1}{U_1}^{\otimes(N-1)}\otimes\ldots\otimes \kketbra{U_N}{U_N}^{\otimes(N-1)},
\end{align}
which, by Eq.~\eqref{eq:output_channel_QCQC(unitary}, is Eq.~\eqref{eq:catalytic_statement_Choi_QCQC}, and concludes the proof.

Finally, as stated in Theorem~\ref{thm::qcqc_app}, this catalytic simulation can be straightforwardly combined with the extended simulation of Sec.~\ref{subsubsec::proof_extendedsimulation_QCQC}, so as to simultaneously perform an extended and a catalytic simulation of the action of $\boldsymbol{S}_\mathsf{QC}$ on unitary channels, still with $N$ queries of each extended unitary channel $\map{U}'_k:\L(\H_{I_k}\otimes\H_{I'_k})\to\L(\H_{O_k}\otimes\H_{O'_k})$. As in Sec.~\ref{subsubsec::proof_catalyticsimulation_QCCC}, it suffices to replace each auxiliary system in position $i$ associated with $\map{U}'_k$ by a pair of auxiliary systems in $\H_{\text{aux}_{k,t_n}^i}\otimes\H_{\text{aux}'^{i}_{k,t_n}}$, isomorphic to $\H_{I_k}\otimes\H_{I'_k}$, whose inputs and outputs are left open, to add the extension wires of Sec.~\ref{subsubsec::proof_extendedsimulation_QCQC}, and to let the routing operators permute these pairs as above, while the target system and the extension wire associated with $\map{U}'_{k_n}$ are swapped with the pair in position 1 associated with $\map{U}'_{k_n}$ as in Sec.~\ref{subsubsec::proof_extendedsimulation_QCQC}. Since the routing operators still only depend on the last label $k_n$ of the control state, the auxiliary state and the extension wires at the end of each time step $n$ again depend on the sequence $(k_1,\ldots,k_n)$ only through the set $K_n$, and the same computation as above then shows that the resulting fixed-order quantum circuit outputs $(\boldsymbol{S}_\mathsf{QC}\otimes\boldsymbol{I})(\map{U}'_1\otimes\ldots\otimes\map{U}'_N)\otimes\map{U}_1^{\prime\otimes(N-1)}\otimes\ldots\otimes\map{U}_N^{\prime\otimes(N-1)}$, each of the $N^2-N$ catalyst copies of the extended unitary channels acting on its own pair of auxiliary inputs and outputs.
\end{proof}
\section{Catalytic quantum-channel transformations}
\label{sec::catalysis}

Originally defined for pure states~\cite{jonathan1999entanglement}, a transformation of a bipartite quantum state $\ket{\psi_1}$ into another bipartite quantum state $\ket{\psi_2}$ via local operations and classical communication (LOCC) is deemed catalytic if 
\begin{align}
    \ket{\psi_1} &\not\mapsto \ket{\psi_2} \\
    &\text{but} \nonumber \\
    \ket{\psi_1}\otimes\ket{\phi} &\mapsto \ket{\psi_2}\otimes\ket{\phi},
\end{align}
where the bipartite quantum state $\ket{\phi}$ is called a quantum catalyst and is generally a function of $\ket{\psi_1}$ and $\ket{\psi_2}$. That is, a catalyst is a quantum state that enables an otherwise impossible transformation, but that is nevertheless preserved in the transformation. Typically, catalyst quantum states are not universal: a quantum state that acts as a catalyst in the transformation of $\ket{\psi_1}$ into $\ket{\psi_2}$ will not necessarily act as a catalyst in the transformation of some other state $\ket{\psi_3}$ into another state $\ket{\psi_4}$. A direct extension to mixed states is straightforward. Moreover, for mixed states, transformations that are enabled by a catalyst whose marginal is preserved but that can nonetheless become classically correlated or entangled with the main system have also been deemed catalytic~\cite{lipka2024catalysis}. A useful feature of catalytic transformations of quantum states is the reusability of the catalyst state: since the catalyst is not consumed during the transformation, but rather preserved intact, in an ideal scenario it can be reused in other identical transformations at later points in time as many times as desired.

Similarly, catalytic transformations of quantum channels have been recently studied in the literature~\cite{Yoshida23,bavaresco25}. In Ref.~\cite{Yoshida23}, which studies the task of transforming unknown qubit unitary channels $\map{U}$ into its inverse $\map{U}^\dagger$---represented at the level of the unitary operator as $f(U)=U^\dagger$---via quantum circuits, an impossible transformation between quantum channels was shown to be enabled by a catalyst quantum state. Namely, it was shown that
\begin{align}
    \map{U}^{\otimes 3} \not\mapsto \map{U}^{-1},
\end{align}
that is, no fixed circuit can implement unitary inversion deterministically and exactly for
every unknown qubit unitary using three calls. At the same time, this transformation is possible via a fixed quantum circuit using a catalyst, according to
\begin{align}
    \map{U}^{\otimes 3}\otimes\ket{\psi_U} \mapsto \map{U}^{-1} \otimes\ket{\psi_U},
\end{align}
where the catalyst state $\ket{\psi_U}\coloneqq(\id\otimes U)\ket{\Psi^-}$, with $\ket{\Psi^-}=(\ket{01}-\ket{10})/\sqrt{2}$, depends on the input of the transformation (as is usual in catalytic protocols), in this case, the unknown unitary $U$.

On the other hand, Ref.~\cite{bavaresco25} discusses how a transformation of quantum channels first defined in Ref.~\cite{Chiribella13} can be seen as catalytic---however, in this case, the catalyst itself is a quantum channel. The scenario there is the same as the one addressed here: a transformation of a quantum channel $\map{A}$ into $f(\map{A})$ that is not possible via a quantum circuit, i.e., 
\begin{align}
    \map{A} \not\mapsto f(\map{A}),
\end{align}
can be enabled with the assistance of a catalyst quantum channel $\map{C}_{\map{A}}$, as in
\begin{align}
    \map{A}\otimes \map{C}_{\map{A}} \mapsto f(\map{A})\otimes \map{C}_{\map{A}},
\end{align}
where $\map{C}_{\map{A}}$ depends on the input quantum channel $\map{A}$.
The quantum channel transformation in question in Ref.~\cite{Chiribella13} is that of the quantum switch: given two unitary channels $\map{U}_A$ and $\map{U}_B$, the output of the quantum switch $\boldsymbol{S}(\map{U}_A\otimes \map{U}_B)$ cannot be prepared, for all input unitaries, by a quantum circuit that queries each of them once; however, given access to a catalyst channel $\map{U}_A$, the transformation $\map{U}_A\otimes \map{U}_B\otimes \map{U}_A\mapsto \boldsymbol{S}(\map{U}_A\otimes \map{U}_B)\otimes \map{U}_A$ can be achieved via a quantum circuit.

\begin{figure*}[t]
\centering
    \includegraphics[width=\textwidth]{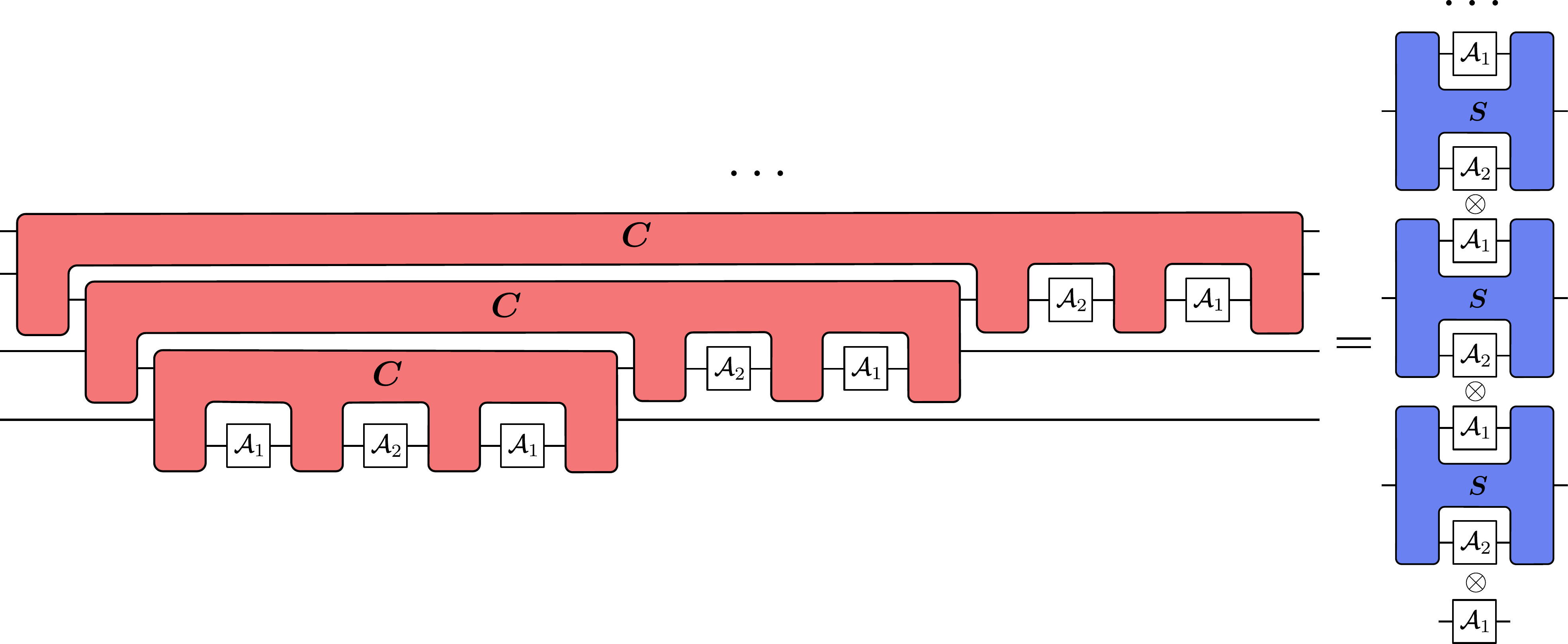}
    \caption{Reuse of a catalyst channel, illustrated for $N=2$ input channels. Each fixed-order quantum circuit $\boldsymbol{C}$ uses the query sequence $(1,2,1)$, i.e., two queries of $\map{A}_1$ and one query of $\map{A}_2$, to simulate the superchannel $\boldsymbol{S}$ catalytically: it outputs $\boldsymbol{S}(\map{A}_1\otimes\map{A}_2)$ together with an unchanged copy of $\map{A}_1$, whose input and output wires are left open. Since this catalyst copy acts as the channel $\map{A}_1$ itself, its open wires can be plugged into the first $\map{A}_1$ slot of a second circuit $\boldsymbol{C}$. That circuit then needs only one additional query of each channel to output another copy of $\boldsymbol{S}(\map{A}_1\otimes\map{A}_2)$ and, again, a catalyst copy of $\map{A}_1$. Iterating this nesting, $k$ copies of $\boldsymbol{S}(\map{A}_1\otimes\map{A}_2)$ are obtained from $k+1$ queries of $\map{A}_1$ and $k$ queries of $\map{A}_2$, with a single catalyst copy of $\map{A}_1$ left over at the end. Unlike a catalyst state, which can be reused at a later time, the catalyst channel is reused within a single larger circuit, all of whose parts must be initialized simultaneously.}
\label{fig::catalysis}
\end{figure*}

While, mathematically, catalytic transformations of quantum channels using catalyst channels are identical to catalytic transformations of quantum channels or states using catalyst states, the main difference concerns the \textit{reusability} of the catalyst channel, particularly in the case where the transformation is implemented by a quantum circuit. While a catalyst state can be reused to implement the same transformation on the subsequent copies of the input \textit{at a later time}, a catalyst quantum channel \textit{can} be used to implement the same transformation on more copies of the input, however, this transformation must be implemented \textit{simultaneously} to the transformation of the first copy. This idea is best illustrated in Fig.~\ref{fig::catalysis}. In order to effectively implement a transformation on a quantum channel, the entire circuit has to be initialized simultaneously. However, this does not forbid one to use the input and output lines corresponding to the catalyst channel as the channel itself, using it, for example, in an additional implementation of the desired function of the input channels. In this manner, the catalyst can be reused as many times as desired, as long as the overall combined circuit is simultaneously initialized on all of its input quantum systems.

\section{Efficiency and optimality}
\label{sec::optimality}

Given the explicit quantum-circuit constructions presented here, a natural question is whether they are efficient and optimal, i.e., what is the scaling of the required resources as a function of the number of channels $N$ and the dimension $d$ of the systems they act upon, and whether they require only the minimum necessary amount of resources for their implementation. We discuss this in the following in terms of the number of queries to the unknown input channels and in terms of the overhead in other resources that go in the quantum circuit implementation such as size of auxiliary systems and number of gates. 

\subsection{Quantum query, circuit, and space complexity}

As previously mentioned, to simulate any $\textup{\textsf{QC-CC}}$ acting on $N$ quantum channels that act on $d$-dimensional quantum systems, or $\textup{\textsf{QC-QC}}$ in the case of $d$-dimensional unitary channels, we prove that at most $N^2$ queries to the input quantum channels are needed. Hence, our simulations show an upper bound on quantum query complexity of $O(N^2)$. In addition to the extra number of queries to the input channels, there is also an overhead in terms of other quantum circuit resources that are necessary in the construction of our simulations. Here, we analyze their cost. As mentioned before, to implement any superchannel in $\textup{\textsf{QC-CC}}$ or $\textup{\textsf{QC-QC}}$ directly, aside from the ability to implement classical and quantum control of order, one must be able to implement all internal operations $\{\tilde{\M}_n\}_{n=1}^{N+1}$ and $\{\tilde V_n\}_{n=1}^{N+1}$, as well as the auxiliary systems $\alpha_n$. These resources are equally needed in the implementation of their simulations. To simulate these superchannels with a fixed-order quantum circuit, however, the incurred overhead to reproduce classical and quantum control of order is accounted for below. 

Both in the classical and the quantum control case, assuming $N$ possible input channels that map $d$-dimensional systems into $d$-dimensional systems, the resources needed for our simulations, in terms of auxiliary systems and controlled-permutation (CPERM) gates, are presented in Table~\ref{tab::circuitcosts}. The quantum query, space, and quantum circuit complexity of our simulations is presented in Table~\ref{tab::complexity}.

\begin{table*}[h!]
  \centering
  {\renewcommand{\arraystretch}{2}
  \begin{tabular}{| c || c | c | c |}
    \hline
    $\quad$Type of simulation$\quad$ & $\quad$Auxiliary dimension$\quad$ & $\quad$Number of CPERM gates$\quad$ & $\quad$Dimension of CPERM gates$\quad$ \\
    \hline
    \hline
    $\boldsymbol{C}\in\textup{\textsf{QC-FO}}$ & $N \times d$ & $2N$ & $ (N+1)\times d$ \\  
    \hline
    $\boldsymbol{C}^\prime\in\textup{\textsf{QC-FO}}$ (extended) & $3N \times d$ & $2N$ & $ (3N+1)\times d$ \\  
    \hline
    $\widehat{\boldsymbol{C}}\in\textup{\textsf{QC-FO}}$ (catalytic) & $(N^2-N) \times d$ & $2N$ & $ (N^2-N+1)\times d$ \\  
    \hline
  \end{tabular}
  }
  \caption{Resources used in our quantum-circuit simulations of any superchannel with classical or quantum control of computational order that acts on $N$ quantum channels (unitary ones, for quantum control) $\map{A}_i:\L(\H_{I_i})\to\L(\H_{O_i})$, with $\dim(\H_{I_i})=\dim(\H_{O_i})=d$.  For extended simulations, there $\map{A}_i:\L(\H_{I_i}\otimes\H_{I'_i}\to\L(\H_{O_i}\otimes\H_{O'_i})$, we also take $\dim(\H_{I'_i})=\dim(\H_{O'_i})=d$.}
  \label{tab::circuitcosts}
\end{table*}

The circuit complexity depends on the cost of implementing the $2N$ CPERM gates. Its simplest version, a controlled swap (CSWAP or Fredkin gate) that acts on two target qubits plus a control qubit can be implemented using two CNOT gates and a Toffoli gate. Hence, assuming a gate set containing Clifford and Tofolli gates, the CSWAP has a gate complexity of $3$, according to 
\begin{equation}
\begin{quantikz}
& \ctrl{2} & \midstick[3,brackets=none]{=}   &           & \ctrl{2} &           & \\
& \targX{} &                                 & \targ{}   & \ctrl{1} & \targ{}   & \\
& \targX{} &                                 & \ctrl{-1} & \targ{}  & \ctrl{-1} &
\end{quantikz}.
\end{equation}

Increasing the number of qubits in each wire from $1$ to $n$ linearly increases the number of necessary gates to $3n$. A general controlled-permutation gate (also called universal programmable permutation) acting on $M$ subsystems and coherently implementing any among $M!$ possible permutations can be implemented with a Beneš network using $O(M\log M)$ CSWAP gates~\cite{benes1964optimal,ratan2007quantum}. 

Hence, for a simulation that requires controlled-permutation gates that permute $M=N+1$ subsystems, the number of required CSWAPs is $O(N\log N)$, each requiring $O(n)=O(\log d)$ elementary gates to be implemented. Our simulation $\boldsymbol{C}$ requires $2N$ controlled-permutation gates, finally amounting to a quantum circuit complexity upper bound of $O(N^2\log N\log d)$. The scaling for extended and catalytic simulations can be analogously computed and is summarized in the table below.

In terms of space complexity, the simulation $\boldsymbol{C}$ requires $N$ auxiliary systems of dimension $d$, each represented by $\lceil\log d\rceil$ qubits. These auxiliary systems are reused throughout the circuit, yielding an overhead in space complexity of $O(N\log d)$. The extended and catalytic simulations require $2N$ and $N(N-1)$ auxiliary systems, respectively, giving space complexity upper bounds of $O(N\log d)$ and $O(N^2\log d)$, assuming that the extension auxiliary systems also have dimension at most $d$. These bounds count the additional auxiliary space introduced by the simulation and are summarized in the table below.

\begin{table*}[h!]
  \centering
  {\renewcommand{\arraystretch}{2}
  \begin{tabular}{| c || c | c | c |}
    \hline
    $\quad$Type of simulation$\quad$ & $\quad$Quantum query complexity$\quad$ & $\quad$Space complexity$\quad$ & $\quad$Quantum circuit complexity$\quad$\\
    \hline
    \hline
    $\boldsymbol{C}\in\textup{\textsf{QC-FO}}$ & $O(N^2)$ & $O(N\log d)$ & $O(N^2\log N\log d)$ \\  
    \hline
    $\boldsymbol{C}^\prime\in\textup{\textsf{QC-FO}}$ (extended) & $O(N^2)$ & $O(N\log d)$ & $O(N^2\log N\log d)$\\  
    \hline
    $\widehat{\boldsymbol{C}}\in\textup{\textsf{QC-FO}}$ (catalytic) & $O(N^2)$ & $O(N^2\log d)$ & $O(N^3\log N\log d)$  \\  
    \hline
  \end{tabular}
  }
  \caption{Summary of upper bounds on the quantum query, space, and quantum circuit complexity overhead of fixed-order quantum-circuit simulations of any superchannel with classical or quantum control of computational order that acts on $N$ quantum channels (unitary ones, for quantum control) $\map{A}_i:\L(\H_{I_i})\to\L(\H_{O_i})$, with $\dim(\H_{I_i})=\dim(\H_{O_i})=d$. For extended simulations, there $\map{A}_i:\L(\H_{I_i}\otimes\H_{I'_i}\to\L(\H_{O_i}\otimes\H_{O'_i})$, we also take $\dim(\H_{I'_i})=\dim(\H_{O'_i})=d$.}
  \label{tab::complexity}
\end{table*}

Consequently, we conclude that from the point of view of the notions of quantum query, space, and circuit complexity, our simulations are efficient, in both the number of slots $N$ of the simulated superchannel and the dimension $d$ of the input quantum channels (or alternatively, also on number of qubits $n=\log d$).

\subsection{Optimality}

The optimality of the quantum circuit complexity bounds and of the space complexity bounds depends on both the optimality of our construction in terms of the controlled-permutation gates and the optimality of the decomposition of these operations into elementary gates, which goes beyond the scope of this work. We now discuss in particular the optimality of our quantum query complexity upper bound. 

Our simulation uses $N^2$ queries to the input channels, establishing an upper bound on the query complexity of the class of superchannels considered here.

We conjecture that the $O(N^{2})$ query complexity of our simulation is asymptotically optimal in the worst case: we expect that any deterministic and exact simulation capable of realizing the simulations in Thms.~\ref{thm::QCCC} and~\ref{thm::qcqc} using a quantum circuit with fixed order requires $\Omega(N^{2})$ queries. This conjecture is consistent with that of Ref.~\cite{facchini15} for the unitary permutation problem and extends the expectation of quadratic query complexity to the broader setting studied here. It concerns the worst-case complexity of simulating the class as a whole, rather than the query complexity of each individual superchannel.

It is easy to see, however, that this conjectured optimality concerns the asymptotic scaling, rather than the exact number $N^{2}$ of queries. For $N=2$, the simulations of Thms.~\ref{thm::QCCC} and~\ref{thm::qcqc} can be implemented using only three queries: two to $\mathcal{A}_1$ and one to $\mathcal{A}_2$. This construction uses query sequence $(1,2,1)$. Likewise, for $N=3$, seven queries suffice: four to $\mathcal{A}_1$, two to $\mathcal{A}_2$, and one to $\mathcal{A}_3$, with query sequence $(1,2,1,3,1,2,1)$, or three to $\mathcal{A}_1$, three to $\mathcal{A}_2$, and one to $\mathcal{A}_3$, with query sequence (1,2,1,3,2,1,2), each of which contains every permutation of the corresponding input labels as a subsequence. 

\end{document}